\documentclass[aps,prb,twocolumn,showpacs,floatfix]{revtex4}
\newcommand{\tw}{t_{\rm w}}
\newcommand{\twone}{t_{\rm w_1}}
\newcommand{\twtwo}{t_{\rm w_2}}
\newcommand{\twthree}{t_{\rm w_3}}
\newcommand{\tws}{t_{\rm ws}}
\newcommand{\taurecw}{t_{\rm rec}^{\rm weak}}
\newcommand{\taurecs}{t_{\rm rec}^{\rm strong}}
\newcommand{\taurec}{t_{\rm rec}}

\newcommand{\Ts}{T_{\rm s}}
\newcommand{\ts}{t_{\rm s}}
\newcommand{\Tsun}{T_{\rm s_1}}
\newcommand{\Tsdeux}{T_{\rm s_2}}

\newcommand{\tpeak}{t_{\rm peak}}
\newcommand{\tmin}{t_{\rm min}}

\newcommand{\teff}{t_{\rm eff}}

\newcommand{\taup}{\tau_{\rm p}}
\newcommand{\tauh}{\tau_{\rm h}}

\newcommand{\Lo}{L_{0}}
\newcommand{\Lmin}{L_{\rm min}}

\newcommand{\Lovlp}{L_{c}(\delta)}
\newcommand{\LovlpT}{L_{\Delta T}}

\newcommand{\LovlpJ}{L_{\Delta J}(p)}
\newcommand{\Leff}{L_{\rm eff}}

\newcommand{\Tg}{T_{\rm g}}
\newcommand{\Tf}{T_{\rm f}}

\newcommand{\Ti}{T_{\rm i}}
\newcommand{\Tm}{T_{\rm m}}

\newcommand{\MZFC}{M_{\rm ZFC}}

\newcommand{\ising}{Fe$_{0.5}$Mn$_{0.5}$TiO$_3$ }
\newcommand{\AgMn}{{Ag(11 at\% Mn) }}
\newcommand{\agmn}{{Ag(11 at\% Mn) }}

\newcommand{\eq}[1]{Eq.~(\ref{#1})}
\newcommand{\kb}{k_{\rm B}}

\newcommand \be {\begin{equation}}
\newcommand \ee {\end{equation}}

\newcommand \bi {\begin{itemize}}
\newcommand \ei {\end{itemize}}

\usepackage{graphicx}
\usepackage{epsf}
\usepackage{amssymb}
\usepackage{array}

\begin{document}

\title{Spin Glasses: A Ghost Story}

\author{
P. E. J{\"o}nsson\cite{newaddressP}
}
\affiliation{
Department of Materials Science, Uppsala University,
Box 534, SE-751 21 Uppsala, Sweden
}

\author{
R. Mathieu\cite{newaddressR}
}
\affiliation{
Department of Materials Science, Uppsala University,
Box 534, SE-751 21 Uppsala, Sweden
}

\author{
H. Yoshino
}
\affiliation{
Department of Earth and Space Science, Faculty of Science,
Osaka University, Toyonaka, 560-0043 Osaka, Japan
}

\author{H. Aruga Katori}
\affiliation{
RIKEN, Hirosawa 2-1, Wako, Saitama, 351-0198, Japan
}
\author{A. Ito}
\affiliation{
RIKEN, Hirosawa 2-1, Wako, Saitama, 351-0198, Japan
}

\author{
P. Nordblad
}
\affiliation{
Department of Materials Science, Uppsala University,
Box 534, SE-751 21 Uppsala, Sweden
}

\date{\today}

\begin{abstract}
Extensive experimental and numerical studies of the non-equilibrium
dynamics of spin glasses subjected to temperature or bond
perturbations have been performed to investigate chaos and memory
effects in selected spin glass systems. Temperature shift and
cycling experiments were performed on the strongly anisotropic
Ising-like system {\ising} and the weakly anisotropic
Heisenberg-like system {\AgMn},  while bond shift and cycling
simulations were carried out on a 4 dimensional Ising
Edwards-Anderson spin glass. These spin glass systems display
qualitatively the same characteristic features and the observed
memory phenomena are found to be consistent with predictions from
the ghost domain scenario of the droplet scaling model.

\end{abstract}

\maketitle

\section{Introduction}

Spin glasses (SG) have been an active field of research for the last
three decades. Experimental and theoretical studies have revealed
an unexpected complexity to a deceivingly simple problem
formulation.\cite{canmyd72,edwand75,shekir75} Many open questions
still remain, not the least concerning the non-equilibrium
dynamics of the spin-glass phase.\cite{bouetal97,bouchaud2000}
Aging,\cite{lunetal83} rejuvenation and memory
\cite{jonetal98,jonetal99} are intriguing characteristics of the
non-equilibrium dynamics in spin glasses. Similar features are to
a certain extent found also in other glassy systems such as
orientational glasses,\cite{douetal99}
polymers,\cite{belcillar2000} strongly interacting nanoparticle
systems,\cite{mametal99,jonhannor2000},
colossal magnetoresistive manganites\cite{matsvenor2000}, and certain ceramic
superconductors.\cite{papetal99,garetal2003}

Aging itself can be found in much simpler systems like standard
phase separating systems. An example is a mixture of oil and
vinegar used in salad dressing. By strongly shaking
the mixture the system can be rejuvenated and slow growth of
equilibrium domains of oil and vinegar is observed afterward.
Then the first non-trivial question is what is growing in glassy systems
during aging. Moreover the aging observed in spin glasses
is very unusual in
several respects. First, the spin glass system can be strongly rejuvenated by an
extremely weak perturbation such as a small change of temperature.
Second, memory effects can be observed even after such strong rejuvenation.
These two aspects are in sharp contrast to the aging in simpler systems.
In the case of the oil $+$ vinegar system, one has to shake the
mixture strongly to rejuvenate it and such a strong
rejuvenation will  completely eliminate memories of the original mixture.

A natural physical picture for the aging, rejuvenation and memory
effects in spin-glasses are proposed by the droplet picture
\cite{bramoo87,fishus88eq,fishus88noneq,fishus91} and its recent
extension---the ghost domain
scenario.\cite{yoslembou2001,sheyosmaa,yos2003} This picture
includes concepts such as aging by domain growth, rejuvenation by
chaos with temperature or bond changes and memory by ghost domains.

The purpose of this study is to find out how chaos effects
are reflected on the nonequilibrium dynamics and to which
extent they  are relevant for spin glasses within the available time window.
It has been proposed in several recent
papers\cite{bouetal2001,dupetal2001,berbou2003,berbou2002}
that the memory effects
observed in experiments can be understood without the concept of
temperature-chaos, but
simply as due to successive freezing of smaller and smaller length
scales on cooling, in other words as the classical
Kovacs effect\cite{kovacs} observed in many glassy systems.
The experiments and simulations reported in this paper are
inspired by and inherit protocols, methods and ideas from Refs.
\onlinecite{yoslembou2001,jonyosnor2002, jonetal2002PRL,
jonyosnor2003,sheyosmaa,yos2003}. Our approach allows
us to distinguish between the classical Kovacs effect and
novel rejuvenation-memory effects due to temperature-chaos.

This article is organized as follows:
In Sec.~\ref{sec-theory} we discuss the consequences of chaotic perturbations
(temperature or bond changes)
on  the non-equilibrium dynamics of spin
glasses based on the ghost domain scenario.
 A special emphasis is put on how spin-glasses that have been
subjected to a strong perturbation gradually recover their original
spin structure. A brief introduction to experiments and simulations
is given in  Sec.~\ref{sec-exp}. Section~\ref{sec-shift} is devoted
to results from detailed temperature-shift experiments on the
{\ising} and {\AgMn} samples and bond-shift simulations on the 4
dimensional  Edwards-Anderson (EA) spin glass. Rejuvenation effects
after perturbations of various strengths are investigated in detail.
Section~\ref{sec-cycling} concerns one and two-step temperature
cycling experiments on the {\AgMn} sample as well as bond-cyclings
on the 4 dimension EA model. The interest is the recovery of memory
after various perturbations. In Sec.~\ref{sec_memory} results from
new memory experiments on {\ising} and {\AgMn} samples are reported
and the influence of cooling/heating rate effects are discussed.

\section{Theory}
\label{sec-theory}

\newcommand{\gstate}{\sigma_{i}^{(T,{\cal J)}}}
\renewcommand{\Lovlp}{L_{c}(\delta)}

In this section we present the theoretical basis for our
experimental and numerical studies on the dynamics of
spin-glasses and related randomly
frustrated systems subjected to perturbations such as
temperature($T$)-shifts/cyclings and bond-shifts/cyclings.

One major basis is the prediction of strong rejuvenation due to the
so-called chaos effects originally found within the droplet,
domain-wall scaling theory due to Bray-Moore and Fisher-Huse.
\cite{bramoo87,fishus88eq,fishus88noneq,fishus91} This theory
predicts that spin glasses are very sensitive to changes of their
environments. Even an infinitesimal change of temperature, or
equivalently an infinitesimal change of the bonds, will reorganize
the spin configuration toward completely different equilibrium
states. The existence of the anticipated chaos effects in the bulk
properties of certain glassy systems have been confirmed by some
theoretical and numerical studies e.g. on the Edwards-Anderson Ising
spin-glass models using the Migdal-Kadanoff renormalization group
method \cite{banbra87,neyhil93,abm02,sheyosmaa} and the mean-field
theory \cite{rizcri2003} as well as on directed polymers in random
media. \cite{salyos2002} However, numerical simulations on the EA
models on ``realistic'' lattices, such as the 3-dimensional cubic
lattice, remain inconclusive about the existence of the
temperature-chaos effect due to the lack of computational
power.\cite{rit94,ney98,bm00,picricrit2001,bm02,takhuk2002}
Furthermore, the link between the chaos effect  and the rejuvenation
effect observed in experiments remains to be clarified. It has been
argued that the chaos effect, even if it exists, may be irrelevant
at the length scales accessible on experimental time scales so that
the mechanism behind the rejuvenation found in experiments is of a
different origin.
\cite{bouchaud2000,komyostak2000A,berbou2002,berbou2003,berhol2002}
In order to shed light on these intriguing issues we study in detail
the crossover from weakly to strongly perturbed regimes of the chaos
effects following recent studies.
\cite{salyos2002,jonyosnor2003,sheyosmaa}

The other major theoretical basis is the ghost domain scenario,
\cite{yoslembou2001,yos2003} which suggests dynamical memory effects
which survive under strong rejuvenation due to the chaos effect.
This scenario explicitly takes into account the remanence of a sort
of symmetry breaking field or {\it bias} left in the spin
configuration of the system by which ``memory'' is imprinted and
retrieved dynamically. We carefully investigate the memory retrieval
process, called {\it healing}  of the original domain structure,
\cite{yos2003}  which takes a macroscopic time. In previous
studies, the importance of this process has not received enough
attention. A traditional interpretation of  the memory phenomena was
based on ``hierarchical phase space pictures''. \cite{vinetal96} In
the ghost domain scenario there is no need for such built-in, static
phase space structures. Some predictions of the ghost domain
scenario are also markedly different from conventional ``real
space'' pictures including earlier phenomenological theory such as
Koper and Hilhorst's \cite{kophil88} and others'
\cite{komyostak2000A,berbou2002} which do not account for the role
of the remanent bias.

It should be remarked that some basic assumptions of the droplet
picture are debated and alternative pictures have been proposed,
which include proposals of anomalously low energy excitations with
an apparent stiffness exponent $\theta=0$
.\cite{houmar2000,krzmar2000,palyou2000} However, in the present
paper we concentrate to work out detailed comparisons between the
theoretical outcomes of the original droplet theory and experimental
and numerical results.

\subsection{Edwards-Anderson model}
\label{subsec-model}

In this section we consider a $\pm J$ Edwards-Anderson  Ising
spin-glass model defined by the Hamiltonian,
\begin{equation}
 H = -\sum_{\langle ij\rangle} J_{ij}S_iS_j -h\sum_iS_i.
\label{eq-EA-model}
\end{equation}
The Ising spin $S_{i}$ is put on a lattice site $i$
($i=1,\ldots,N$) on a $d$-dimensional (hyper-)cubic lattice.
The interactions $J_{ij}$ are quenched random variables drawn with equal
probability among $\pm J$ with $J > 0$ and $h$ is an external field.
In the following we choose the Boltzmann's constant to be $\kb=1$
for simplicity.

The original form of the droplet theory only concerns Ising spin
glasses, but we assume that essentially the same picture also
applies for vector spin glasses. Recently it has e.g. been found
from a Migdal-Kadanoff renormalization group analysis that vector
spin-glasses exhibit qualitatively similar but quantitatively much
stronger chaos effects than Ising spin-glasses. \cite{krz2003}

\subsection{Overlaps between equilibrium states of different environments}
\label{subsec-overlap}

We assume that an equilibrium spin glass state $\Gamma^{(T,{\cal
J})}$ is represented by its {\it typical} spin configuration
specified as $\sqrt{q_{\rm EA}}(T)\sigma^{(T,{\cal J})}_{i}$ where
$i=1..N$. Here $T$ is a temperature below the spin-glass transition
temperature $T_{\rm g}$ and ${\cal J}$ represents a  given set of
bonds $J_{ij}$. The parameter $q_{\rm EA}(T)$ is the
Edwards-Anderson (EA) order parameter which is $1$ as $T \to 0$ and
decreases by increasing $T$ due to thermal fluctuations. Apart from
the thermal fluctuations parametrized by $q_{\rm EA}$, the typical
spin configuration is represented by the {\it backbone} spin
configuration  represented by quenched random variables $\{
\sigma^{(T,{\cal J})}_{i}\}$ which takes Ising values  $\pm 1$.
Furthermore, we assume that the only other possible equilibrium
state at the same environment $(T,{\cal J})$ is
$\bar{\Gamma}^{(T,{\cal J})}$ whose configuration is given by
$-\sqrt{q_{\rm EA}}(T)\sigma^{(T,{\cal J})}_{i}$.

The scaling theory \cite{bramoo87,fishus88eq,fishus88noneq} suggests
the chaos effect: The backbone spin configuration $\{
\sigma^{(T,{\cal J})}_{i}\}$ changes significantly by slight changes
of the environment $(T,{\cal J})$. Our interest in the present paper
is to investigate how this effect is reflected on the dynamics. Let
us consider two sets of different environments $A=(T,{\cal J})$ and
$B=(T',{\cal J}')$ which are specified from the following:
\begin{enumerate}
\item Temperature-change

The temperature-change simply means $T'=T+\Delta T$ with an {\it
infinitesimal} $\Delta T$ and  ${\cal J}={\cal J}'$.

\item Bond-change

The set of bonds ${\cal J}'$ is created from ${\cal J}$ by changing
the {\it sign} of an {\it infinitesimal} fraction  $p$ of ${\cal J}$
randomly while $T=T'$. As noted in Ref.~\onlinecite{ney98}  this
amounts to a perturbation of strength $\Delta J \sim J \sqrt{p}$.

\end{enumerate}

\begin{figure}[t]
\includegraphics[width=0.6\columnwidth]{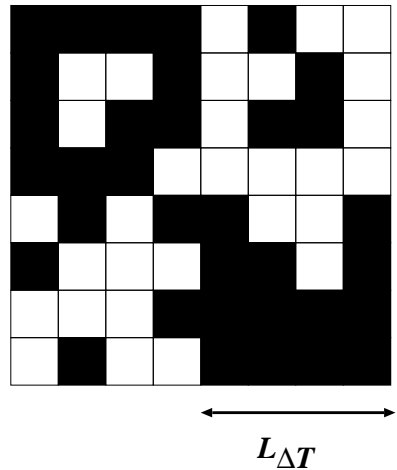}
\caption{A schematic picture of the configuration of the
local overlap $O^{AB}_{i}$ between the equilibrium state $\Gamma_{A}$
and  $\Gamma_{B}$.
The black and white represents $\pm 1$ values of the local overlap.
As a first approximation, the configuration of the local overlap can be
described as a collection of ``blocks''
of linear size  of the overlap length $\LovlpT$
within which the sign of the overlap
is biased to positive (black) or negative (white).
However there are minorities within the blocks
which have the opposite sign of
those of the majorities. 
}
\label{fig-overlap-conf}
\end{figure}

The relative differences between the backbone spin configurations
may be detected by introducing the {\it local overlap}
\begin{equation}
O^{AB}_{i}=\sigma^{A}_{i}\sigma^{B}_{i}
\label{eq-overlap}
\end{equation}
which takes Ising $\pm 1$ values. In Fig.~\ref{fig-overlap-conf}, a
schematic picture of the configuration of the local overlap $\{
O^{AB}_{i}\}$ is shown. The spatial pattern of the local overlap can
be decomposed into {\it blocks}: The value of the local overlap
$O^{AB}$ is essentially {\it uniform} (either $+1$ or $-1$) within a
given block while the values on different blocks are completely
uncorrelated. The correlation length of the local overlaps, which is
the typical size of the blocks, corresponds to what is called the
{\it overlap length} between $\Gamma_{A}$ and $\Gamma_{B}$. In the
case of temperature changes the overlap length is given by,
\begin{equation}
\LovlpT = \Lo \left(\frac{s(T) |\Delta T|}{\Upsilon(T)}\right)^{-1/\zeta}
\label{eq-lovlp-temp}
\end{equation}
where $\Lo$, $\Upsilon(T)$ and $\zeta(>0)$ are the unit length scale
the stiffness constant and the chaos exponent, respectively. The
factor $s(T)$ describes the temperature dependence of the 
droplet entropy \cite{abm02}. For the case of bond perturbations,
the overlap length associated with a bond perturbation of strength
$\Delta J$ is expected to scale as,
\begin{equation}
L_{\Delta J}= \Lo (|\Delta J|/J)^{-1/\zeta}
\label{eq-lovlp-bond}
\end{equation}
with\cite{ney98} $|\Delta J |\sim J \sqrt{p}$. Note that the chaos
exponent $\zeta$ is expected  to be the same for both temperature
and bond chaos. We emphasize that a strong bond-chaos effect induced
by an {\it infinitesimal} change of the bonds is as non-trivial as
the temperature chaos effect. It should be noted that such a
sensitive response to a perturbation does not happen in
non-frustrated systems such as simple ferromagnets.

The above picture is the simplest one which only takes into account
{\it typical} aspects of the chaos effect. Let us present an
improved discussion on how the bulk of the equilibrium spin glass
state is affected by perturbations on various length scales. It will
lead us to find a crossover from a {\it weakly perturbed regime} $L
\ll \LovlpT$ to a {\it strongly perturbed regime} $L \gg \LovlpT$.
For simplicity, let us decompose a block of size $\LovlpT^{d}$ into
smaller sub-blocks of linear size $L_{n} = \LovlpT/2^{n}$ with
$n=1,2,\ldots$. The {\it majority} of the sub-blocks will have a
common value of the local overlap $O^{AB}$ (either $+1$ or $-1$).
However, there will be {\it minority} sub-blocks which have the
opposite sign of the overlap with respect to the majority. The
probability $p_{\rm minor}$ that a sub-block belongs to such a
minority group is expected to be a function of the scaled length
$y=L_{n}/\LovlpT=2^{-n}$.
In the weakly perturbed regime $y \ll 1$
the probability scales as, \cite{salyos2002,sheyosmaa}
\be
p_{\rm minor}(y) \propto y^{\zeta} \qquad  y \ll 1
\label{eq-prob-event}
\ee
Note that $p_{\rm minor}$ is simply proportional to $|\Delta T|$ or
$|\Delta J|$ as can be seen by inserting \eq{eq-lovlp-temp} or
\eq{eq-lovlp-bond} in \eq{eq-prob-event}. The presence of the
minority phase is due to marginal droplets with vanishingly small
free-energy gap which easily responds to perturbations.
\cite{salyos2002,jonyosnor2003,sheyosmaa} For a detailed discussion
see Refs.~\onlinecite{salyos2002,sheyosmaa}. Note that the
probability $p_{\rm minor}(y)$ increases with increasing length
($y$) and saturates to $1$ as $y \to 1$.

\subsection{Relaxation in a fixed working environment}
\label{subsec-theory-iso}

Consider aging in a given working environment $A=(T_{A},{\cal
J}_{A})$ after a rapid temperature quench from above the spin glass
transition temperature $\Tg$ down to a working temperature $T_{A}$
below $\Tg$. We suppose that aging can be understood in terms of
{\it domain growth}. Let us start by giving a general definition of
a {\it domain}. First, we average out short time thermal
fluctuations on the temporal spin configuration $S_{i}(t)$. Then the
spin configuration at time $t$ may be represented by $\sqrt{q_{\rm
EA}(T_{A})}s_{i}(t)$. Here $s_{i}(t)$ is an Ising variable which
represents a coarse-grained temporal spin configuration. Second, we
project this spatio-temporal spin configuration onto any desired
reference equilibrium state $\Gamma_{\rm R}$ at an environment
$R=(T_{R},{\cal J}_{R})$, \be
\tilde{s}^{R}_{i}(t)=\sigma_{i}^{R}s_{i}(t). \label{eq-projection}
\ee
Now we define a {\it domain} with respect to the general reference
state $\Gamma_{\rm R}$. It has the following two essential
properties:
\begin{itemize}
\item A domain belonging to $\Gamma_{\rm R}$ (or  $\bar{\Gamma}_{R}$)
is a local region in the space within which the {\it sign} of the
projection $\tilde{s}^{R}_{i}$ is {\it biased} to either positive or
negative. The spatial variation of the sign of the bias defines the
geometrical organization of domains, i.e. domain wall configuration.
\item The {\it amplitude of the bias} in the interior of a domain
is the {\it order parameter} defined as
\be
\rho_{R}(t) = \left| \frac{1}{{\cal N}_{R}}
\sum_{i \in \mbox{ a domain of } \Gamma_{R}}\tilde{s}^{R}_{i}(t) \right|
\label{eq-def-op}
\ee
where ${\cal N}_{R}$ is the number of
spins belonging to the domain.
The amplitude of the bias can be smaller than $1$
indicating that the interior of the domain is ``ghostlike'' (see Fig. \ref{fig-ghosts})
\end{itemize}
To avoid confusion, let us note that the order parameter defined
above is different from the EA order parameter $q_{\rm EA}$ which
parametrizes the  thermal fluctuations on top of the backbone spin
configuration.

The natural choice for the reference equilibrium state is  $R=A$,
i.e. the equilibrium state $\Gamma_{\rm A}$, of the working
environment $A=(T_{A},{\cal J}_{A})$ itself. After time $t$ from the
temperature quench, domains belonging to $\Gamma_A$ and
$\bar{\Gamma}_A$ will have a certain typical size $L_{T_{A}}(t)$.
The domain size $L_{T_{A}}(t)$ will grow very slowly by activated
dynamics. Here the subscript $T_{A}$ is used to emphasize the 
temperature dependence of the  growth law [see \eq{eq-growth}]. (The
growth law is discussed in detail in appendix \ref{app-growthlaw}.)
Note in this context that the order parameter $\rho_{A}$
[\eq{eq-def-op}] within the domains belonging to $\Gamma_{\rm A}$
and $\bar{\Gamma}_A$ takes the maximum value $1$ constantly during
isothermal aging.

\begin{figure}[htb]
\includegraphics[width=0.9\columnwidth]{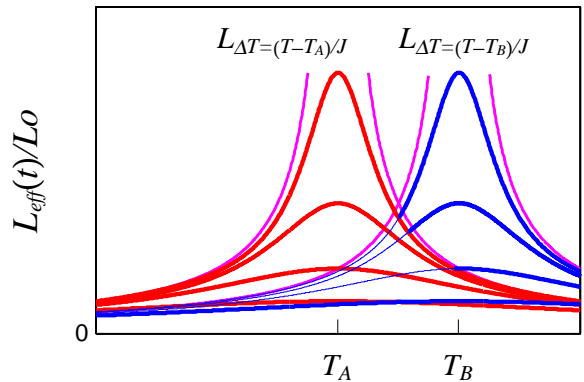}
\caption{(Color online) A schematic illustration of the 
temperature profiles of the effective domain size around
the working temperatures $T_{A}$ and $T_{B}$. The solid and dotted
lines represent the effective domain size $\Leff(t)$ at 
logarithmically separated times $t$ 
during isothermal aging at $T_{A}$ and $T_{B}$
respectively. 
}
\label{fig-leff}
\end{figure}

Let us now consider more generally what is happening on the
reference states associated with different environments during
isothermal aging, i.e. $R\neq A$ . As illustrated schematically
in Fig.~\ref{fig-leff}, the domains of the reference states at different
environments $R \neq A$ will grow with time up to the overlap length
between $\Gamma^{A}$ and $\Gamma^{R}$. Concerning this point it is
useful to note the relation between the projections
$\tilde{s}^{R}_{i}(t)$ and $\tilde{s}^{A}_{i}(t)$ given by, \be
\tilde{s}^{R}_{i}(t)=O^{AR}_{i}\tilde{s}^{A}_{i}(t), \label{eq-AB}
\ee which immediately follows from \eq{eq-projection}. Here
$O_{i}^{AR}$ is the overlap between the equilibrium state
$\Gamma_{A}$ and $\Gamma_{R}$ defined in \eq{eq-overlap}. The
spatial pattern of overlap $O_{i}^{AR}$ is roughly uniform over the
length scale of the overlap length $\LovlpT$ between $\Gamma_{A}$
and $\Gamma_{R}$. Beyond $\LovlpT$, however, the configuration of
the overlap $O_{i}^{AR}$ is random.  Then it follows that the growth
of domains belonging to $\Gamma_{A}$ ( and $\bar{\Gamma}_{A}$) only
contributes to the growth of domains belonging to $\Gamma_{R}$ (and
$\bar{\Gamma}_{R}$) up to the overlap length. Thus we expect that at
environment $R$ the typical domain size, which we denote as
effective domain size $\Leff(t)$, should scale
as,\cite{jonyosnor2002,jonyosnor2003}
\begin{equation}
\Leff(t) = \LovlpT F \left(\frac{L_{T_{A}}(t)}{\LovlpT}\right).
\label{eq-scaling-leff-1}
\end{equation}
The scaling function $F(x)$ reflects the following three regimes as
the age of the system, $\tw$, increases:
\begin{enumerate}
\item {\em Accumulative regime }

At length scales much shorter than the overlap length $\LovlpT$, the growth of the domains belonging to
$\Gamma_{A}$ leads to the growth of the domains belonging to
$\Gamma_{R}$ such that $\Leff(\tw)=L_{T_{A}}(\tw)$.
This yields
\begin{equation}
\lim_{x\to 0}F(x)= x.
\label{eq-scaling-leff-2}
\end{equation}

\item  {\it Weakly perturbed regime}

As we discussed in the previous subsection, the chaos effect emerges
gradually at length scales shorter than the overlap length. We now
want to determine the first correction term to
\eq{eq-scaling-leff-2} due to defects at the scale of
$L_{T_{A}}(t)$. Intuitively we assume that $\Leff$ is analytic and
an even function of $\Delta T$ (and $\Delta J$). We also expect that
the correction term is an analytic function of $p_{\rm minor}$ [See
\eq{eq-prob-event}] which is proportional to $|\Delta T|$ (and
$|\Delta J|$). Combining these we find that the fist correction term
should be $O(p^{2}_{\rm minor})$ leading
to\cite{foot-correction-term}
\begin{equation}
F(x)=x (1- c x^{2\zeta}) \qquad x \ll 1.
\label{eq-scaling-leff-3}
\end{equation}
 We note that because of defects at length scales $L \ll  L_{T_{A}}(t)$ the order parameter $\rho_R$ [\eq{eq-def-op}] is smaller than $\rho_A=1$.

\item {\em Strongly perturbed regime}

The domains belonging to $\Gamma_{R}$ can by aging at $A$ only grow
to the size of the upper bound $\LovlpT$, yielding
$\Leff(\tw)=\LovlpT$. This requires
\begin{equation}
F(x)=1 \qquad x \gg 1.
\label{eq-scaling-leff-4}
\end{equation}

\end{enumerate}

\subsection{Relaxation after shift of working environments}
\label{subsec-theory-shift}

Let us now consider a shift of working environment $A \to B$, after
aging the system in the environment $A$ for a time $\tw$.
$\Gamma_{B}$ is the reference state since the working environment is
now $B$. Just after the $T$-shift, the sizes of the domains
$\Gamma_{B}$ associated with the environment $B$ (at temperature
$T_{B}$) is given by the effective domain size $\Leff(\tw)$ in
\eq{eq-scaling-leff-1} (see Fig. \ref{fig-leff}). Thus the spin
configuration just after the temperature shift $T_{A} \to T_{B}$ is
equivalent to that after usual isothermal aging done at $B$ for a
certain waiting time $t_{\rm eff}$ (after direct temperature quench
from above $\Tg$ down to $T_{B}$). The {\it effective time}, $t_{\rm
eff}$, is defined through
\begin{equation}
L_{T_{B}}(t_{\rm eff})=\Leff(\tw).
\label{eq-leff-teff}
\end{equation}
Since $\Leff(\tw)$ is limited by the overlap length $\LovlpT$ {\it
rejuvenation} occurs after the temperature (or bond)- shifts: the
system looks {\it  younger}  [i.e. $L_{T_{B}}(t_{\rm
eff})<L_{T_{A}}(\tw)$] than it would have been if the aging was
fully accumulative [i.e. $L_{T_{B}}(t_{\rm eff})=L_{T_{A}}(\tw)$].

The effective time $\teff$ can be determined experimentally by
measuring the ZFC relaxation after temperature and bond shifts (see
Sec.~\ref{sec-shift}). As in the isothermal case, \cite{lunetal83} a
crossover occurs between quasi-equilibrium and out-of-equilibrium
dynamics at $L_{T_B}(t) \sim L_{T_{B}}(t_{\rm eff})$, where
$L_{T_B}(t)$ is the length scale on which the system is observed at
time $t$ after the temperature (or bond) shift.

\begin{figure*}[thb]
\includegraphics[width=0.9\textwidth]{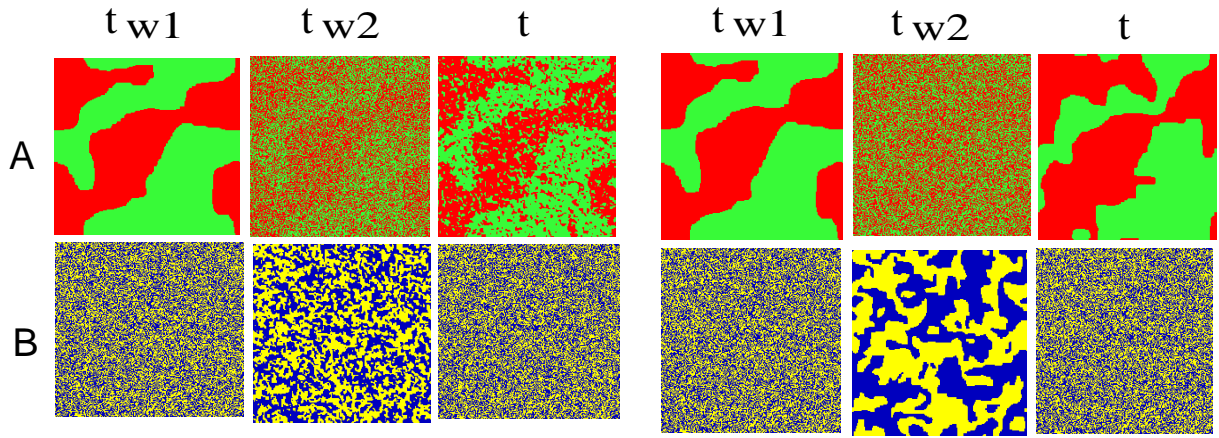}
\caption{(Color online) An illustration of the evolution of ghost domains in a
``perturbation-healing'' protocol at large length scales.
The above pictures are generated by an artificial simulation:
Monte Carlo simulation is performed on a set of Ising spins
on a square lattice by cycling the working Hamiltonian given by
Mattis models whose
ground states are exactly given by $\sigma^{A}$ and $\sigma^{B}$
(see the note\cite{footmattis} for the details).
The different colors represent the sign $+$
and $-$ of the projections $\tilde{s}^{A}_{i}(t)$ (top figures)
and  $\tilde{s}^{B}_{i}(t)$ (bottom figures) at the end of
initial aging state of duration $\tw$ , at the end of the
perturbation stage of duration $\taup$ and after time $\tauh$
in the healing stage.
Here the backbone spin configuration of the target equilibrium states
$\sigma^{A}$ and $\sigma^{B}$ are
completely unrelated random configurations so that the local overlap
$O^{AB}_{i}=\tilde{s}^{A}_{i}(t)\tilde{s}^{B}_{i}(t)=\sigma^{A}\sigma^{B}$
(see \eq{eq-AB}) is a completely random configuration:
the overlap length is just $1$ in the unit lattice.
The remanent bias at the end of the perturbation
stage is found as (left) $\rho_{\rm rem} \sim 0.15$ and (right)
$\rho_{\rm rem} \sim 0.04$. In the former the human eyes just
barely distinguish the ghost domains, whereas in the latter the
domains have become indistinguishable. In both cases, the strength
of the bias increases up to $O(1)$ in the healing stage: (left)
$\rho_{\rm rec} \sim 0.59$ and (right) $\rho_{\rm rec} \sim 0.58$.
} \label{fig-ghosts}
\end{figure*}

Now let us turn our attention to what happens to the amplitude of
order parameters within the domains after $T$-shifts,
\begin{itemize}

\item The order parameter $\rho_{B}$ within domains belonging to $\Gamma_{B}$
evolves as follows. Although the system is essentially equilibrated
with respect to $B$ up to the effective domain size $\Leff(\tw)$, it
contains some defects which cause a certain reduction of the order
parameter. These defects are progressively eliminated after the
$T$-shift and  $\rho_{B}$ approaches the full amplitude $1$. At time
$t$ after the shift of environment, such defects smaller than
$L_{T_{B}}(t)$ are eliminated, but the larger ones still remain.
Thus, this process finishes once defects as large as the domain size
$\Leff$ are removed. The time scale needed to finish this transient
process is given by \be L_{T_{B}}(t_{\rm trans})=\Leff
\label{eq-t-trans} \ee thus $t_{\rm trans} \sim \teff$ because of
\eq{eq-leff-teff}.

\item The order parameter within domains belonging to equilibrium states
at other temperatures $T \neq T_{B}$ evolves as follows. At length
scales greater than $L_{\Delta T=(T-T_{B})/T}$ (see
Fig.~\ref{fig-leff}) the domains grown before the $T$-shift suffer a
progressive reduction of the order parameter. However, the spatial
pattern of the {\it sign of the bias} remains the same as before the
$T$-shift (see Fig.~\ref{fig-ghosts}). This is a very important
point that we discuss in detail in sec. \ref{subsec-theory-cycle}.

\end{itemize}

Let us also remark that the population of thermally active droplets
at a certain scale $L$, which is proportional to  \cite{fishus88eq}
$(\kb T/J) (L/L_{0})^{-\theta}$, cannot follow sudden changes of the
working temperatures, but needs a certain time to be switched off
(or switched on). Progressive adjustments of the population of
thermally active droplets take place on the time scale $t_{\rm
trans}$ defined in Eq.~\ref{eq-t-trans}.

In experiments, the progressive elimination of the remanent defects
on domains belonging to $\Gamma_{B}$ after the shifts should give
rise to certain excessive contributions to the relaxation of the
magnetic susceptibility with the duration time $t_{\rm trans}$ given
above. The amplitude of the excessive response is expected to be
proportional to $\Delta T$ or $\Delta J$. This is because both the
population of the isolated defects due to the chaos effect in the
weakly perturbed regime (see \eq{eq-prob-event}) and the excessive
thermal droplets are proportional to $\Delta T$ or $\Delta J$.

\subsection{Relaxation under cyclings of working environments}
\label{subsec-theory-cycle}

We are now in position to discuss dynamics under cycling of working
environments such as temperature cycling. Here we consider the simplest protocol.
1) {\it Initial aging stage} - after a
temperature quench from above $\Tg$, the system is aged for a time $\twone$
within environment $A$ - ($T_{A},{\cal J}_{A}$).
2) {\it Perturbation stage} - Then a ``perturbation'' is applied and
the system is aged a time $\twtwo$ at the new environment $B$ -
($T_{B},{\cal J}_{B}$).  3) {\it Healing stage} - finally the
working environment is brought back to $A$ where the spin structure obtained in
the initial aging stage  slowly ``heals'' from the effects of the
perturbation.

\subsubsection{Weakly perturbed regime}
\label{subsubsec-cycle-weak}

Let us focus on what is happening on length scales shorter than the
overlap length $\LovlpT$ during the one-step cycling introduced
above. This  will be relevant for observations at correspondingly
short time scales in experiments and simulations. The effect of the
change of working environments should be {\it perturbative} in the
weakly perturbed regime $L < \LovlpT$ as we discuss below.

During the perturbation stage, the growth of domains belonging to
$\Gamma_B$ introduces rare defects of size $L$  on top of the
domains belonging to $\Gamma_A$ with a probability proportional to
$(L/\LovlpT)^{\zeta} \ll 1$ (see  \eq{eq-prob-event}). Since the
probability is low, these defects are isolated from each other.
During the healing stage, these island like objects are
progressively removed one by one from the small ones up to larger
ones. Since the maximum size of such a defect is $L_{B}(\twtwo)$, we
find that the removal of the defects will be completed in a time
scale $\taurecw$ given by \be L_{A}(\taurecw) = L_{B}(\twtwo).
\label{eq-tau-rec-weak} \ee Here the super-script ``weak'' indicates
that the formula is valid only in the weakly perturbed regime. After
this recovery time the original state of interior of the domains
belonging to $\Gamma_{B}$ is restored. Finally let us also note that
domain size at A continues to grow during the aging at B and
vice-versa within the weakly perturbed regime.

\subsubsection{Strongly perturbed regime}
\label{subsubsec-cycle-strong}

Suppose that the length scales explored during the three stages
$L_{T_{A}}(\twone)$, $L_{T_{B}}(\twtwo)$ and $L_{T_{A}}(t)$ are all
greater than the overlap lengths $\LovlpT$. Then, the strongly
perturbed regime of the chaos effect (see Sec.~\ref{subsec-overlap})
which appears at length scales greater than the overlap length
$\LovlpT$ should come into play.
Figure \ref{fig-ghosts} gives an illustration of  how projections
$\tilde{s}_{i}^{A}(t)$ and $\tilde{s}_{i}^{B}(t)$ evolves at length
scales greater than the overlap length. The unit of length scale is
chosen to be the overlap length $\LovlpT$. Thus the pattern of the
local overlaps $O^{AB}_{i}$ is completely random beyond the unit
length scale.

In the initial aging stage, the domains belonging to $\Gamma^{A}$
grow up to the size $L_{T_{A}}(\twone)$. The amplitude of the order
parameter within the domains has the full amplitude ($1$)
everywhere.

In the perturbation stage, the system is relaxing in the environment
$B$. Since the spin configuration before the perturbation stage is
completely random with respect to $B$ beyond the overlap length
$\LovlpT$, the domains belonging to $\Gamma^{B}$ grow independently
of the previous aging. In the meantime, the domains belonging to
$\Gamma^{A}$ become ``ghost'' like, i.e. they keep their overall
``shape'' while their interior becomes {\it noisy} due to the growth
of domains belonging to the reference state $B$.  The remanent
strength of the order parameter of $A$ defined in \eq{eq-def-op} \be
\rho_{\rm rem}(t) = \left| \frac{1}{{\cal N}_{A}} \sum_{i \in \mbox{
a domain of } \Gamma_{A}}\tilde{s}^{A}_{i}(t) \right| \ee slowly
decays with time $t$. Here ${\cal N}_{A} \sim
(L_{T_{A}}(\tw)/\Lo)^{d}$ is the number of
spins belonging to a domain of $\Gamma_{A}$.
After time $t$ from the beginning of the perturbation stage it
becomes, \cite{yoslembou2001,yos2003} \be \rho_{\rm rem}(t) \sim
\left( \frac{L_{T_B}(t)}{\LovlpT} \right)^{-\bar{\lambda}}
\label{eq-bias} \ee where $\bar{\lambda}$ is a dynamical exponent.
The crucially important point is that the spatial pattern of the
{\it sign} of the local bias (order parameter) - with decreasing
amplitude of strength $\rho_{\rm rem}(t)$ - is preserved during the
perturbation stages. We call this spatial structure a {\it ghost
domain}. Such ghost domains are responsible for the memory effect
allowed after full rejuvenation induced by the chaos effect.

In the healing stage, the domains belonging to $\Gamma_{A}$ start to
grow from the unit length scale $\LovlpT$ all over again. However,
the initial condition for this healing stage is {\it not} completely
random, since ghost domains with bias $\rho_{\rm rem}(\twtwo)$
exist. We thus need to consider domain growth with slightly biased
initial condition.\cite{BK92} The bias now {\it increases} with time
as,\cite{yos2003} \be \rho_{\rm rec}(t; \rho_{\rm rem}) \sim
\rho_{\rm rem}(\twtwo) \left(\frac{L_{T_A}(t)}{\LovlpT}
\right)^{\lambda}. \label{eq-q-increase} \ee Since the size of the
ghost domain itself is finite, it also continues to grow during the
healing stage. Following Ref.~\onlinecite{yoslembou2001} we call the
growth of the  bias inside a  ghost domain during the healing stage
{\it inner-coarsening} and the further growth of the size of the
ghost domain itself {\it outer-coarsening}.

The growth of the bias stops when the bias (order parameter)
saturates to the full amplitude.
This defines the {\it recovery time} $\taurecs$,
\be
\rho_{\rm rec}(\taurecs;\rho_{\rm rem}) \sim 1 \qquad L(\taurecs)/\LovlpT
\sim \rho_{\rm rem}^{-1/\lambda}.
\label{eq-q-tau-rec}
\ee
Combining this with \eq{eq-bias} the relation \cite{yos2003}
\be
\frac{L_{T_A}(\taurecs)}{\LovlpT}=\left(\frac{L_{T_B}(\twtwo)}{\LovlpT}
\right)^{\bar{\lambda}/\lambda}
\label{eq-tau-rec}
\ee
is obtained. Here, the super-script ``strong'' indicates that the
formula is valid only in the strongly perturbed regime.

An important remark is that the two exponents $\lambda$ and
$\bar{\lambda}$ are in general different. In
Ref.~\onlinecite{BK92}, a scaling relation
$\bar{\lambda}+\lambda=d$ was found. Only in the special case
$\lambda=\bar{\lambda}$, which happens e.g. in the spherical model
considered in Ref.~\onlinecite{yoslembou2001}, the relation
\eq{eq-tau-rec} is accidentally simplified to $L_{T_A}(\tau^{\rm
strong}_{rec})=L_{T_B}(\taup)$. Concerning the exponent
$\bar{\lambda}$ the inequality is proposed as,
\cite{fishus88noneq} $ d/2 \leq  \bar{\lambda} <  d. $
Combining
that with $\bar{\lambda}+\lambda=d$, a useful inequality is
obtained, \cite{yos2003}
\be
\bar{\lambda}/\lambda \geq 1.
\label{eq-lambda-ratio}
\ee
In Ref.~\onlinecite{yos2003}, the 4d
Ising EA model was studied and it was found that $\bar{\lambda}
\sim 3.2$ and $\lambda \sim 0.8$. In this case and also in general
for 3d systems the recovery time is very large $\taurec^{\rm strong} \gg \taurec^{weak}$.

The above results are markedly different from what one would expect
from conventional ``real space'' or ``phase space'' arguments which
neglect the role of bias. If such a mechanism causing a symmetry
breaking is absent, the ``new domains'' grown during the healing
could often have the wrong sign of bias with respect to the original
one leading to total erasure of memory. In contrast to the
hierarchical phase space models, \cite{vinetal96} the ghost domain
scenario predicts memory also in the positive $T$-cycling ($T_{A} <
T_{B}$) case. Indeed such examples are already reported (see Fig 6.
in Ref.~\onlinecite{vinetal96} and Fig. 3 in Ref.
~\onlinecite{Gbergetal90}).

\subsubsection{Multiplicative rejuvenation effect}
\label{subsub-multicucl}

The above considerations for the one step cycling can be extended to
multi-step cycling cases. At large length/time scales beyond the
overlap length, some quite counter-intuitive predictions follow due
to the multiplicative nature of the noise
effect.\cite{yoslembou2001} For example, in a $T$-cycling $T_{A}
(\twone) \to T_{B} (\twtwo) \to T_{C} (\twthree) \to T_{A}$, the
amplitude of the order parameter of $A$ is reduced to $ \rho_{\rm
rem} \sim  \rho_{B} \times \rho_{C} $ in the end of the 2nd
perturbation. Here $\rho_{B}$ and $\rho_{C}$ represent reductions
due to domain growth at $B$ and $C$. Thus, the recovery time
$\taurecs$ of the order parameter of $A$ becomes, \be
\frac{L_{T_A}(\taurecs)}{\LovlpT} =
\left(\frac{L_{T_B}(\twtwo)}{\LovlpT}\frac{L_{T_C}(\twthree)}{\LovlpT}\right)^{\bar{\lambda}/\lambda}.
\label{eq-tau-rec-multi} \ee This tells that sequential short time
perturbations can cause huge recovery times. Let us note that this
multiplicative rejuvenation effect is absent at length/time scales
smaller than the overlap length $\LovlpT$. At such short length
scales, changes of temperature (or bonds) only amount to put
isolated island like defects on the domains which rarely overlap
with each other. The multiplicative effect discussed above is hardly
expected within conventional pictures which do not contain the time
evolution of the bias (order parameter).

\subsection{Freezing of aging  by slow changes
of working environments - heating/cooling  rate effects}
\label{sec-heating-cooling}

The effect of finite heating/cooling rate is a very important
problem from an experimental point of view. Even the fastest
cooling/heating rates such as those used in ``temperature quench''
experiments are always extremely slow compared to the atomic spin
flip time $\tau_{\rm m} \sim \hbar/J \sim 10^{-13}$ s. Typically the
maximum experimental heating/cooling rate is $v_{T} \sim O(10^{-1})$
K/s and $\Tg \sim O(10)$ K (see Appendix ~\ref{app-exp}), which in
simulations would be equivalent to an extremely slow heating/cooling
rate of $v_{T} \sim 10^{-15} \Tg/{\rm MCS}$. Thus, the instantaneous
changes of the working environments assumed in the previous sections
are very unrealistic, at least experimentally.

Nonetheless, we expect that the effect of a finite heating/cooling
rate can be taken into account by introducing a characteristic
length scale $L_{\rm min}(v_T,T)$. As discussed in
Ref.~\onlinecite{yos2003}, it is natural to expect that the
competition between accumulative and chaotic (rejuvenation)
processes during a slow change of working environments, either
temperature or bond changes, results in a sort of freezing of aging,
such that the effective domain size with respect to the temporal
working environment becomes a constant  in time $L_{\rm
min}(v_T,T)$. This domain size is expected to decrease when the rate
of the changes increases, e.g. the heating or cooling rate $v_{T}$.
The characteristic length $\Lmin(v_{T},T)$ can be seen as a {\it
renormalized overlap length} in the following senses:
\begin{itemize}
\item The frozen age $\tmin$ given by $L_{T}(\tmin)=\Lmin(v_{T},T)$
sets the minimum
time scale for the start of the domain growth when the target
temperature $T$ is just reached after continuous temperature changes.
This effect can be readily seen experimentally
as we discuss in Secs.~\ref{sec-T-shift} and \ref{app-teff}.

\item  The overlap length  $\LovlpT$ appearing
in equations \eq{eq-tau-rec} and \eq{eq-tau-rec-multi} should be
replaced by $\Lmin(v_{T},T)$. The latter should be greater than the
overlap length $\LovlpT$ between the equilibrium states at two
temperatures connected by a continuous temperature change. Thus, the
finiteness of the heating/cooling rate can significantly reduce the
recovery times for the memories.
\end{itemize}

Indeed, it is known from previous experimental studies that
cooling/heating rate effects are non-accumulative and that it is
only the rate very close to the target temperature that affects the
observed isothermal aging behavior.\cite{jonetal98,jonetal99}
However, within this narrow temperature region around the target
temperature, heating/cooling rate effects were found to be relevant,
but as long as the employed observation time is made long even a
cooling with $v_{T} \sim O(10^{-1})$~K/s can be regarded as a
``quench'' that allows experimental observations of non-equilibrium
dynamical scaling properties of isothermal aging starting from a
random spin configuration. \cite{vinetal96,jonetal2002PRL} An
interpretation of this feature would be that the observation times
used in the experiments are actually larger than the effective age
$\tmin$ related to the renormalized overlap length. This remarkable
feature is quite different from that of other glassy systems
governed by simple thermal slowing down, such as super-cooled
liquids,\cite{lehnag98} in which the effect of aging at different
temperatures during the heating/cooling only add up accumulatively.
It may be argued that the unusual cooling rate {\it independence} of
the dynamics in spin glasses already in itself supports the
relevance of the temperature-chaos concept.

\subsection{Comparisons to the classical Kovacs effect}
\label{sec-kovacs}

The mechanisms and nature of the memory effects involving length
scales shorter and longer than the overlap length are markedly
different within the ghost domain scenario.
After a  change of working environments involving only
 short length scales the chaos
effect is weak (perturbative) and the system is not rejuvenated.
Such  non-chaotic {\it perturbative} effects yield, as discussed in
Sec.~\ref{subsubsec-cycle-weak}, the trivial recovery time given by
\eq{eq-tau-rec-weak}. This recovery time fits well with conventional
intuition based on ``length scales'' or ``phase space''. Such
effects may be understood as classical memory effects in a {\it
fixed energy landscape} known as Kovacs effects in polymer glasses,\cite{kovacs}
which do not accompany a real rejuvenation. 
It is reasonable to expect that Kovacs effects exist
in a broad class of systems \cite{bouchaud2000,berbou2002,beretal,bbdg2003}
as far as the chaos effects are absent or irrelevant.
The fixed energy landscape picture is valid only at
length scales shorter than the overlap length where defects are
approximately independent of each other. For perturbations involving
length scales greater than $\LovlpT$ this picture becomes totally
invalid and the scenario outlined above for the strongly perturbed
regime is required.

\section{Introduction to experiments and simulations}

{\it Experiments. ---}  The non-equilibrium dynamics of two
canonical spin glasses was investigated using superconducting
quantum interference device (SQUID) magnetometry. The samples are
a single crystal of \ising ($\Tg \approx 21.3$~K), the standard
Ising spin-glass,\cite{itoetal86,itoetal90} and  a polycrystalline
sample of the Heisenberg \agmn spin-glass ($\Tg \approx
32.8$~K). The experiments were performed in non-commercial SQUID
magnetometers,\cite{magetal97} designed for low-field measurements
and optimum temperature control (see Appendix~\ref{app-exp} for
details).

Experimental results for the two samples will be presented using the following thermal procedures:

\begin{enumerate}
\item[(i)]
In an isothermal aging experiment the sample is quenched (cooled as
rapidly as possible) from a temperature above the transition
temperature to the measurement temperature $\Tm$ in the spin glass
phase. In dc measurements, the relaxation of the ZFC magnetization
is recorded in a small probing field after aging the sample in zero
field for a time $\tw$ at the measurement temperature.
\item[(ii)]
In a $T$-shift experiment (Sec.~\ref{sec-T-shift}) the sample is,
after a quench, first aged at $\Ti$ before measuring the relaxation
at $\Tm=\Ti+\Delta T$.
\item[(iii)]
In a $T$-cycling experiment (Sec.~\ref{sec_T_cycling}) the sample is
first aged at  $\Tm$, then at $\Tm + \Delta T$ and subsequently the
relaxation is recorded at  $\Tm$.
\item[(iv)]
In a temperature memory experiment\cite{jonetal98,matetal2001}
(Sec.~\ref{sec_memory}) the system is cooled from a temperature
above the transition temperature to the lowest temperature. The
cooling is interrupted at one or several temperatures in the spin
glass phase. The magnetization is subsequently recorded on
re-heating.
\end{enumerate}

In experiments (i)--(iii) all temperature changes are made with
the maximum cooling rate ($v_{\rm cool} \sim 0.05$~K/s) or maximum
heating rate ($v_{\rm heat} \sim 0.5$~K/s). A temperature quench
therefore refers to a cooling with the maximum cooling rate. The
magnetic fields employed in the experiments are small enough ($h
\approx 0.1 - 1$ Oe) to ensure a linear response\cite{matetal2001} of the system.

\label{sec-exp}
\begin{figure}[thb]
\includegraphics[width=0.9\columnwidth]{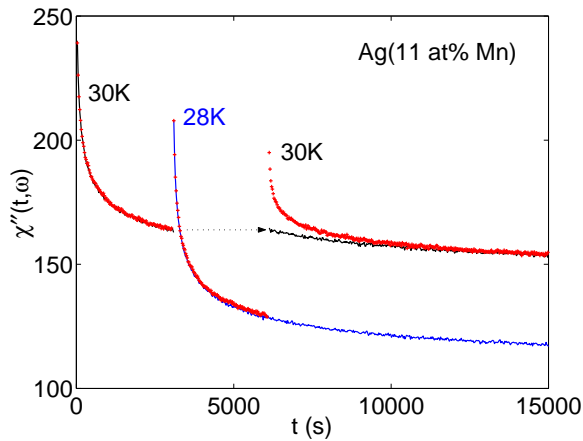}
\caption{(Color online) Ac susceptibility vs time after a
temperature quench. A temperature cycling from 30~K to 28~K and back
to 30~K is shown as pluses. The solid lines are the isothermal aging
at 28~K and 30~K. After the negative $T$-shift from 30~K to 28~K the
system is {\em completely rejuvenated}. Still, after the
 $T$-cycling it keeps a {\em memory} of the first aging at
 30~K. $f=510$~mHz.
}
\label{rm}
\end{figure}
To shortly illustrate the rejuvenation and memory phenomena, that
are examined in detail in this article, we show in Fig.~\ref{rm} a
negative $T$-cycling experiment on the \agmn sample. The
ac-susceptibility is measured as function of time after a
temperature quench. The system is aged at 30~K for 3000~s,
afterward the temperature is changed to 28~K for 3000~s, and
finally it is changed back to 30~K. The isothermal aging at 30~K
and 28~K are shown as references. We first
notice that the aging at 28~K is the same in the cycling
experiment as after a direct quench. Hence, at 28 K, the system
appears unaffected by the previous  aging at 30~K---the system is
{\em completely rejuvenated} at the lower temperature.
However, some time ($\taurec$) after changing the temperature back
to 30 K  the ac signal is the same as if the aging at 28~K had not
taken place; despite the rejuvenation at 28~K the system keeps a
{\em memory} of the aging at 30~K. One of our major interests is to
clarify if  the memory recovery process with duration $\taurec$ can
be understood in terms of the ghost domain picture
(Sec.~\ref{sec-theory}).

{\it Simulations. --} Standard heat-bath Monte Carlo simulations of
the EA Ising model (introduced in Sec.~\ref{subsec-model}) on the 4
dimensional hyper cubic lattice ($T_{\rm g}=2.0J$)\cite{bercam97}
were performed following protocols (i)-(iii), but using a bond
change ${\cal J}_{A} \to {\cal J}_{B}$ instead of a temperature
change. The initial conditions are random spin configurations to
mimic  a direct temperature quench (with $v_T=\infty$) to $T<T_{\rm
g}$. Detailed comparisons of the effects of bond perturbations and
temperature changes allows us to clarify the common mechanism of
rejuvenation and memory effects. An important advantage of the
numerical approach is that the dynamical length scale $L_{T}(t)$ can
be obtained directly in the
simulations.\cite{rieger95,yoshuktak2002} Details of the model and
the simulation methods are given in Appendix \ref{app-simulation}.

\section{Rejuvenation effects after temperature and bond shifts}
\label{sec-shift}

In this section, we present a quantitative study of
the rejuvenation effect using $T$-shift experiments and bond-shift simulations.

\subsection{Temperature-shift experiments}
\label{sec-T-shift}

\begin{figure}[thb]
\includegraphics[width=0.9\columnwidth]{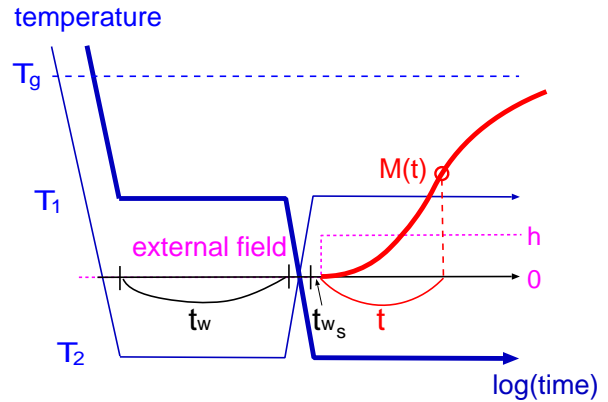}
\caption{(Color online) Schematic representation of a {\it twin} $T$-shift
experiment. In the experiment represented using the thick line,
$(\Ti,\Tm)=(T_{1},T_{2})$, while $(\Ti,\Tm)=(T_{2},T_{1})$ in the
experiment represented by the thin line. In ac experiments,
$\chi(t)$ is also monitored during the halt at $\Ti$.
In order to ensure thermal stability at $\Tm$, a short waiting time $\tws=3$~s is needed  before applying the magnetic field and recording the ZFC relaxation.\cite{footstab}
\label{twinTshift}}
\end{figure}

In a $T$-shift experiment, the system is quenched to the initial
temperature $\Ti (< \Tg)$, where it is aged a certain time $\tw$;
the temperature is changed to the measurement temperature
$\Tm=\Ti+\Delta T$, and ZFC-relaxation $M(t)$ (or ac
susceptibility relaxation) is recorded. Conjugate experiments with
$\Ti$ and $\Tm$ being interchanged are called {\em twin} $T$-shift
experiments (see Fig.~\ref{twinTshift}).

The effective age ($\teff$) of the SG system at $\Tm$, due to the
previous aging at $\Ti$, can be determined either from the maximum
in the relaxation rate $S(t)=h^{-1} d M(t)/d \log t$ of ZFC relaxation measurements,
\cite{graetal88,sanetal88,jonyosnor2002} or by the amount of time
that an
ac susceptibility relaxation curve measured
after a $T$-shift, needs to be shifted to merge with the reference relaxation curve
measured without a $T$-shift.
\cite{mametal99,dupetal2001,takhuk2002}
We will here show that
both ways to derive $\teff$ are consistent, discuss the
experimental limitations that determine the accuracy of the
estimations and indicate in which time window the derived
effective age gives non-trivial information. Finally, we will use
the extracted $\teff$ data to quantitatively analyze the emergence
of the chaos effect.

\subsubsection{ZFC relaxation after $T$-shifts}

\begin{figure*}[htb]
\includegraphics[height=9cm]{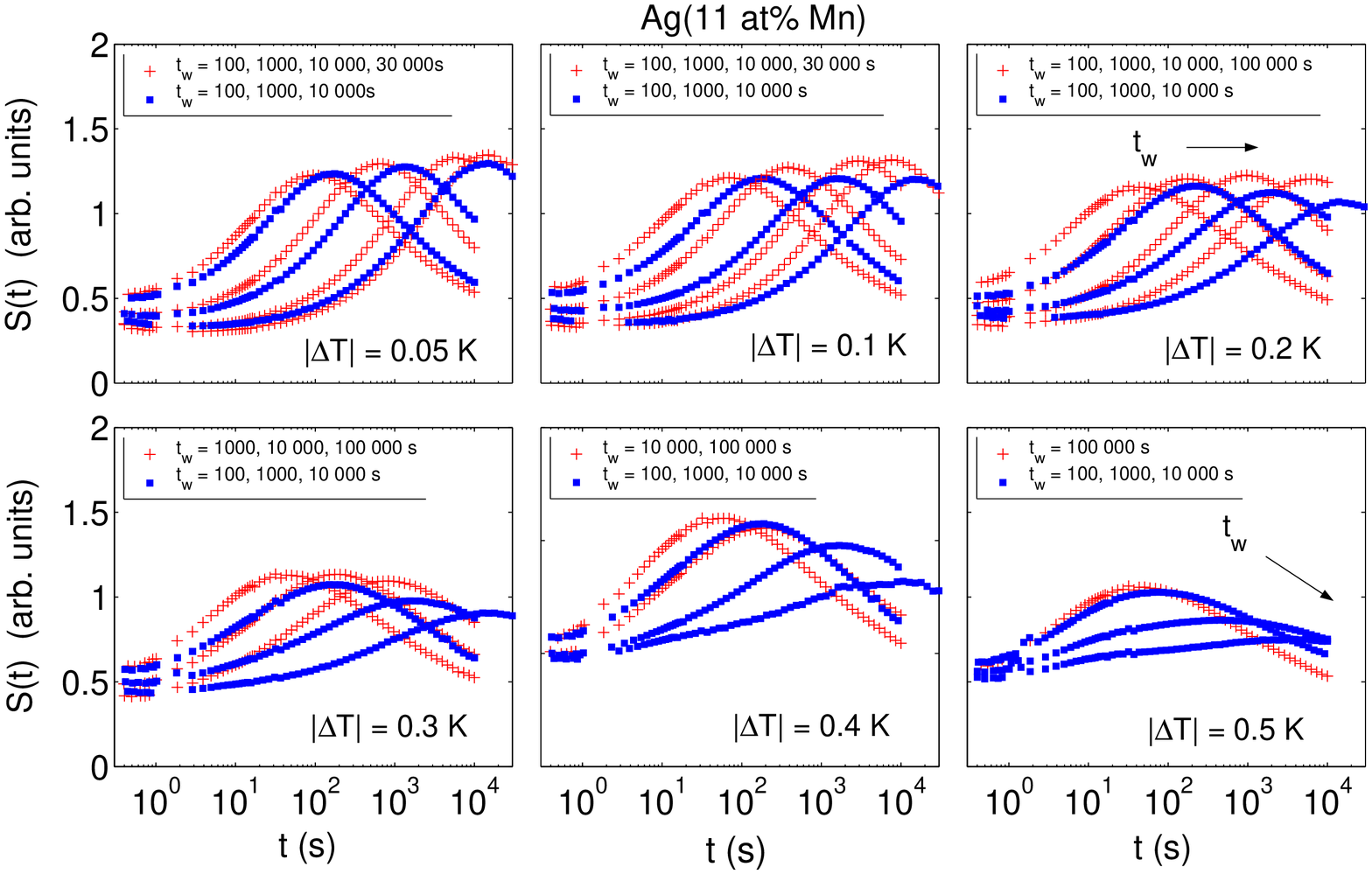}
\includegraphics[height=9cm]{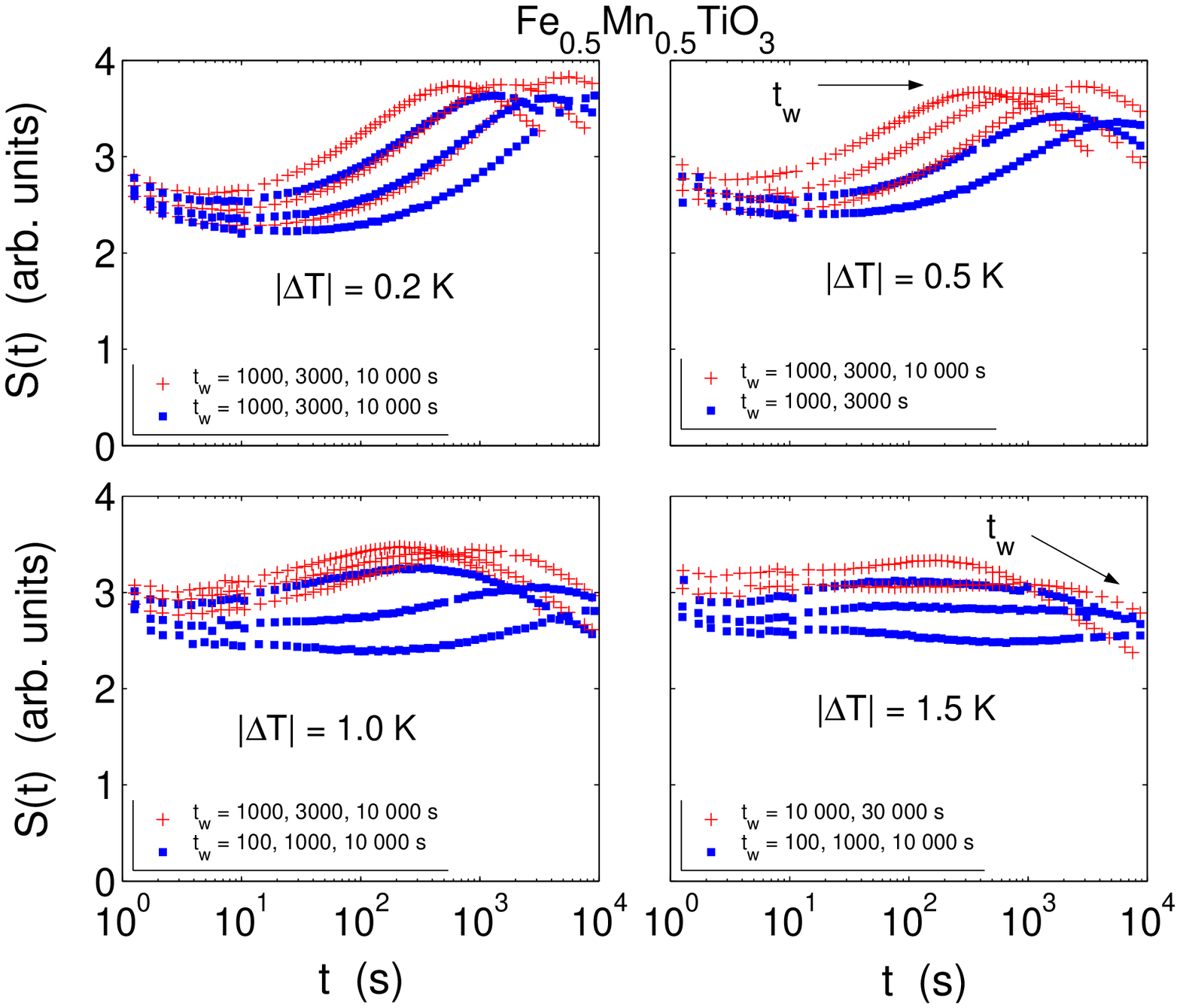}
\caption{(Color online) Relaxation rate vs time after positive and
negative twin $T$-shift measurements. Pluses are positive $T$-shifts
$(T_1,T_2)$, while squares are negative $T$-shifts $(T_2,T_1)$. (a)
\agmn: $T_1 = 30 - |\Delta T|$ K and $T_2 = 30$ K. (b) \ising: $T_1
= 19 - |\Delta T|$ K and $T_2 = 19$ K. Note that for each $\Delta t$
the peak position $t_{\rm max}$ increases with $\tw$ and the maximum
in $S(t)$ broadens, which suggests that these data reflect the
weakly perturbed regime of the temperature-chaos effect. (See later
Fig.~\ref{s_4dea.fig} for the corresponding data of bond-shifts in
the 4d EA model.)} \label{fig-S-AgMn-Ising}
\end{figure*}

The relaxation rate after twin $T$-shift experiments are shown in
Fig.~\ref{fig-S-AgMn-Ising} for the two samples. The effective age
($\teff$) of the system after a $T$-shift is determined from the
time $\tpeak$ corresponding to the maximum in $S(t)$ (see
Appendix~\ref{app-teff} for details).
The shape of the $S(t)$ curves changes after both negative and
positive $T$-shifts; the maximum in $S(t)$ becomes broader with
increasing $|\Delta T|$ (and for a certain $|\Delta T|$ also with
increasing $\tw$). Due  to this broadening $\tpeak$ becomes less
well-defined and thus  $\teff$ less accurate. A broadening of $S(t)$
with increasing $|\Delta T|$ is observed for both samples, but it
should be noted that the temperature shifts employed for the \ising
sample are much larger than those for the \AgMn sample. That the
data in Fig.~\ref{fig-S-AgMn-Ising} shows an increase of $\tpeak$
with $\tw$ indicates that the $\Delta T$'s are small enough to
belong to the weakly perturbed regime, i.e. $L_{T_i}(\tw) \lesssim
\LovlpT$.
The broadening of the relaxation rate can hence be explained by
the weak chaos effect (see Sec.~\ref{subsec-theory-iso}); during
the aging at $\Ti$ ghost domains grow at nearby temperatures up to
the overlap length. The interior of
these ghost domains contain defects or noise and do therefore not
have full amplitude of the order parameter [Eq.~\ref{eq-def-op}].
These defects are progressively eliminated after the $T$-shift
giving rise to extra responses (see
Sec.~\ref{subsec-theory-shift}). Such extra responses are
reflected in the ZFC magnetization measurements as a broadening of
$S(t)$.

\begin{figure}[htb]
\includegraphics[width=0.9\columnwidth]{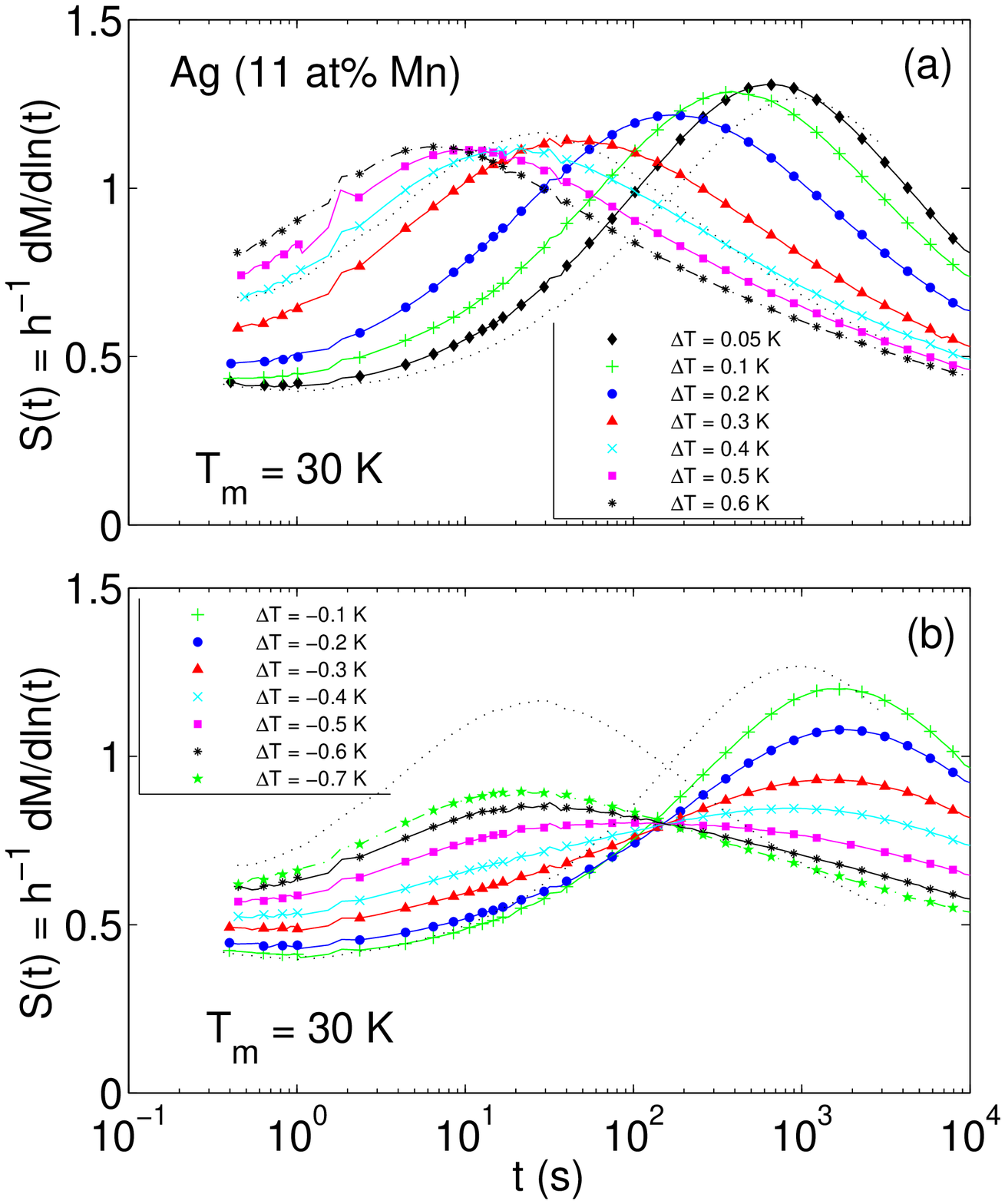}
\caption{(Color online) Relaxation rate vs time after aging the \agmn sample at
(a)
 $\Ti =29.95$, 29.9, 29.8, 29.7, 29.6, 29.5, 29.4 K  (b) $\Ti=30.1$,
 30.2, 30.3, 30.4, 30.5, 30.6, 30.7 K using $\tw$ = 1000 s and $\Tm$ =
 30 K. Dotted lines are ordinary isothermal aging curves with $\tw$ = 3
 and 1000 s. (See Fig.~3 in Ref.~\onlinecite{djujonnor99} for
the case of a Cu(Mn) spin glass.) } \label{fig-S2-AgMn}
\end{figure}

For the \agmn sample, if $|\Delta T|$ is increased even further the
maximum in $S(t)$ again becomes narrower, as can be seen in
Fig.~\ref{fig-S2-AgMn}, where positive and negative $T$-shift with
$\Tm=30$~K and $\tw=1000$~s are shown for various values of $\Ti$.
For larger enough $|\Delta T|$ the peak position $\tpeak$ piles up
around $\tmin$. This $\tmin$ ($\Lmin =L_{T}(\tmin)$) is the shortest
effective age (domain size) in the system after a $T$-change and it
depends on the cooling/heating rate as discussed in
Sec.~\ref{sec-heating-cooling} and Appendix~\ref{app-teff}. A
similar behavior has been observed in a Cu(Mn) spin glass, as shown
in Fig.~3 of Ref.~\onlinecite{djujonnor99}. That $\tpeak$ saturates
to $\tmin$ indicates that $\LovlpT < \Lmin$; the experimentally
accessible times $\tw$ [or lengths $L_{T_i}(\tw)$] lie in the
strongly perturbed regime. However since the overlap length is
``hidden'' behind the domains grown during the $T$-change, these
measurements do not give any direct information about the overlap
length.

For the \ising sample the strongly chaotic regime could not be observed within
the temperature range used in the experiments reported in the
present section. However, we expect to see a narrowing of $S(t)$
also for the \ising sample if $|\Delta T|$ is made large enough.

\begin{figure*}[htb]
\includegraphics[width=0.95\textwidth]{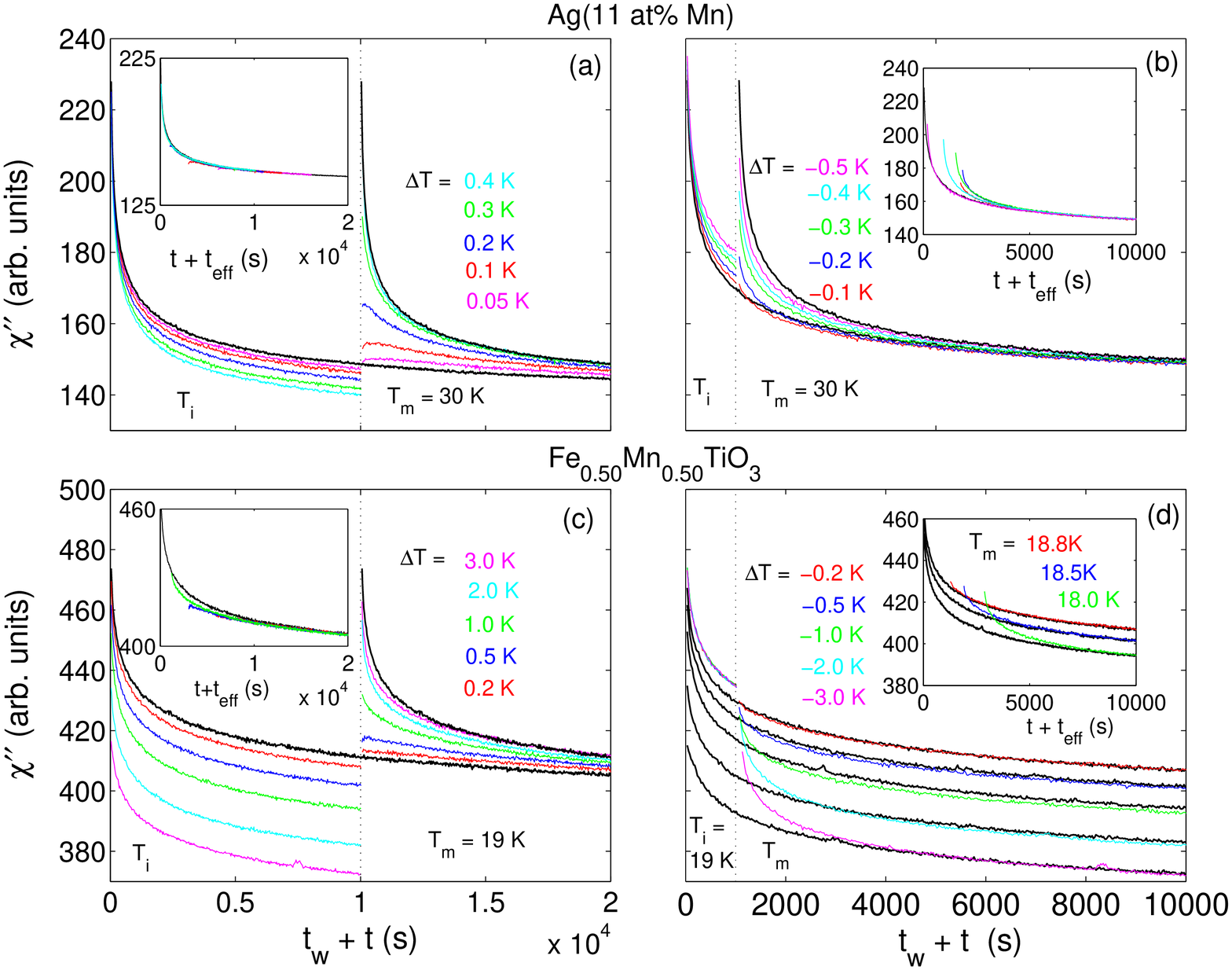}
\caption{(Color online) $\chi''$ vs time measured for positive and negative $T$-shift experiments. (a) The \agmn sample is aged at $T_i =29.95$, 29.9,
29.8, 29.7, 29.6 K for $\tw = 10\,000$~s before changing the  temperature
to  $T_m=30$~K.
(b) It is aged at $T_i = 30.1$, 30.2,
30.3, 30.4, 30.5 K for $\tw = 1000$~s before changing the temperature
to  $T_m=30$~K.
(c) The \ising sample is aged at $\Ti =
18.8$, 18.5, 18.0, 17.0, 16.0 K  for $t_w = 10\,000$~s before changing the
temperature to $T_m=19$~K.
(d) It is aged at $\Ti =
19$~K  for $t_w = 1000$ s before changing the temperature to
$\Tm=18.8$, 18.5, 18.0, 17.0, 16.0 K.
The thick lines are the
(reference) isothermal aging curves starting from $\tw + t =0$ and
$ \tw + t =\tw$. The insets show $\chi''(\Tm)$ vs $t+\teff$ with
$\teff$ determined from ZFC relaxation experiments. $f=510$~mHz. }
\label{fig_ac_shift}
\end{figure*}

\subsubsection{ac susceptibility relaxation after $T$-shifts}

In an isothermal aging experiment
$\chi''(t)$ is recorded vs. time immediately after the
quench (only allowing some time for thermal stabilization and
stabilization of the ac signal). In a $T$-shift experiment,
$\chi''(t)$ is continuously measured after the initial temperature
quench to $\Ti$. In the following we choose $t=0$ immediately
after the $T$-shift to $\Tm$.

$T$-shift ac relaxation measurements with $\Ti < \Tm$ (positive
T-shift) are shown in Fig.~\ref{fig_ac_shift}(a,c). For large enough
$\Delta T$ the relaxation curves after the $T$-shifts  become
identical to the reference isothermal aging curve. The $\Delta T /
\Tg$ needed to make the aging at the lower temperature negligible
is however much larger for the \ising sample than for the \agmn
sample. If the time scale of these $\chi''(t)$ curves are shifted
by the effective time determined from the corresponding ZFC
relaxation experiments reported before, they merge with the
reference isothermal $\chi''(t)$ curve. This is true for both
samples [see the insets of Fig.~\ref{fig_ac_shift}(a,c)], however a
transient part of the curves at short times lies below the
isothermal aging curve.

$T$-shift ac susceptibility relaxation measurements with $T_i >T_m$
(negative T-shift) are shown in Fig.~\ref{fig_ac_shift}(b,d). For
the \agmn sample, the $T$-shift relaxation curves lie below the
reference isothermal aging curve for small $|\Delta T|$. However for
larger $|\Delta T|$ the $T$-shifted relaxation curves lie above the
reference curve and for large enough $|\Delta T|$ the relaxation
curves are identical to the isothermal aging curve (complete
rejuvenation). For the \ising sample, no complete rejuvenation can
be observed even for $\Delta T=-3$~K ($\Delta T /T_g = -0.14$),
whereas for the \AgMn sample complete rejuvenation is already
observed for $|\Delta T|/T_g \gtrsim 0.03$ on the examined time
scales. Shifting the time scale by $\teff$, determined from ZFC
relaxation experiments, makes the $T$-shift curve merge with the
reference curve at long time scales as shown in the insets of the
figures. At short time scales, a long transient relaxation exists
during which  $\chi''(t)$ lies above the reference curve.

The transient part of the susceptibility $\Delta \chi''$ shows
nonmonotonic behavior with increasing $\Delta T$ in the case of
negative T-shifts; $\Delta \chi''$ initially becomes larger but it
eventually disappears for large $\Delta T$'s as $\teff \to 0$. These
two features can be observed for the \agmn sample in
Fig.~\ref{fig_ac_shift}(b) and complete rejuvenation ($\teff=0$ and
$\Delta \chi''=0$) is shown in Fig.~\ref{rm}. However, only the
initial increase of $\Delta \chi''$ can be observed for the \ising
sample in Fig.~\ref{fig_ac_shift}(d). The nonmonotonic behavior of
$\Delta \chi''$ is consistent with the picture presented in
Sec.~\ref{subsec-theory-shift}; in the weakly perturbed regime, the
excess transient responses should be proportional to $\Delta T$ and
the duration of the transient response $t_{\rm trans}$ is roughly
the same as the effective time $\teff$, while in the strongly
perturbed regime $\Delta \chi'' \to 0$ as $\teff \to 0$, which is
the limit of complete rejuvenation.

\subsubsection{Non-accumulative aging}

The previous sections described how to determine the effective
age ($\teff$) of the SG systems after a $T$-shift.
The extracted effective age can now be used to examine if the aging is
fully accumulative or not.
The advantage of the twin $T$-shift method \cite{jonyosnor2002}
is that it allows one to distinguish between the two cases although
the growth law $L_T(t)$ is unknown.
By plotting $t_2$ ($\tw$ or $\teff$ at $T_2$) vs $t_1$ ($\tw$ or $\teff$
at $T_1$), the sets of data from the positive $T$-shifts and
the corresponding negative $T$-shifts should fall on the same
line corresponding to  $t_2=L_{T_2}^{-1}[L_{T_1}(t_1)]$
if the aging is fully accumulative, while any deviation suggests
emergence of rejuvenation (non-accumulative aging).

\begin{figure}[htb]
\includegraphics[width=0.8\columnwidth]{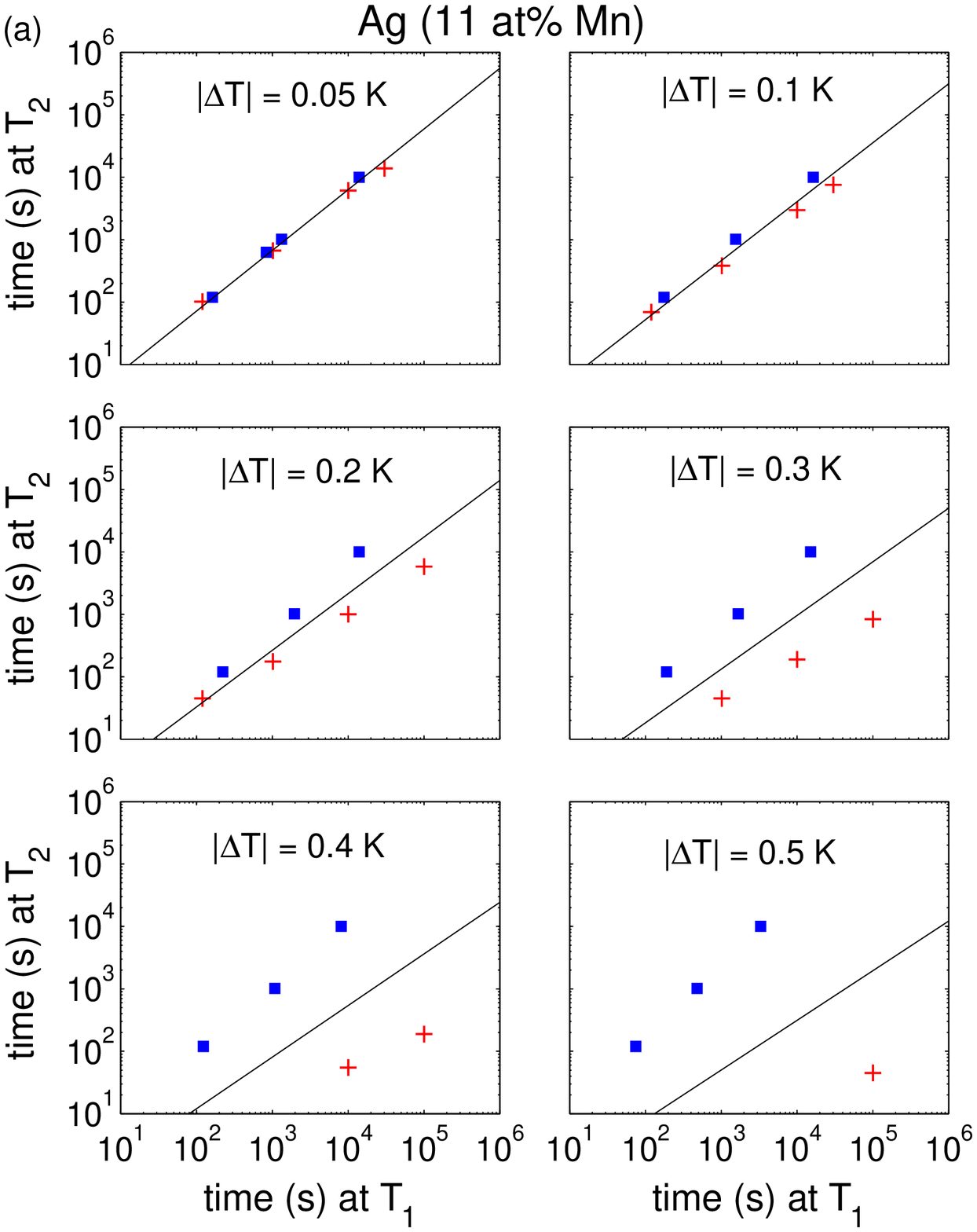}
\includegraphics[width=0.8\columnwidth]{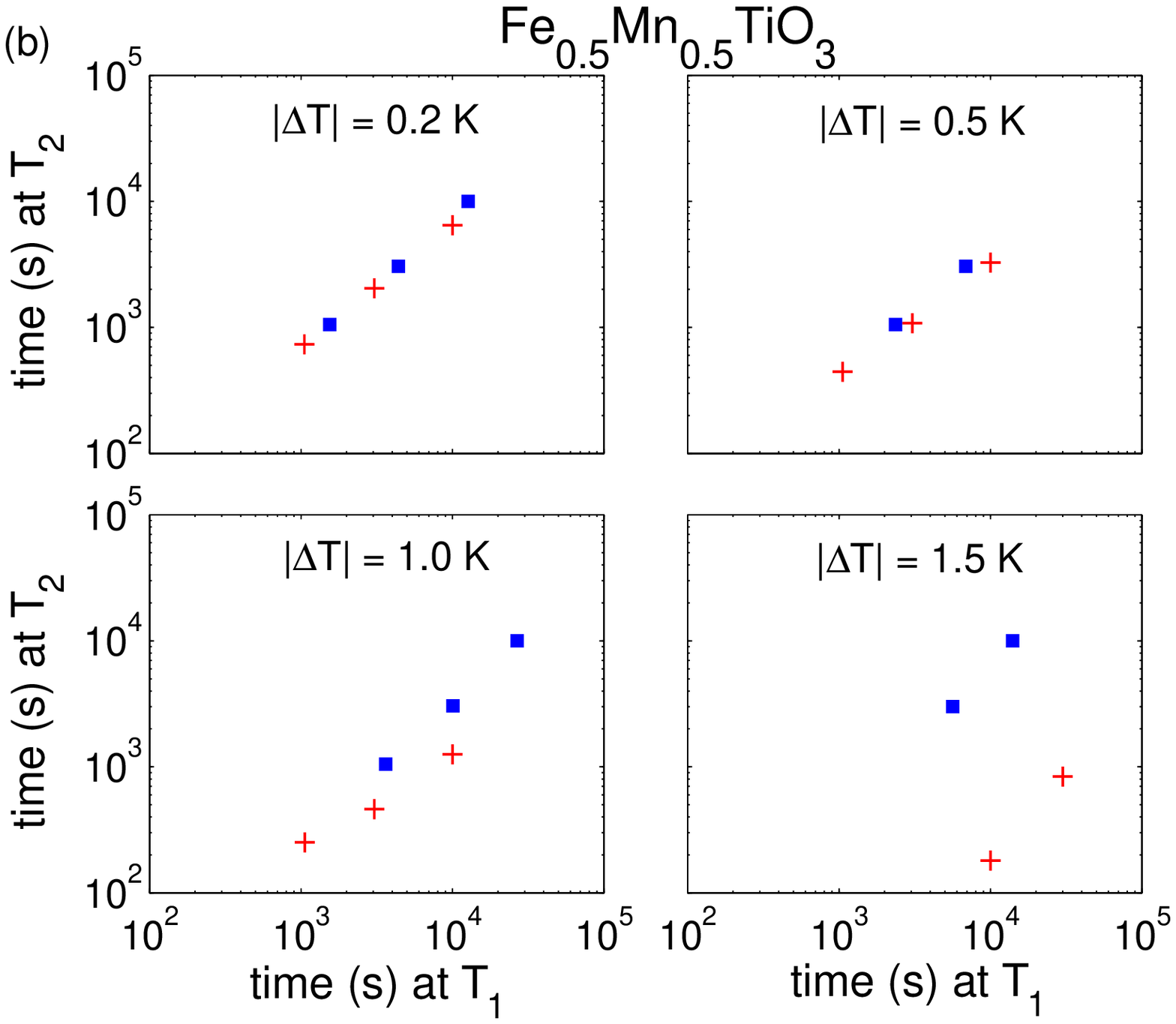}
\caption{(Color online) Relation between $\tw$ and $\teff$ in twin experiments -
 ($T_1$,$T_2$) shown as pluses and ($T_2$,$T_1$) shown as squares. (a)
 For the  \AgMn sample, $T_1 = (30 - |\Delta T|)$~K and $T_2=30$~K with
 $|\Delta T| =  0.05$, 0.1, 0.2, 0.3, 0.4, 0.5~K ($0.015 \Tg$).
The solid line indicates fully accumulative aging $L_{T_1}(t_1)=L_{T_2}(t_2)$,
calculated assuming logarithmic domain growth [Eq.~\ref{eq-growth}].
(b) for the \ising sample, $T_1 = (19 - |\Delta T|)$~K and $T_2=19$~K
 with $|\Delta T| =  0.2$, 0.5, 1.0, 1.5~K ($0.064\Tg$).
\label{fig_tapp}}
\end{figure}

Plots of $t_1$ vs $t_2$ are shown in Fig.~\ref{fig_tapp}. $\teff$
is determined from the twin $T$-shift ZFC relaxation experiments
shown in Fig.~\ref{fig-S-AgMn-Ising}.
For the \AgMn sample, the line of accumulative aging is shown in
addition; a logarithmic domain growth law [\eq{eq-growth}] has been used with
parameters taken from previous
studies.\cite{jonetal2002PRL,jonyosnor2002}
It can be seen for both
samples that the aging is accumulative for small values of
$|\Delta T|$ while for larger $|\Delta T|$ non-accumulative aging
is observed (the data for positive and negative $T$-shifts do not
any longer fall on the same line). There is however a large
difference between the two samples as to how large values of
$|\Delta T|/\Tg$ are required to introduce non-accumulative aging.
A similar analysis of $\teff$ after twin $T$-shift experiments
was recently performed on the 3d EA Ising
model.\cite{takhuk2002} Only very weak rejuvenation effects
could be observed for the Ising
model on the time/length scales accessible by the numerical
simulations even for such large $\Delta T/\Tg$ as 0.3.

\subsubsection{Evidence for temperature-chaos}
\label{subsubsec-Tchaos} Is the non-accumulative aging consistent
with temperature chaos  as predicted by the droplet model? In order
to answer that question we need to transform timescales into length
scales. In other words we need to know the functional form of the
domain growth law. In Ref.~\onlinecite{jonetal2002PRL} it was shown
that the ac relaxation data measured at different frequencies and
temperatures are consistent with a logarithmic domain growth law
[Eq.~(\ref{eq-growth})]. The temperatures used in that study include
those in Fig.~\ref{fig-S-AgMn-Ising} for the \agmn sample, but not
for the \ising sample. We therefore restrict the search of
temperature-chaos to the \agmn sample.

\begin{figure}[h!]
\includegraphics[width=0.8\columnwidth]{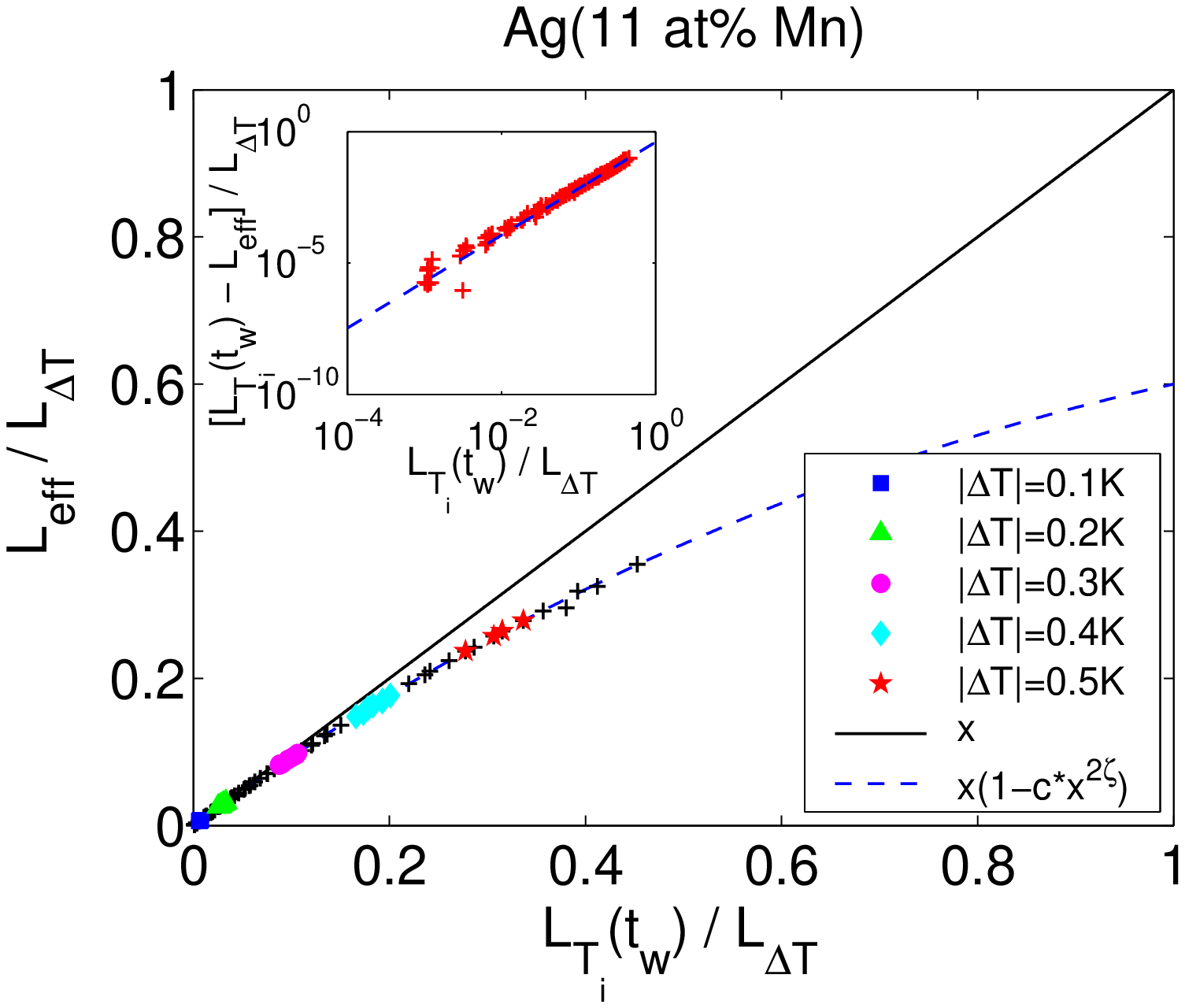}
\caption{(Color online) Scaling plot of $\Leff$ with $\LovlpT = L_0 (a |\Delta
T|/\Tg)^{-1/\zeta}$ with $a=6$ and $1/\zeta=2.6$. The fit is due to
the scaling ansatz for the weakly perturbed regime [Eq.~(\ref{eq-scaling-leff-3})] with c=0.4.
The deviation from accumulative aging is shown in the inset and the line is the fit to the scaling ansatz for the weakly perturbed regime.
(See Fig.  \ref{scale-lteff-4dea.fig} for the
corresponding scaling plot of the data of bond-shifts in the 4 dim
EA model.)
\label{scale-leff-AgMn}}
\end{figure}

For the \agmn sample, we showed in Ref.~\onlinecite{jonyosnor2002}
that positive and negative twin $T$-shifts can be made equivalent by
plotting $L_{\Tm}(\teff)$ vs $L_{\Ti}(\tw)$. This supports the
expectation of the droplet theory \cite{bramoo87,fishus88eq} that
there is a unique overlap length between a given pair of
temperatures. By scaling $L_{\Tm}(\teff)$ and $L_{\Ti}(\tw)$ with
the overlap length $\LovlpT = L_0 (a |\Delta T|/\Tg)^{-1/\zeta}$,
all data corresponding to different $\Ti$, $\Tm$, and $\tw$ merge on
a master curve if $1/\zeta=2.6$ (see Fig.~\ref{scale-leff-AgMn}).
This master curve is consistent with the scaling ansatz in the
weakly perturbed regime
[Eq.~(\ref{eq-scaling-leff-1}-\ref{eq-scaling-leff-3})] as shown in
the figure.

In the limit of strong chaos it is theoretically expected that
$L_{\Tm}(\teff)=\LovlpT$ 
[Eq.~(\ref{eq-scaling-leff-1}) and (\ref{eq-scaling-leff-4})], and
hence that the effective age of the system should only be given by
the overlap length $\LovlpT$ and not by the wait time $\tw$. Our
study is however strongly limited by the experimental time window.
The upper limit $10^4 -10^5$~s is set by how long we can wait for
our experiments to finish, while the lower limit $\tmin$ is
set by the cooling/heating rate and the time needed to stabilize the
temperature (see Appendix~\ref{app-teff}).
For large $\Delta T \gtrsim 0.7$, $\tpeak$ is saturated to $\tmin$
(corresponding to $\Lmin$) and obviously such values of $\tpeak$
cannot be used in the scaling plot in Fig.~\ref{scale-leff-AgMn}.
$\LovlpT$ cannot be observed directly in this regime, but the fact
that $\LovlpT < \Lmin$ in itself give evidence for strong chaos
within the experimental time (length) window.

Finally, we will make some comments on the validity of the scaling
presented in Fig.~\ref{scale-leff-AgMn}. The quality of the scaling
does not change significantly when altering the values of the
parameters ($\tau_m=10^{-13}$~s, $\Tg=32.8$~K, $z\nu=7.2$,
$\psi=1.2$, $J=\Tg$, and   $\nu=1.1$  from
Ref.~\onlinecite{jonyosnor2002}), with one exception---symmetry
between positive and negative $T$-shifts is obtained only if $\psi
\nu \approx 1.3$. The x- and y-scale are completely arbitrary. We
have intentionally chosen the constant $a$ so that the data appear
to be rather close to the strong chaos regime.

\subsection{Bond shift simulations}
\label{subsec-bond-shift}

\begin{figure*}[t]
\includegraphics[width=0.95\textwidth]{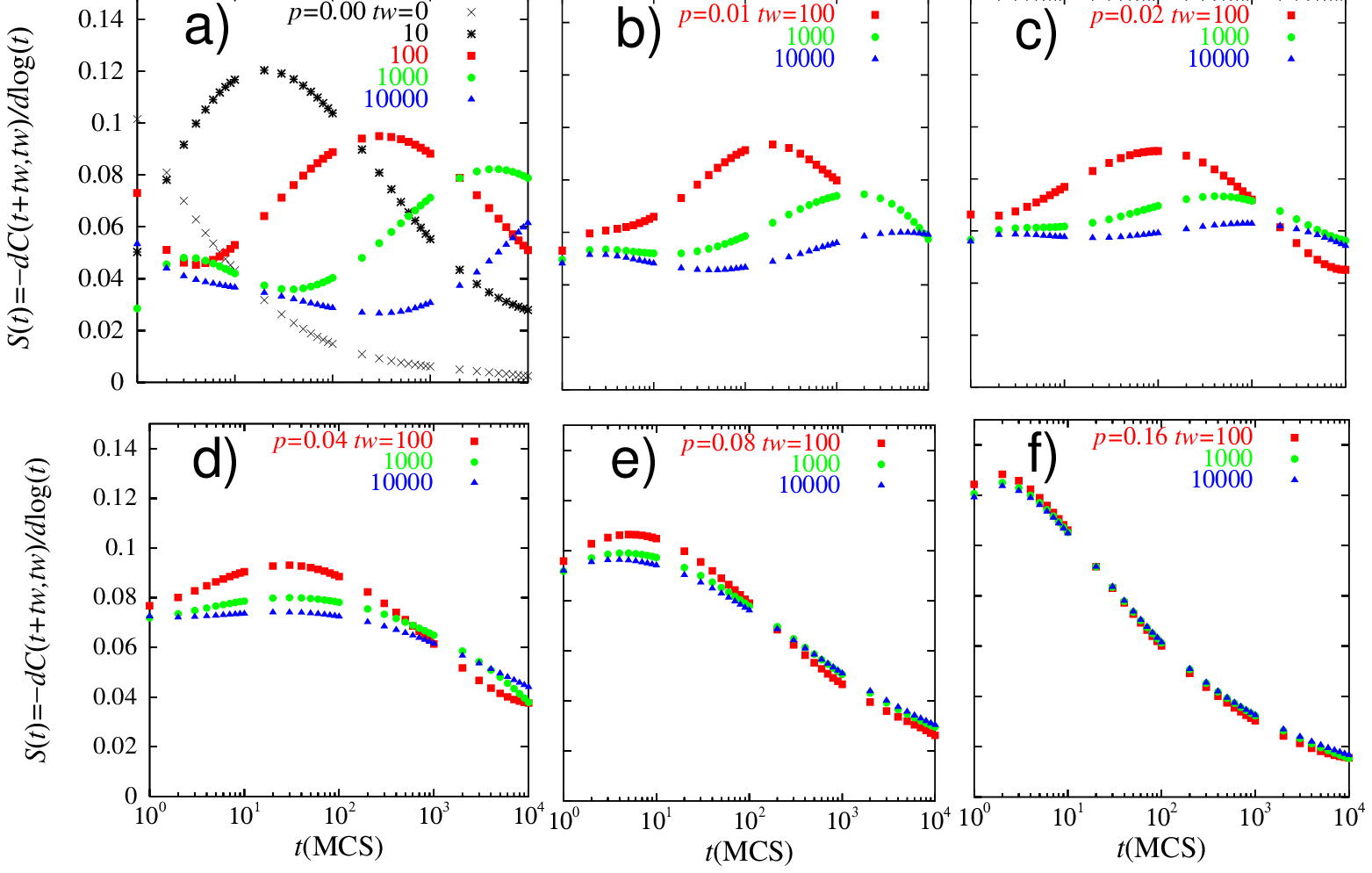}
\caption{(Color online) The relaxation rate $S(t)=- d C(t+\tw,\tw)/\log(t)$ of
the 4DEA model after different waiting times $\tw$ with  bond
shifts of various strength $p$ at temperature $T=0.6T_{g}$.
\label{s_4dea.fig}}
\end{figure*}

We now present results of bond-shift simulations on the $\pm J$ EA
model introduced in section \ref{subsec-model}. Two sets of bonds
${\cal J}_{A}$ and ${\cal J}_{B}$ are prepared as explained in
section \ref{subsec-overlap}. Namely the set of bonds ${\cal J}_{B}$
is created from the set ${\cal J}_{A}$ by changing the {\it sign} of
a small fraction $p$ of the bonds randomly. The
protocol of the simulation is the following: a system with a certain
set of bonds ${\cal J}^{A}$ is aged for a  time $\tw$, after which
the bonds are replaced by ${\cal J}^{B}$ and the relaxation of the
spin auto-correlation function \be C_{\rm
ZFC}(t+\tw,\tw)=(1/N) \sum_{i}\langle S_{i}(\tw)S_{i}(t+\tw) \rangle \,, 
\ee where $i$ runs over the $N$ Ising spins $S_{i}$ in the system, is
recorded. The bracket $\langle \ldots \rangle $ denotes the averages
over different realizations of initial conditions, thermal noises
and random bonds.
The subscript ``ZFC'' is meant to emphasize that this auto
correlation function is conjugate to the ZFC susceptibility: if the
fluctuation dissipation theorem (FDT) holds $(1-C_{\rm
ZFC}(t))(J/T)$ becomes identical to the ZFC susceptibility. The
relaxation rate $S(t)$ is extracted by computing the logarithmic
derivatives numerically, \be S(t)= - d C_{\rm
ZFC}(t+\tw,\tw)/\log(t). \ee Usually the ZFC susceptibility is used
to obtain the relaxation rate. Here, we used the auto-correlation
function since it has much less statistical fluctuations. We note
that FDT is well satisfied within what is called the
quasi-equilibrium regime \cite{yoshuktak2002} so that $S(t)$ defined
above yields almost the same effective time $\teff$ as obtained from
the ZFC susceptibility. We remark that the autocorrelation function
can be measured experimentally by noise-measurement technique
\cite{HO02}.

In Fig.~\ref{s_4dea.fig}, we show some data of relaxation rates
obtained at $T/J=1.2$ ($T/T_{g}=0.6$) for  bond-shift simulations
with wait times $\tw=10^2,10^3,10^4$ and various strength of the
perturbations $\Delta J/J \sim \sqrt{p}$ with $p$ in the range
0--0.16. For relatively small $p$ $(\lesssim 0.05)$, it can be seen
that the peaks in the relaxation rate $S(t)$ become {\it broader}
either by increasing $p$ or $\tw$. Simultaneously, with increasing
$p$, they become somewhat suppressed and their positions are
slightly shifted to shorter times compared to the peaks of the
reference curves for $p=0.0$ [Fig.~\ref{s_4dea.fig}(a)]. At larger
$p$ $( \gtrsim 0.05)$, different features can be noticed. The
relaxation rate $S(t)$ still becomes broader by increasing $\tw$ but
tends to saturate to some limiting curve. By increasing the strength
of perturbation $p$ further, not only the peak position of $S(t)$ is
shifted to shorter times, but the shape of $S(t)$ again becomes {\it
narrower}. Remarkably, these features are very similar to
experimental data where $\Delta T$ is varied. The effects of
temperature perturbations shown in Fig.~\ref{fig-S-AgMn-Ising}
showed considerable broadening of $S(t)$ while
Fig.~\ref{fig-S2-AgMn} showed narrowing of $S(t)$.

The results can be qualitatively interpreted according to the
picture presented in Sec.~\ref{subsec-theory-shift}. We know that
the dynamical length scales $L_{T}(t)$ do not vary much within the
feasible range of time scales $\tw$ (see Fig. \ref{fig-lt}), while
the overlap length $\LovlpJ \propto \sqrt{p}^{-1/\zeta}$ given in
\eq{eq-lovlp-bond} varies appreciably with the strength of the
perturbation $p$. 
The saturation of the $\tw$ dependence can be
interpreted to show that with large enough $p$ the strongly
perturbed regime of the chaos effect enters the numerical time
(length) window. The {\it broadening} of $S(t)$ suggests that the
order parameter [Eq.~\ref{eq-def-op}] does not have its full
amplitude, in other words the domains are noisy or ghost like. This
noise is progressively eliminated starting from the small defects
and give rise to the excessive response, as discussed in
Sec.~\ref{subsec-theory-shift}. On the other hand, the {\it
narrowing} of $S(t)$ at larger $p > 0.05$ suggests that the overlap
length $\LovlpJ$ has become smaller than $L_{T}(\tw)$ and even
approaches the minimum length scale $\Lo \sim 1$. Thus the excessive
response decreases as the effective domain size
[Eq.~(\ref{eq-scaling-leff-1}]) is reduced with increasing $p$ and
finally disappears in the strongly perturbed regime. Interestingly,
this behavior resembles the results from our $T$-shift experiments
on the \AgMn sample. There, a considerable narrowing of the peak in
$S(t)$ occurred for large temperature shifts $\Delta T \gtrsim 0.6$
K (see Fig. \ref{fig-S2-AgMn}), from which we concluded that the
overlap length had become as small as the lower limit
$\Lmin$ induced by the finiteness of the heating/cooling
rate  and the time required to stabilize a temperature in experiments. 

\begin{figure}[h]
\includegraphics[width=0.45\textwidth]{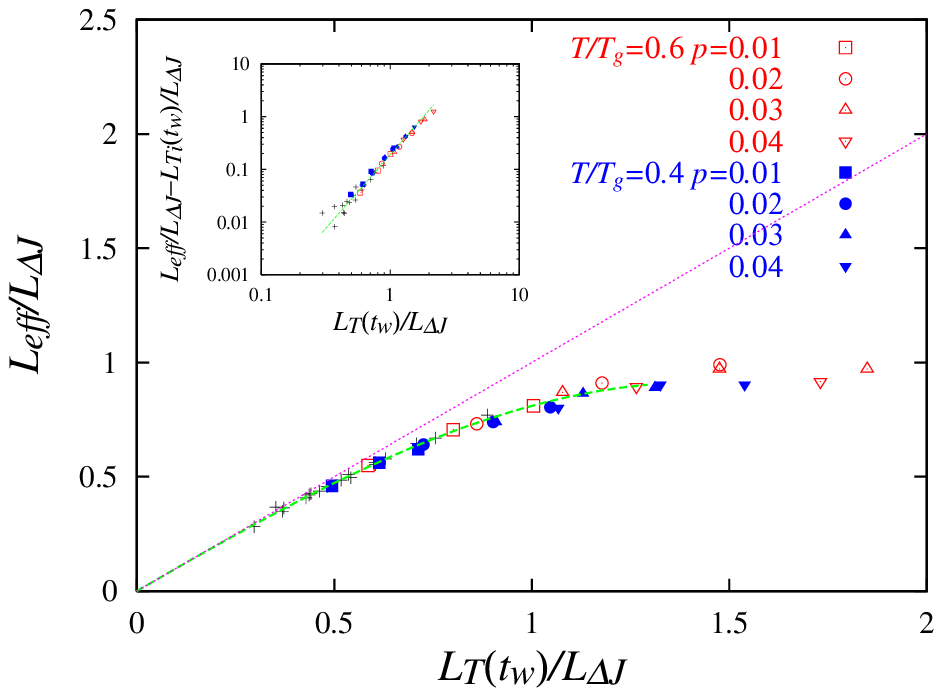}
\caption{(Color online) Test of the scaling of $\Leff$ after bond-shifts in the
4DEA model. The straight line in the main figure is the no-chaos
limit. The fit is due to the scaling ansatz for the weakly
perturbed regime \eq{eq-scaling-leff-1} with
$\LovlpJ=0.1\sqrt{p}^{-1/\zeta}$ with $\zeta=0.9$.
The dynamical length $L_{T}(t)$ is due
to the data obtained in Refs.
\onlinecite{hukyostak2000,yoshuktak2002}. Thus, there are
essentially zero fitting parameters. The correction part to the
no-chaos limit is shown in the inset, where the straight line is
the power law with exponent $2\zeta$.
See the corresponding scaling plot
in the case of the temperature-shift experiments on the \agmn sample
shown in  Fig. \ref{scale-leff-AgMn}. \label{scale-lteff-4dea.fig}}
\end{figure}

Let us turn to a more quantitative analysis similar to that of the
temperature perturbation results on \agmn in Fig.~\ref{scale-leff-AgMn}. 
We first determine the effective time $\teff$
from the peak position $\tpeak$ of $S(t)$ (see footnote for details
\cite{foot_tpeak_tw}). Then the effective domain size is obtained as
$\Leff=L_{T}(\teff)$, using the data of the dynamical length
$L_{T}(t)$ obtained in Refs.
\onlinecite{hukyostak2000,yoshuktak2002} to convert time scales to
length scales (see Fig.~\ref{fig-lt}). The overlap length was
calculated from  $\LovlpJ/\Lo= 0.1 \sqrt{p}^{-1/\zeta}$
(\eq{eq-lovlp-bond}) assuming the chaos exponent to be $\zeta=0.9$.
The value of the chaos exponent $\zeta$
of the present  4-dim EA Ising model reported in  earlier works
varies as $0.85$ \cite{ney98} and $1.3$ \cite{hukiba2003}.
Here we assume $\zeta=0.9$ but small variations of the 
chaos exponent do not affect significantly the results reported
below.

We are now in position to examine the theory presented in section
\ref{subsec-theory-iso} and \ref{subsec-theory-shift}.
Equation~(\ref{eq-scaling-leff-1}) suggests that all data  should collapse onto a master curve by plotting $L_{\rm eff}/\LovlpJ$ vs $L_T(\tw)/\LovlpJ$.
Fig.~\ref{scale-lteff-4dea.fig} shows the resulting scaling plot.
The scaling works perfectly well including all data from the two
different temperatures $T/J=1.2$ and $0.8$ at $p< 0.05$, which
implies that the data collapse on a universal master curve. A fit
according to the scaling ansatz for the weakly perturbed regime
$L_{T}(\tw)/L_{0} \ll \LovlpJ$ is also shown. The correction part
$L_{\rm eff}/\LovlpJ-L_{T}(\tw)/\LovlpJ$ to the accumulative limit is
plotted in the inset. It can be seen that the correction term is
proportional to $(L_{T}/(\tw)\LovlpJ)^{2\zeta}$ with $\zeta=0.9$,
which is in perfect agreement with the expected scaling behavior
for the weakly perturbed regime $L \ll \LovlpJ$ (see
\eq{eq-scaling-leff-3}). In the above scaling plot we excluded
data for $p> 0.05$ where $\teff$ has become too small. As was
shown in Fig.~\ref{s_4dea.fig}, $S(t)$ becomes {\it narrower} at
these larger values of $p$, which suggests that the lower limit of
the length scales  $\Lo$ now comes into play explicitly. The
scaling ansatz Eq.~(\ref{eq-scaling-leff-1}-\ref{eq-scaling-leff-4})
should {\it not} work in such a regime.

\section{Temperature and Bond Cyclings}
\label{sec-cycling} In this section we will investigate
the ``memory'' that survives even under strong temperature or bond
perturbations.

\subsection{Temperature-cycling experiments}
\label{sec_T_cycling}

In Sec.~\ref{sec-T-shift} we found strong
evidence that for the \agmn sample our experimental time window
lies in the strongly perturbed regime, i.e. $\LovlpT < \Lmin$, in the case of large enough
temperature shifts ($|\Delta T|$  $\gtrsim$ 1~K). For the \ising
sample on the other hand, the strongly perturbed regime was not
reached within the temperature range and time window of our experiments.
With this knowledge in mind, we
can for the \agmn sample, focus on how the SG {\em heals} after a
negative $T$-cycling into the strongly perturbed regime and make
comparisons to the  ghost domain scenario presented in
Sec.~\ref{sec-theory}.

\subsubsection{One-step temperature-cycling}

\begin{figure}[htb]
\includegraphics[width=0.9\columnwidth]{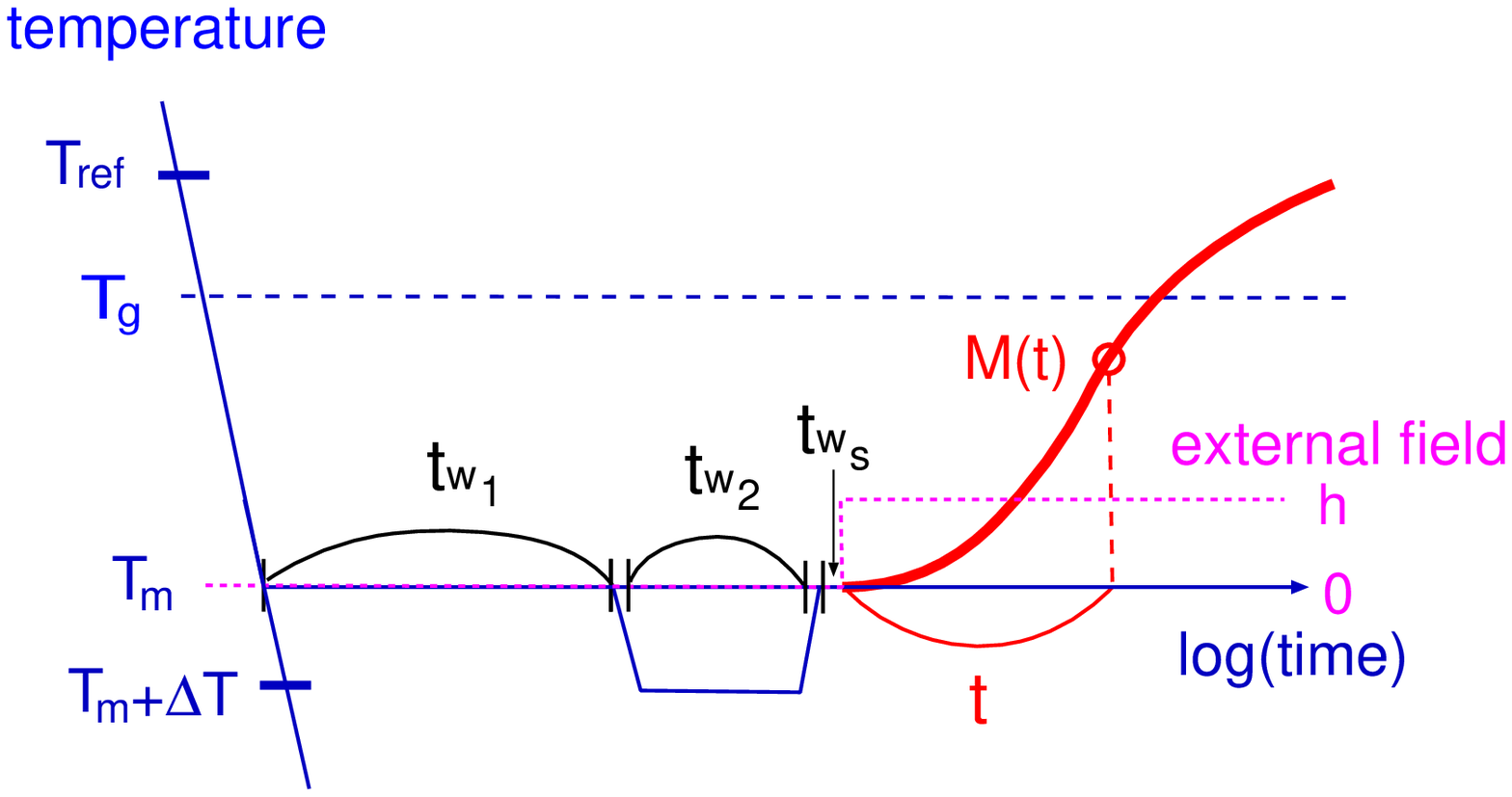}
\caption{(Color online) Schematic representation of a one-step
temperature cycling experiments according to the protocol $\Tm
(\twone) \to \Tm+\Delta T (\twtwo) \to \Tm$. When returning to $\Tm$
a short waiting time $\tws=3$~s is needed, before applying the
magnetic field and recording the ZFC relaxation, in order to ensure
thermal stability. In ac experiments $\chi(t)$ is also recorded
during the cycling. } \label{fig_schem_cyc}
\end{figure}

In a one-step temperature cycling experiments (see
Fig.~\ref{fig_schem_cyc}), the sample is first cooled to $\Tm$ and
aged there a time $\twone$, the temperature is subsequently
changed to $\Tm + \Delta T$ where the sample is aged for a time
$\twtwo$ (perturbation stage) and then the temperature is put back
to $\Tm$ (healing stage). After a short wait time $\tws$ to ensure
thermal stability at $\Tm$, the magnetic field is switched on and
the magnetization $\MZFC(t)$ is recorded, or the ac susceptibility
is recorded continuously during the cycling process.

\begin{figure}[htb]
\includegraphics[width=0.9\columnwidth]{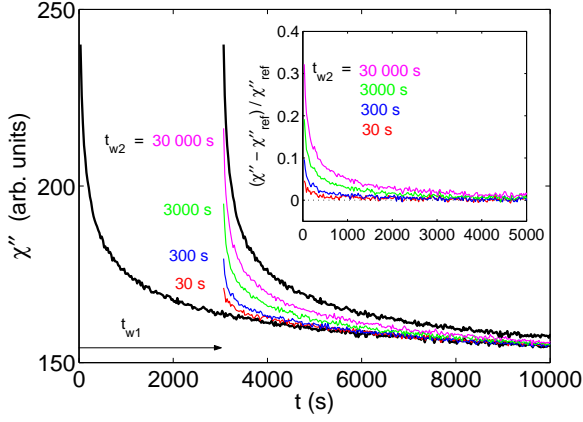}
\caption{(Color online) $\chi''(t)$ vs time at $\Tm$ for experiments
with a negative T-cycling of  $\Delta T = -2$~K ($\Tm +\Delta T$ =
28 K); $\twone$ = 3000 s and $\twtwo$ = 30, 300,  3000, 30\,000 s
(bottom to top). The thick (black) lines represent the reference
isothermal aging at $\Tm$, one of these curves is shifted a quantity
$\twone$ in time. The inset shows $[\chi''(t)-\chi''_{\rm
ref}(t)]/\chi''_{\rm ref}(t)$ for the transient part of the
susceptibility after the $T$-cycling. $\chi''(t)$ during the all
stages of the $T$-cycling are shown in Fig.~\ref{rm} for
$\twtwo=3000$~s.
 $\omega/2\pi=510$~mHz.}
\label{AgMn-Tcycl-ac}
\end{figure}

A negative one-step $T$-cycling experiments ($\Delta T = -2$~K)
measuring the out of phase component of the ac susceptibility is
shown in Fig.~\ref{rm}.
 Results of such measurements with $\twone
= 3000$~s are shown in Fig.~\ref{AgMn-Tcycl-ac}. The aging at
$Tm+\Delta T$ has been cut away in this figure and only the aging at
$\Tm$ is shown. It can be clearly seen that the aging at $\Tm+\Delta
T$ introduces a considerable amount of excessive response, which
increases with increasing $\twtwo$. In the ghost domain scenario the
excessive response is attributed  to the introduction of noise in
the ghost domains by the aging at the temperature $T+\Delta T$. This
yields a reduction of the order parameter [c.f. Eq.~\ref{eq-bias}]
so that the recovery time $t_{\rm rec}$ corresponds to the time
scale at which the order parameter is recovered and hence the
excessive susceptibility disappears [c.f. Eq.~\ref{eq-tau-rec}].
From Fig.~\ref{AgMn-Tcycl-ac}, the recovery times $t_{\rm rec}$ are
found to be of order $O(10^{3}-10^{4})$~s. If the $T$-cyclings would
have kept the system in the weakly perturbed regime, the recovery
times would become $\taurecw=0.04,\; 0.1,\; 0.3,\; 0.9$~s for
$\twtwo=30,\; 300,\; 3000,\; 30\,000$~s according to
\eq{eq-tau-rec-weak} and the growth law \eq{eq-growth} with the same
parameters as in Sec.~\ref{subsubsec-Tchaos}. Consistent with the
prediction in Sec.~{\ref{subsubsec-cycle-strong}}, the observed
values of $t_{\rm rec}$ are several orders of magnitude larger than
$\taurecw$.
However, the $\twtwo$ dependence of $t_{\rm rec}$ is weaker than expected from Eq.~\ref{eq-tau-rec}, indicating that corrections to the assymptotic formula are important on the experimental length scales.\cite{foot_trec}
Finally, let us  note that a similar observation of anomalously
large recovery times was made in a recent study by Sasaki et
al.\cite{sasetal2002}

\begin{figure*}[thb]
\includegraphics[width=0.95\textwidth]{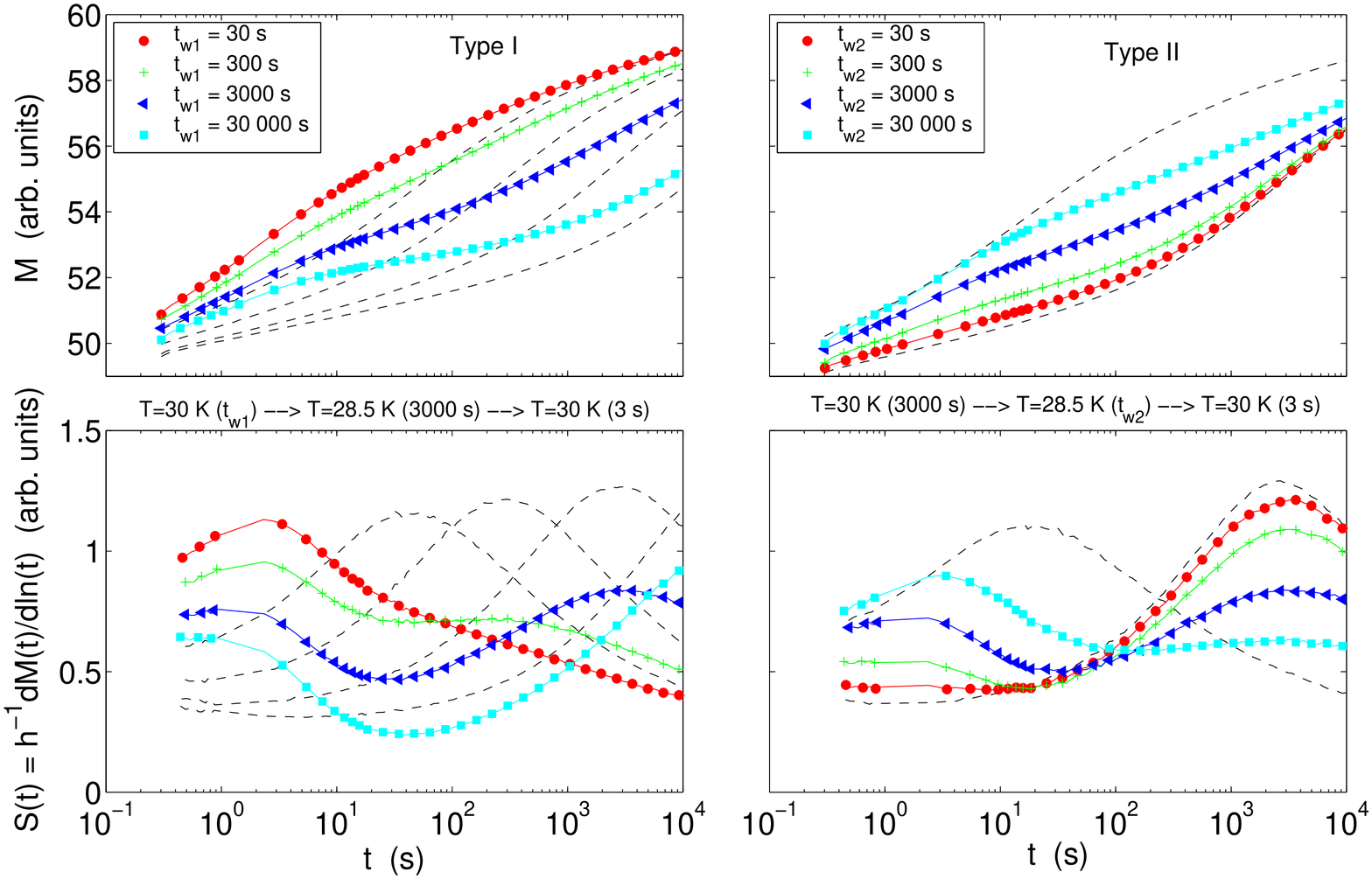}
\caption{(Color online) ZFC relaxation curves ($\MZFC(t)$ and $S(t)$) obtained
after a negative $T$-cycling of $\Delta T$ = -1.5K ($\Tm$ = 30 K;
$\Tm + \Delta T$ = 28.5 K); Type I: $\twone$ = 30, 300, 3000, and 30\,000 s and
$\twtwo$ = 3000 s;
Type II: $\twone$ = 3000 s and $\twtwo$ = 30, 300,  3000, 30\,000
 s.
Isothermal aging recorded at $\Tm$ = 30 K  are added for comparison
in dashed lines; Type I: $\tw$ = $\twone$; Type II: $\tw$ = 3 s and
$\tw$ = 3000 s. (See later Fig.~\ref{bond-cycle} for the
corresponding data of the one-step bond-cycling in the 4d EA model.)
} \label{AgMn_cycl}
\end{figure*}

The ZFC magnetization has been measured after negative one-step
$T$-cyclings on the \agmn sample with $\Tm=30$~K and $\Delta T =
-1.5$~K. Figure~\ref{AgMn_cycl} shows the ZFC relaxation data; Type
I: the initial wait time $\twone$ is varied while the duration of
the perturbation $\twtwo=3000$~s is fixed. Type II: the initial wait
time is fixed to $\twone=3000$~s while the duration of the
perturbation $\twtwo$ is varied. Isothermal aging data are also
shown in these figures for comparison. As seen in the figures, the
magnetization always exhibits an enhanced growth rate at observation
times around $\tmin$ after the cycling and in addition a second
enhanced growth rate at observation times around $\twone$ which are
manifested as two peaks in the relaxation rate $S(t)$. The peak of
$S(t)$ around $\tmin$ can be attributed to rejuvenation effects and
the peak  around $\twone$ to memory effects. It can be seen that by
increasing the duration of the perturbation $\twtwo$ the height of
the 2nd peak of $S(t)$ at around  $\twone$ dramatically decreases,
but the peak position itself does not change appreciably. By
studying the magnetization curve and comparing it to the reference
curve of isothermal aging (with wait time $\tw=\twone$), it can be
appreciated that the cycling data eventually merges with that of the
isothermal aging curve (although it is outside the time scales of
our measurements). The time scale of the merging appears to become
larger for longer duration of the perturbation $\twtwo$.

Within the ghost domain picture discussed in
Sec.~\ref{sec-theory}, this two stage enhancement of the
relaxation rate can be understood as follows. The aging at the
second stage introduces strong noise and reduces the amplitude of
the order parameter in the ghost domains $L_{\Tm}(\twone)$ grown
at $\Tm$ during the initial aging. The system is hence strongly
disordered with respect to the equilibrium state at $T_{m}$, but
the original domain structure is conserved as a bias effect (the
order parameter $\rho_{\rm rem}>0$). During the third stage
(healing stage) the aging called inner-coarsening starts, i.e. new
domain growth starting from an almost random state although with
weakly biased initial condition. Hence, when the magnetic field is
switched on in the healing stage, the inner-coarsening has already
proceeded for the time $\tmin$. This is reflected by the 1st peak
of $S(t)$. During the healing stage, the strength of the bias
keeps increasing so that the noise (minority phases) within ghost
domains is progressively removed. The 2nd peak of $S(t)$
corresponds to the size of the ghost domain $L_{\Tm}(\twone)$,
which itself continues to slowly expand in the healing stage
(outer-coarsening). Thus, this represents the memory of the
thermal history before the perturbation. An illustration of this
noise-imprinting-healing scenario is given in
Fig.~\ref{fig-ghosts}. However, the increase of the bias is so
slow that the recovery time $\taurecs$ of the order parameter can
be extremely large as given in \eq{eq-tau-rec}. Then, the order
parameter within a ghost domain may not be recovered fully until
the ghost domain itself has grown appreciably in the healing
stage. Probably this explains the fact that the 2nd peak of $S(t)$
is greatly suppressed (but not erased) by increasing the duration
of the perturbation $\twtwo$, which also increases the recovery
time $\taurecs$.

\subsubsection{Two-step temperature-cycling}

\begin{figure}[hbt]
\includegraphics[width=\columnwidth]{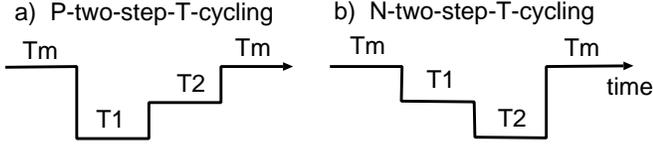}
\caption{Schematic illustration of a)  positive ($T_2>T_1$) and b) negative ($T_2<T_1$) two-step-T-cycling.}
\label{fig-PN-2stepc}
\end{figure}

Finally, two-step $T$-cyclings are studied in order to investigate
the anomalous multiplicative effects of noise anticipated by the
ghost domain picture [see Sec.~\ref{subsub-multicucl}]. In this
protocol an extra $T$-shift is added to the one-step temperature
cycling procedure shown in Fig.~\ref{fig_schem_cyc}; the sample is
cooled to $\Tm$ where it is aged a wait time $\twone$ (initial
aging stage), subsequently the temperature is shifted to
$T_{1}=\Tm+\Delta T_1$ for a time $\twtwo$ (1st perturbation
stage), then shifted to $T_{2}=\Tm+\Delta T_2$ for a time
$\twthree$ (2nd perturbation stage) and finally changed back to
$\Tm$ (healing stage) where $\MZFC(t)$ is recorded. As illustrated
in Fig.~\ref{fig-PN-2stepc} one can make two different kinds of
2-step-$T$-cycling experiments (with $T_1,T_2 <\Tm$); involving
either a positive ($T_2>T_1$) or a negative ($T_2<T_1$) $T$-shift between $T_1$
and $T_2$.

\begin{figure}[hbt]
\includegraphics[width=\columnwidth]{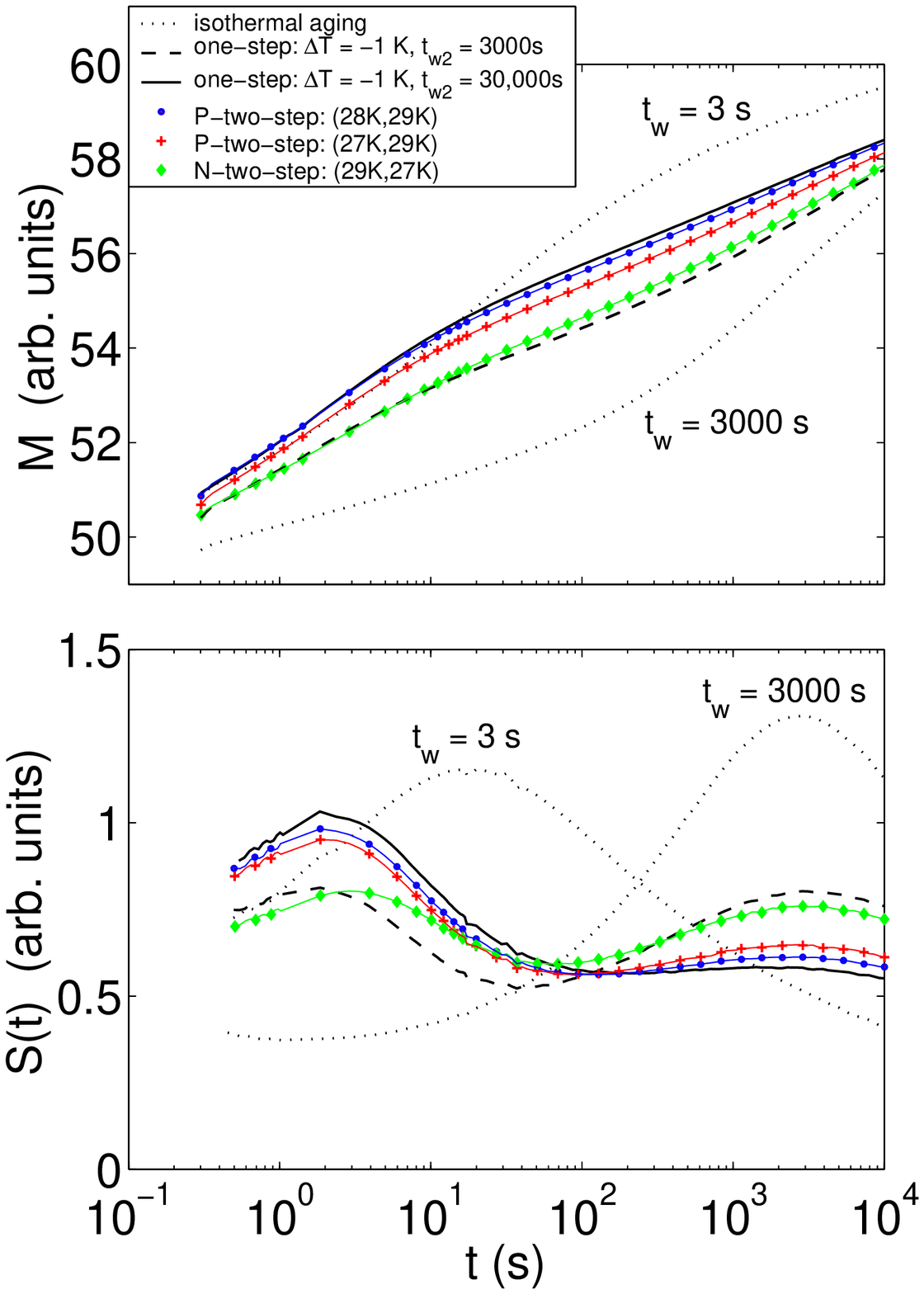}
\caption{(Color online) ZFC relaxation curves  obtained after
negative two-step $T$-cyclings.
Two-step-$T$-cycling experiments
$T_{m} (\twone) \to T_1$  $(\twtwo) \to T_2$ $(\twthree) \to T_{m}$
with $T_{m}=30$~K and $\twone=\twtwo=\twthree=3000$~s are shown.
A negative $T$-shift in $T_{1} \to T_{2}$ is made with
$(T_{1},T_{2})=(29\,{\rm K},27\,{\rm K})$ (diamonds),
and positive $T$-shifts in $T_{1} \to T_{2}$ are made with
$(T_{1},T_{2})=(28\,{\rm K},29\,{\rm K})$ (circles), $(T_{1},T_{2})=(27\,{\rm K},29\,{\rm K})$ (pulses).
The isothermal aging with $\tw = 3$ and $3000$~s (dotted lines)
at $T_{m}=30$~K and one-step $T$-cycling experiments
$T_{m} (\twone) \to \Tm+\Delta T$ $(\twtwo)  \to T_{m}$
with $T_{m}=30$~K, $\Tm+\Delta T=29$~K, $\twtwo = 3000$ and $30\,000$~s (dashed lines)
are shown for comparison.
(See Fig.  \ref{bond-two-stepcycle} for the corresponding data
of the two-step bond-cycling in the 4d EA model.)
\label{AgMn_2step-cycl}}
\end{figure}

In Fig.~\ref{AgMn_2step-cycl}, a set of data using the two-step temperature-cycling protocol with
$\Tm=30$~K and $\twone=\twtwo=\twthree=3000$~s are shown, together with
reference data for one-step cycling experiments and
isothermal aging experiments at $\Tm$. It can be seen that the
data of the N-two-step experiment with $(T_{1},T_{2})=(29\,{\rm K},27\,{\rm
K})$ is slightly more rejuvenated than the corresponding one-step
cycling with  $\Tm+\Delta T = 29$~K and $\twtwo=3000$~s, while the data of the
P-two-step experiment with $(T_1,T_2)=(27\,{\rm K},29\,{\rm K})$ is much more
rejuvenated. Furthermore, the data of the P-two-step experiment with
$(T_1,T_2)=(28\,{\rm K},29\,{\rm K})$ is strikingly more
rejuvenated, so that the $\MZFC(t)$ curve is close the one-step
cycling with  $\Tm+\Delta T = 29$~K and $\twtwo=30\,000$~s.
This may appear surprising, since
the total duration of the actual perturbation is only $\twtwo+\twthree=6000$~s.

The anomalously strong rejuvenation effect after the 2-step
cycling can be understood by the ghost domain picture as the
multiplicative effect of noise given in \eq{eq-tau-rec-multi}. One
should however take into account the renormalization of the
overlap length $\LovlpT \to \Lmin(v_{T},T)$ due to  finiteness
of heating/cooling rates discussed in Sec.
\ref{sec-heating-cooling}. Naturally, the latter effect can
significantly or even completely hide the multiplicative nature of
the noise introduction as appears to have happened in a previous
study.\cite{bouetal2001} To reduce such ``disturbing'' effects as
much as possible, one should use fast enough heating/cooling
rates. It should also be recalled that in the present experiments,
the heating is almost $10$ times faster than cooling which implies
that the renormalized overlap length $\Lmin(v_{T},T)$ is larger
for cooling than for heating. This may explain the apparent
asymmetry of the negative and positive two-step $T$-cyclings in
Fig.~\ref{AgMn_2step-cycl}. The weaker rejuvenation effect
observed in the P-two-step measurement $(T_{1},T_{2})=(27\,{\rm
K},29\,{\rm K})$  compared with  $(28\,{\rm K},29\,{\rm K})$ can
simply be attributed to the temperature dependence of the domain
growth law \eq{eq-growth}.

\subsection{Bond-cycling simulations}

\subsubsection{One-step bond cycling}

For a one-step bond-cycling simulation on the $\pm J$ EA model, we
prepare two sets of realizations of interaction bonds ${\cal J}^{A}$
and ${\cal J}^{B}$ as before. The set of bonds ${\cal J}_{B}$ is
created from ${\cal J}_{A}$ by changing the {\it sign} of a small
fraction $p$ of the latter randomly. In order to work in the
strongly perturbed regime we choose
$p=0.2$, 
since the analysis in the previous section (see Fig.
\ref{scale-lteff-4dea.fig}) ensures that almost the whole time
window of the simulations lies well within the strongly perturbed
regime with $p=0.2$. We also made simulations with weaker strength
of the perturbation $p=0.05$, for which time scales greater than
$10^{3}$ MCS belongs to the strongly perturbed regime as can be seen
in Fig.~\ref{scale-lteff-4dea.fig}, and obtained qualitatively the
same results but with larger statistical errors.
The working temperature is fixed to $T/T_{g}=0.6$ throughout the
simulations and the strength of the probing field is $h/J=0.1$,
which is small enough to observe linear response within the present
time scales.\cite{yoshuktak2002}

The procedure of the one-step bond-cycling is the following. First
the system is let to evolve under ${\cal J}^{A}$ for a time $\twone$
at temperature $T$ starting from a random spin configuration. Then,
the bonds are replaced by ${\cal J}^{B}$ and the system is let to
evolve another time interval $\twtwo$. Finally, the bonds are put
back to ${\cal J}^{A}$ and the system is let to evolve. After a time
interval $\twthree$, a small magnetic field $h$ is switched on and
the growth of the magnetization is measured to obtain the ZFC
susceptibility \be \chi_{\rm
ZFC}(t)=(J/h)(1/N))\sum_{i}\langle S_{i}(t+\twthree+\twtwo+\twone) \rangle \ee
where $t$ is the time elapsed after the magnetic field is switched
on. We also measured the conjugate auto-correlation function to the
ZFC susceptibility in the same protocol but without applying a
magnetic field, \be C_{\rm ZFC}(t) =
C(t+\twthree+\twtwo+\twone,\twthree+\twtwo+\twone), \ee where
$C(t_1,t_2)=(1/N)\sum_{i}S_{i}(t_1)S_{i}(t_2)$ is the spin
autocorrelation function. Again, the subscript ``ZFC'' is put  to
emphasize that this auto correlation function is conjugate to the
ZFC susceptibility if the FDT holds. We expect both quantities to
reflect the inner- and outer-coarsening discussed in Sec.
\ref{sec-theory}, but there are also apparent differences. Firstly,
the FDT is expected to be violated in the outer-coarsening
regime.\cite{yoshuktak2002} Secondly, the linear susceptibility
$\chi_{\rm ZFC}(t)$ should have the so called  weak long term memory
(WLTM) property \cite{bouetal97}, stating that an integral of linear
responses during a finite interval of time finally disappears at
later times, whereas the auto correlation does not have such a
property.

\begin{figure*}[t]
\includegraphics[width=0.9\columnwidth]{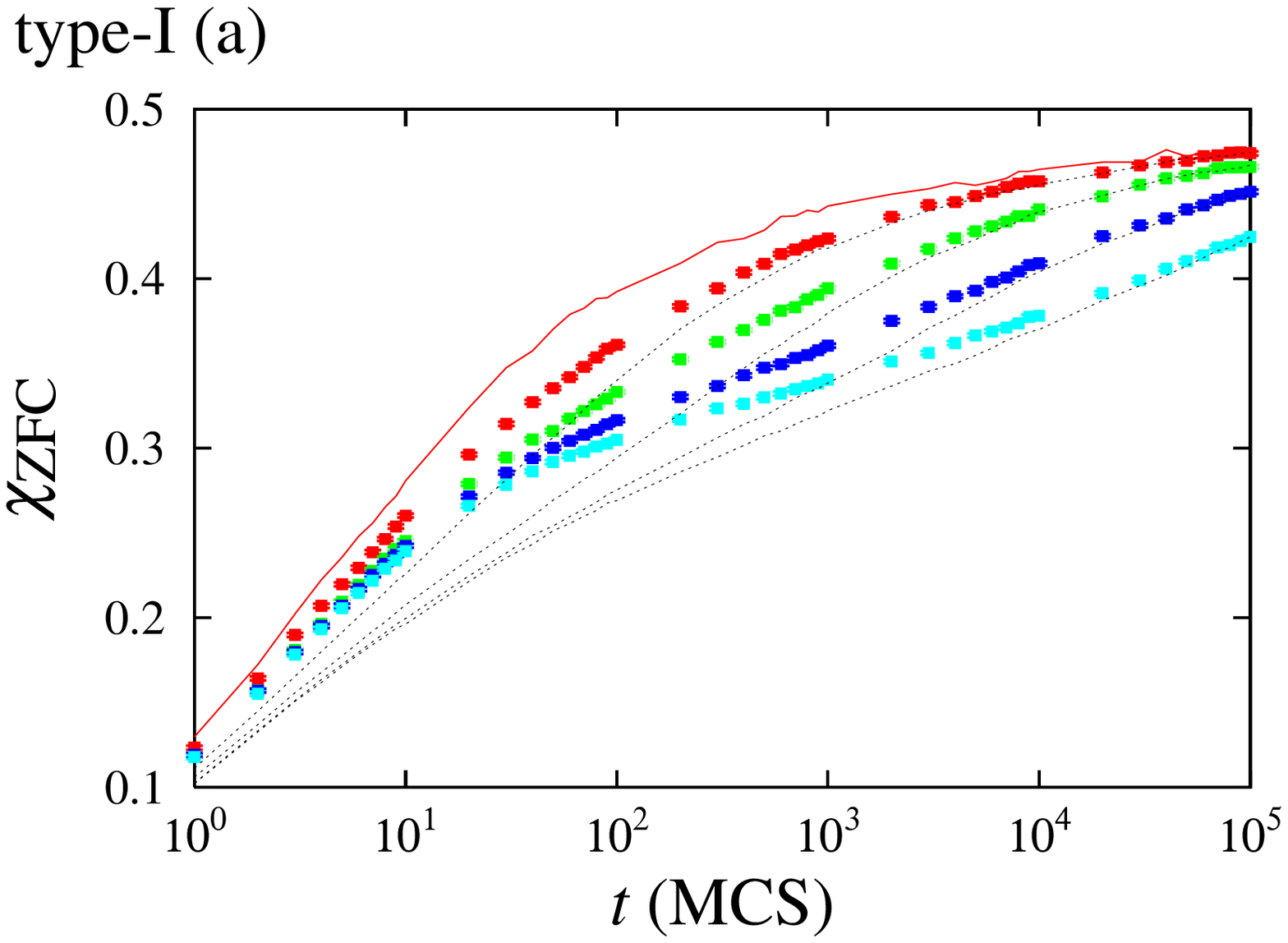}
\includegraphics[width=0.9\columnwidth]{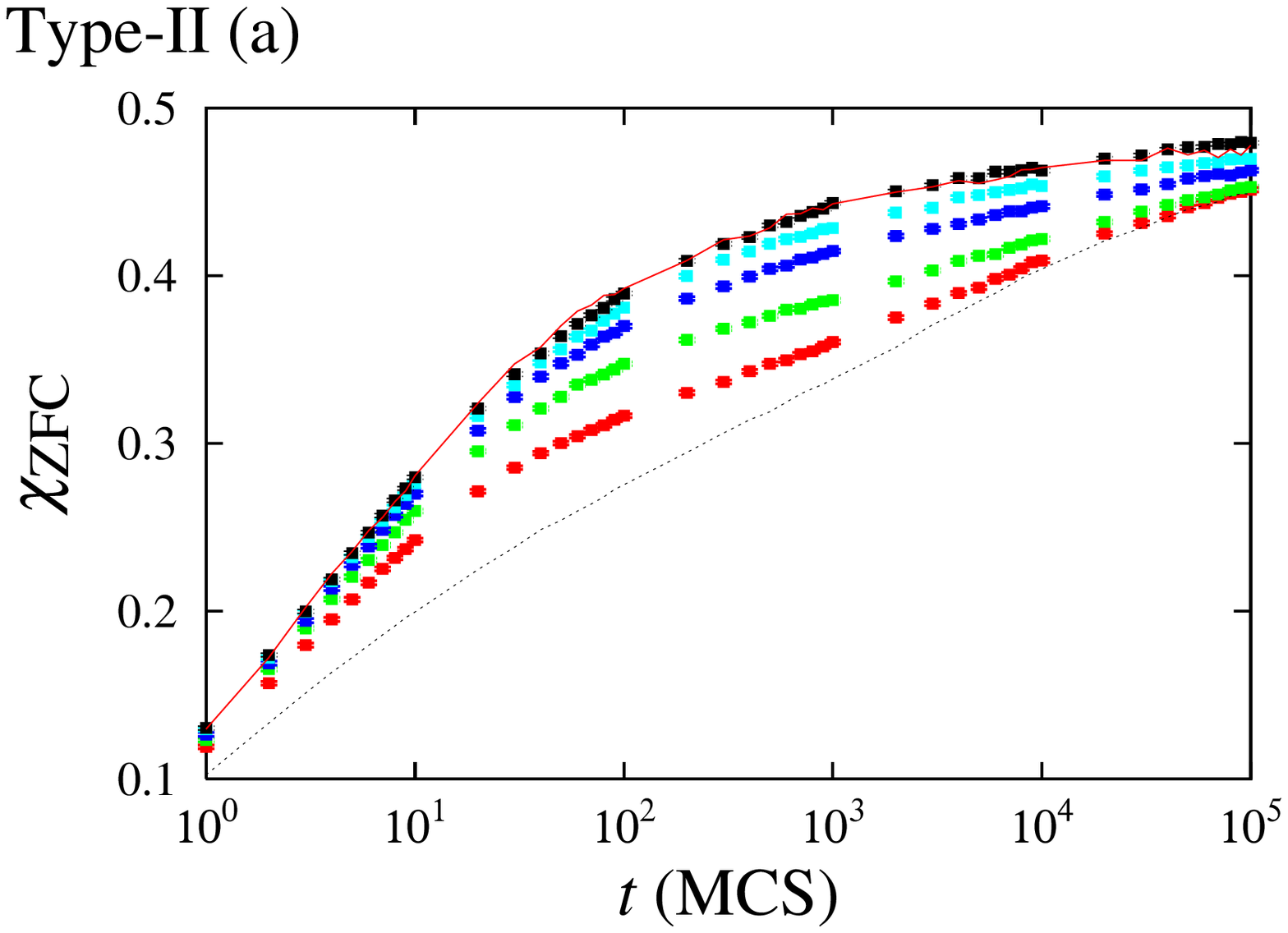}
\includegraphics[width=0.9\columnwidth]{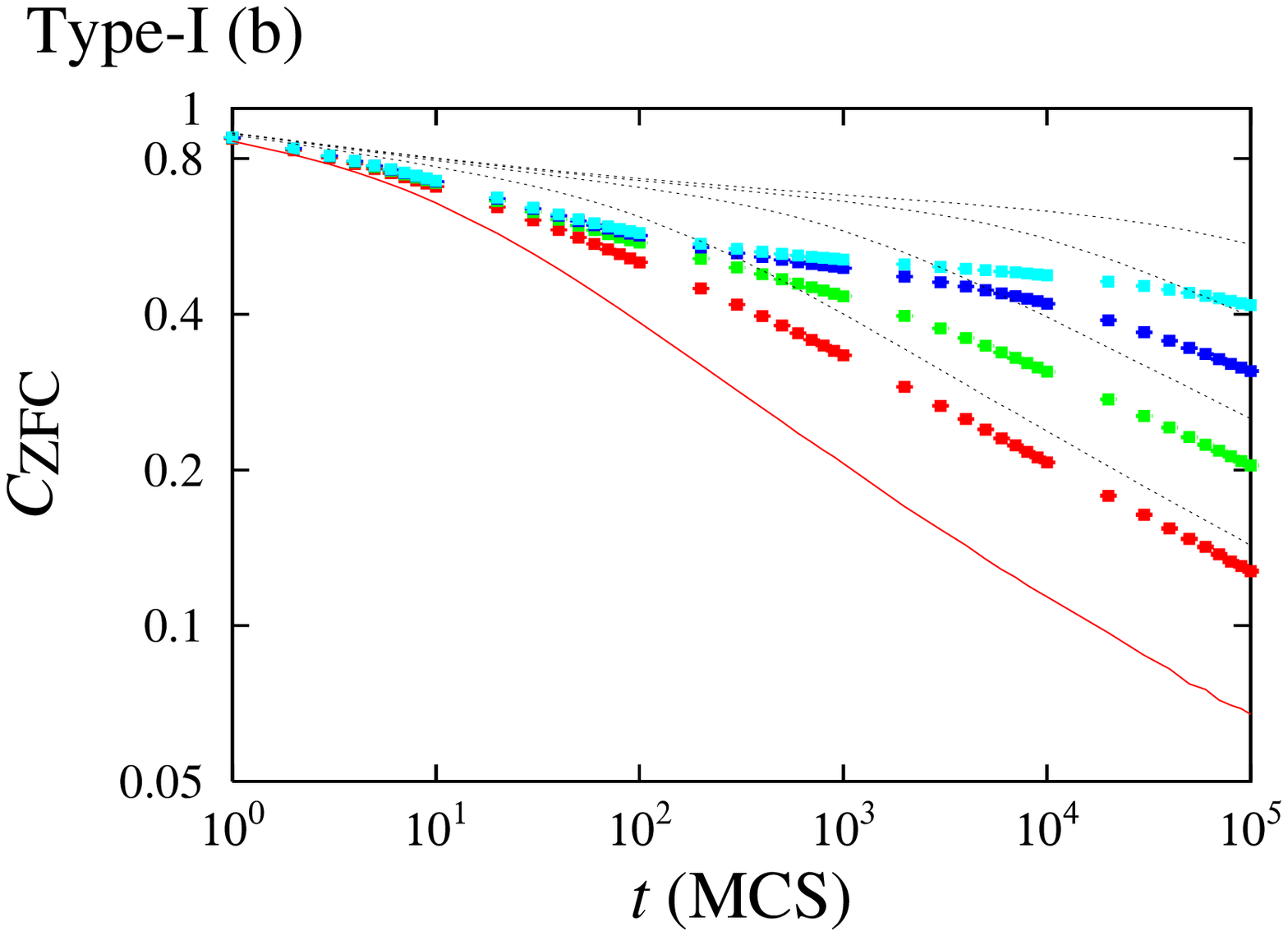}
\includegraphics[width=0.9\columnwidth]{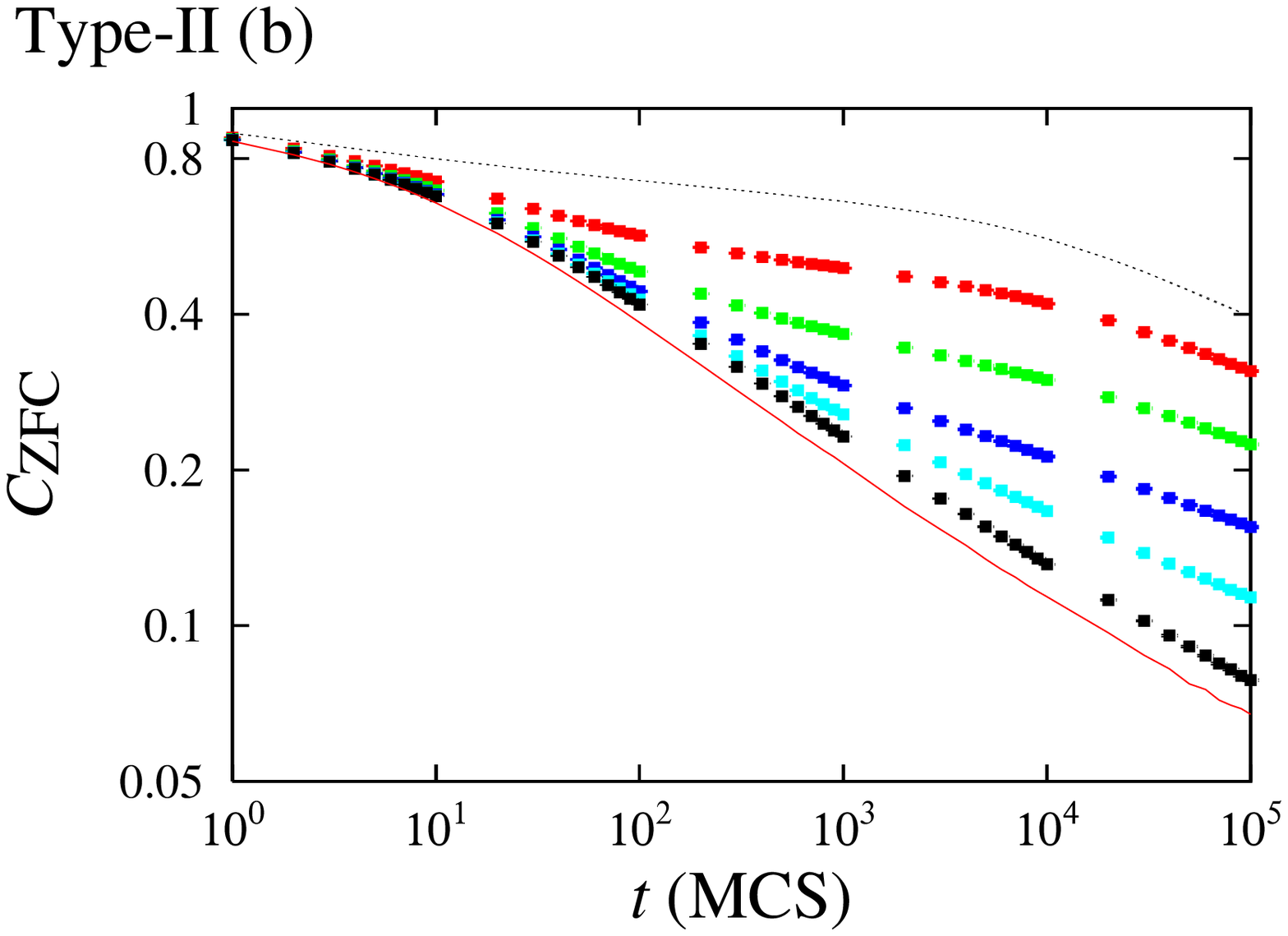}
\caption{(Color online) ZFC susceptibility and the corresponding
auto-correlation function in the 4DEA model after bond-cyclings. 
The filled squared are the data obtained after bond-cyclings with Type I:
$\twone=10^2,10^3,10^4,10^5$ (MCS) ($\chi_{\rm ZFC}$ from top to bottom $C_{\rm ZFC}$ from bottom to top),  $\twtwo=10$,
and  $\twthree=10$ (MCS); Type II: $\twone=10^4$ (MCS),
$\twtwo=10,40,160,640$ and $10^{5}$ (MCS) 
 ($\chi_{\rm ZFC}$ from top to bottom $C_{\rm ZFC}$ from bottom to top),
and
$\twthree=10$ (MCS).
The solid lines are the reference isothermal
aging data (a) $\chi_{0}(t,\tw)$ (b) $C_0(t,\tw)$ with $\tw=\twthree$ and the dotted lines those with 
$\tw=\twone+\twthree$. 
(See Fig. \ref{AgMn_cycl} for the
corresponding temperature-cycling experiment.)}
\label{bond-cycle}
\end{figure*}

In Fig.~\ref{bond-cycle}, we show a data set labeled ``type-I'',
where  the initial wait time $\twone$ is varied while the other time
scales are fixed.
Another data set labeled ``type-II'' is also shown, where the
duration of the perturbation $\twtwo$ is varied while the other time
scales are fixed. The corresponding data in the case of
temperature-cyclings are  shown in Fig.~\ref{AgMn_cycl}. In all data
in Fig.~\ref{bond-cycle}, there is a common feature that the ZFC
susceptibility $\chi_{\rm ZFC}(t)$ and the conjugate auto
correlation function $C_{\rm ZFC}(t)$ exhibit two-step relaxation
processes which can be naturally understood as the inner- and
outer-coarsening discussed in section \ref{sec-theory}. In the figures
the data of the reference isothermal aging without perturbations of
the ZFC susceptibility $\chi_{0}(t,\tw)$ and the autocorrelation
function $C_{0}(t,\tw)$ are shown for comparison.

Quite remarkably, the generic features of the behavior of the ZFC
susceptibility $\chi_{\rm ZFC}(t)$ is the same as in
temperature-cycling experiments. First, there is an initial growth
of the susceptibility which appears to be close to the curve of the
reference isothermal aging $\chi_{0}(t,\tw)$
with wait time $\tw = \twthree$. This feature can be naturally
understood to be due to the inner-coarsening. Then, there is a
crossover to $\chi_{0}(t,\tw)$ with wait time $\tw =
\twone+\twthree$ which can naturally be understood to be due to the
outer-coarsening.

Correspondingly, and as expected, the auto-correlation function
$C_{\rm ZFC}(t)$ exhibits a similar two-step relaxation as
$\chi_{\rm ZFC}(t)$. It initially follows the reference curve
$C_{0}(t,\tw)$ of $\tw=\twthree$ (inner-coarsening) and makes a
crossover to slower decay which depends on $\twone+\twthree$
(outer-coarsening) at later times. This auto-correlation function
was studied in bond-cycling simulations of the spherical and Ising
Mattis model in Ref.~\onlinecite{yoslembou2001} with rejuvenation
put in by hand, where essentially the same result as here was
obtained. Thus, the present results support that the same mechanism
of memory is at work in the spin-glass model.

\subsubsection{Two-step bond-cycling}

For the two-step bond-cycling, we consider that  the procedure
${\cal J}^{A} \to {\cal J}^{B} \to {\cal J}^{C} \to {\cal J}^{A}$
is the same as the one-step bond cycling except that some time
$\twthree$ is spent at ${\cal J}^{C}$ before coming back to ${\cal
J}^{A}$. The bonds of ${\cal J}^{A}$ and ${\cal J}^{B}$ are
prepared as in the one-step cycling case, i.e the set of bonds ${\cal
J}^{B}$ is created from ${\cal J}^{A}$ by a perturbation of
strength $p$. The 3rd set of bonds ${\cal J}^{C}$ is created from
${\cal J}^{B}$ by a perturbation of strength $p$. Note, that this
corresponds to create ${\cal J}^{C}$ out of ${\cal J}^{A}$ by a
perturbation of strength $p'=2p(1-p)$. Here, we again use $p=0.2$
and thus $p'=0.32$ in order to work in the strongly perturbed
regime. The temperature is again fixed to $T/T_{g}=0.6$ throughout
the simulations and the strength of the probing field is
$h/J=0.1$.

\begin{figure}[h]
\includegraphics[width=0.9\columnwidth]{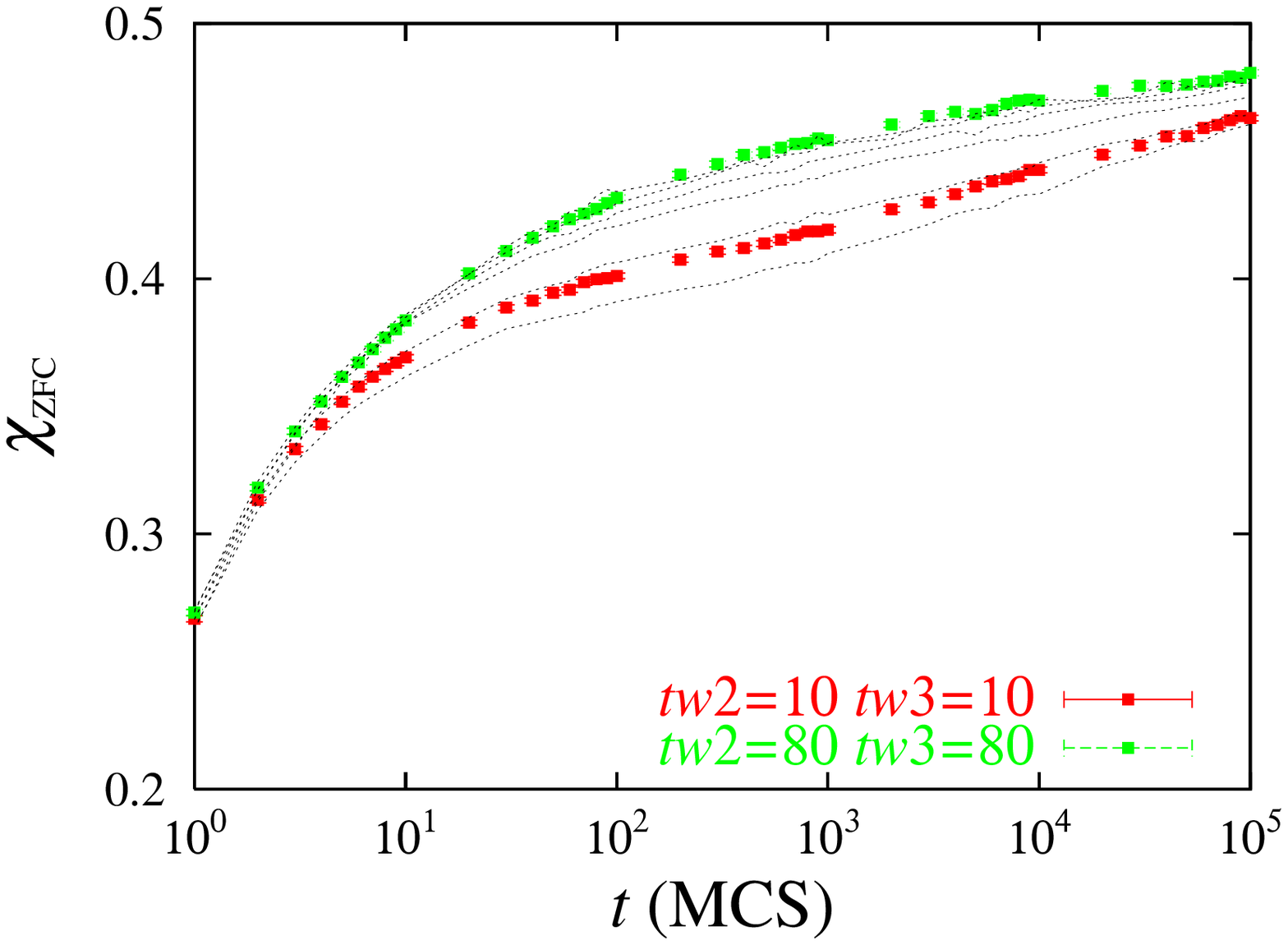}
\includegraphics[width=0.9\columnwidth]{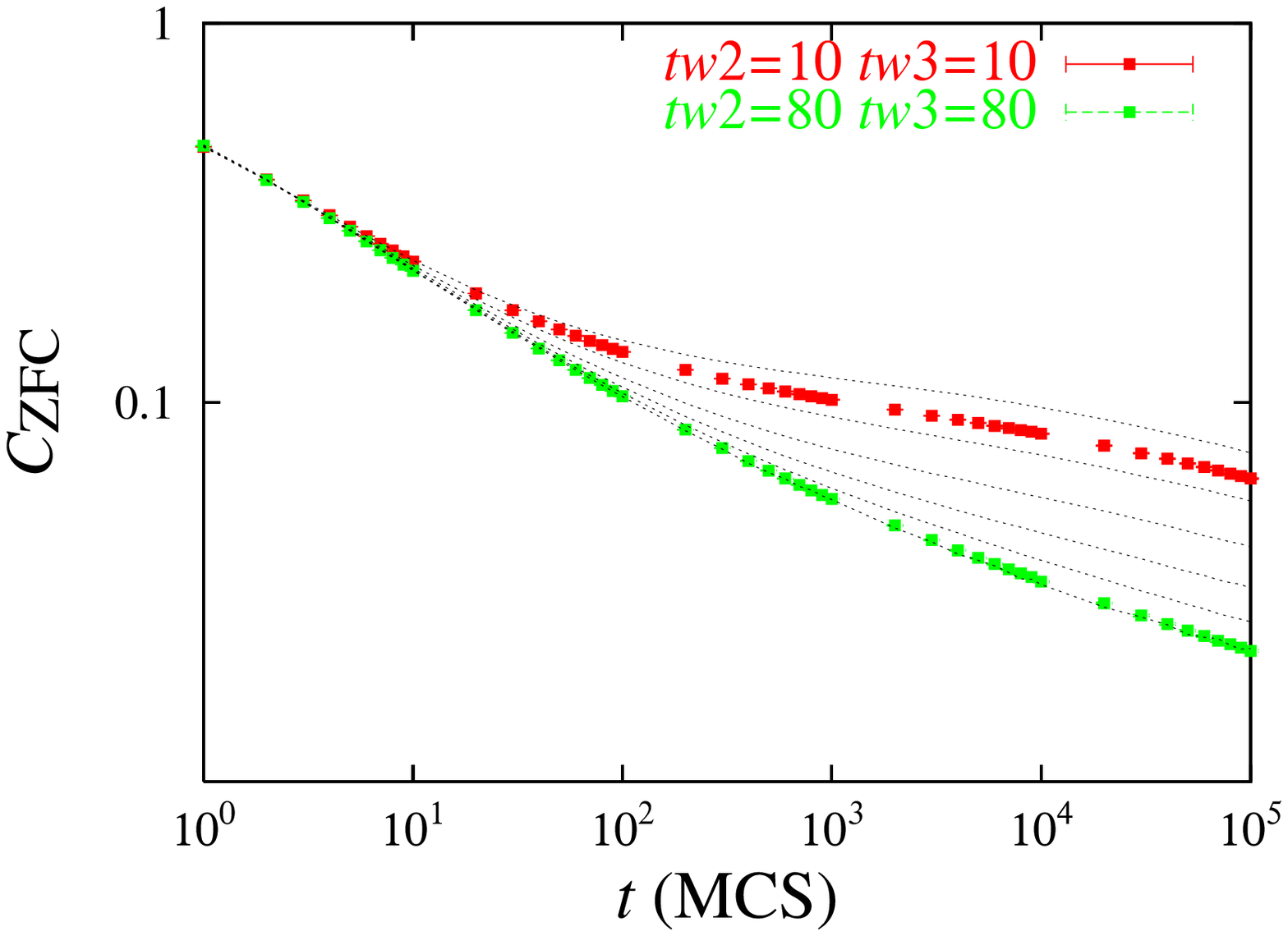}
\caption{(Color online) The relaxation after two-step bond-cycling
${\cal J}^{A} \to {\cal J}^{B} \to {\cal J}^{C} \to {\cal J}^{A}$.
The target equilibrium state is cycled as  ${\cal J}^{A}$ ($10^{4}$
(MCS)) $\to$ ${\cal J}^{B}$ ($\twtwo$ (MCS)) $\to$ ${\cal J}^{C}$
($\twthree$ (MCS))
 $\to$ ${\cal J}^{A}$ ($t$ (MCS)).
In the figures the dotted lines are the reference data of one-step bond-cycling
${\cal J}^{A}$ ($10^{4}$ (MCS))  $\to$ ${\cal J}^{C}$ ($\twtwo$ (MCS))
 $\to$ ${\cal J}^{A}$ ($t$ (MCS)) with  $\twtwo=20,40,80,160,320,640$ (MCS)
which vary from bottom to top in (a) and from top to bottom in (b).
(See Fig. \ref{AgMn_2step-cycl} for the corresponding two-step
temperature-cycling experiment.) \label{bond-two-stepcycle} }
\end{figure}

In Fig.~\ref{bond-two-stepcycle}, we show the ZFC susceptibility
$\chi_{\rm ZFC}(t)$ and the conjugate auto correlation function
$C_{\rm ZFC}(t)$ after the two-step bond cyclings ${\cal J}^{A}
\to {\cal J}^{B} \to {\cal J}^{C} \to {\cal J}^{A}$ which is
compared with the data of one-step bond cyclings ${\cal J}^{A} \to
{\cal J}^{C} \to {\cal J}^{A}$. It can be seen that the effect of
the two step perturbations with $\twtwo=\twthree=10$ (MCS) is
stronger than the one step perturbation with $\twtwo=20$ (MCS).
Moreover, two step perturbations with $\twtwo=\twthree=80$ (MCS)
is as strong as the one step perturbation with $\twtwo=640$ (MCS).

The above result can hardly be understood by a naive length scale
argument. It resembles the result of the corresponding two-step
temperature cycling experiments discussed in
Sec.~\ref{sec_T_cycling} and is consistent with the expectation in
the ghost domain scenario, which predicts multiplicative effects of
perturbations in the strongly perturbed regime (see
\eq{eq-tau-rec-multi}).

\section{Memory experiments}
\label{sec_memory}

This section discusses memory experiments, which probes
nonequilibrium dynamic under continuous temperature changes and
halts, using  low frequency ac susceptibility and ZFC magnetization
measurements.
Such experiments have become a popular tool for investigations of
memory and rejuvenation effects in various glassy systems, for
example interacting nanoparticles,\cite{jonhannor2000} polymer
glasses,\cite{belcillar2002} and granular
superconductors\cite{garetal2003}. It can however be difficult to
interpret the experiments, since memory and rejuvenation effects are
mixed with cooling/heating rate effects in a nontrivial way.

\subsection{ac memory}

\begin{figure}[htb]
\includegraphics[width=0.4\textwidth]{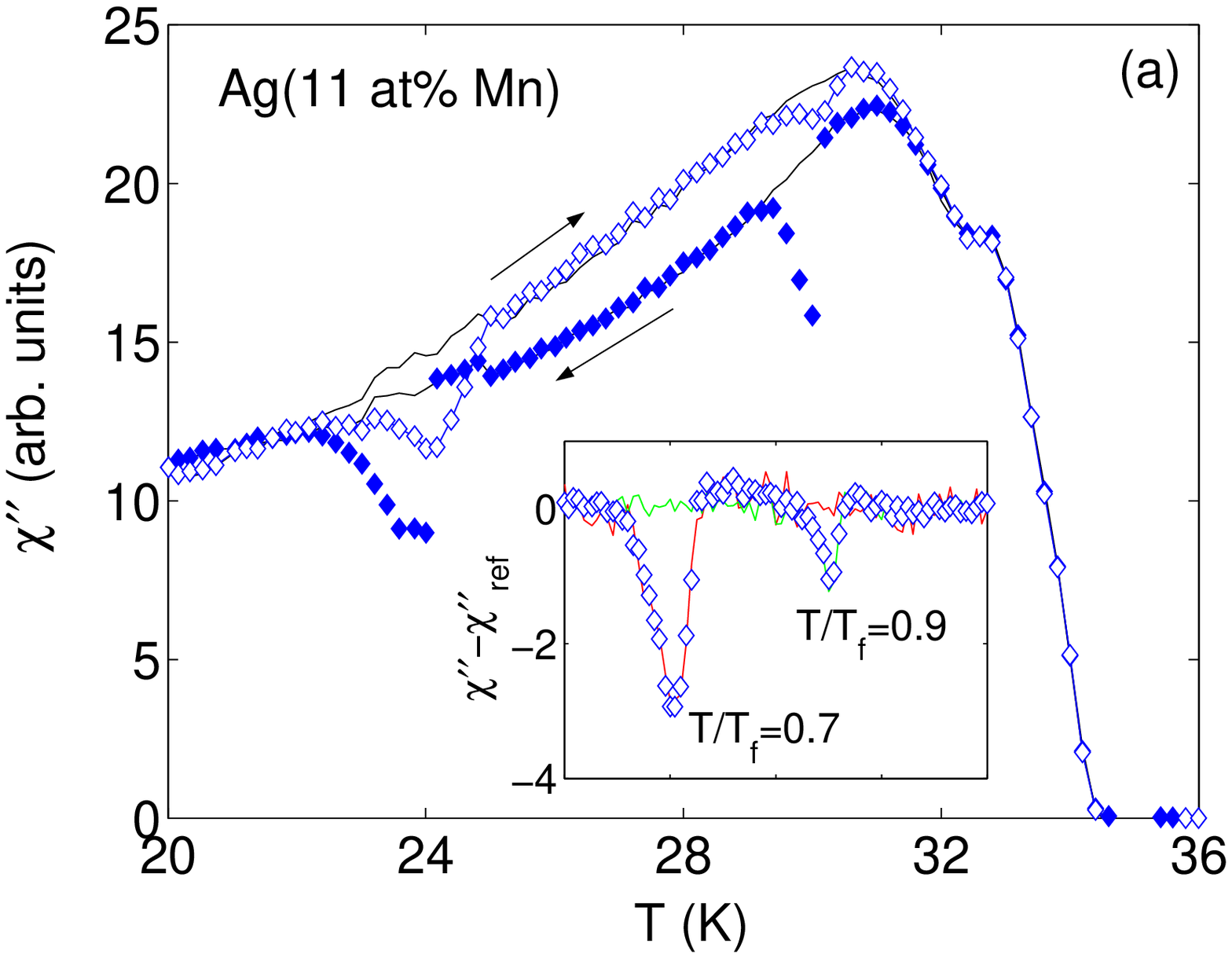}
\includegraphics[width=0.4\textwidth]{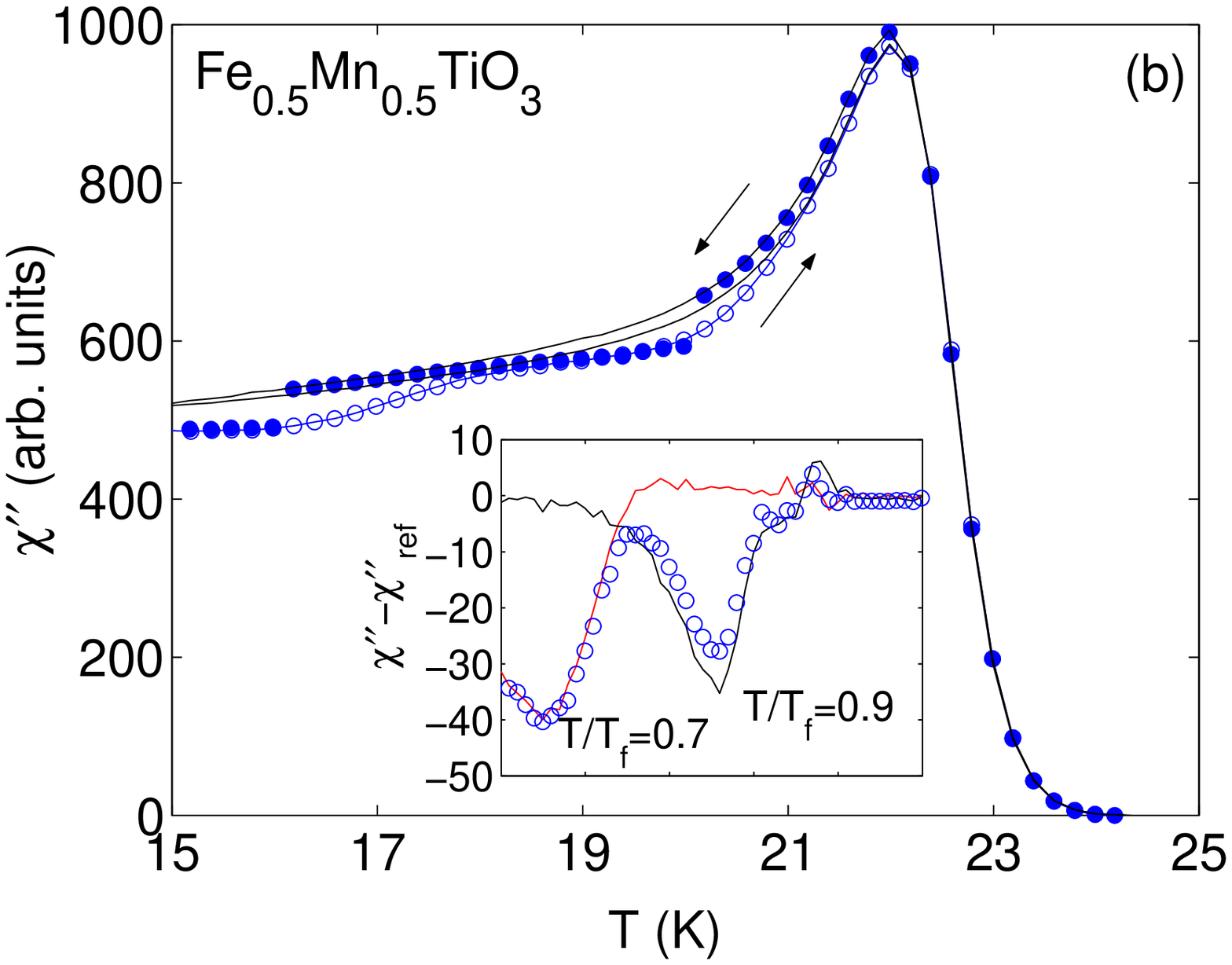}
\caption{(Color online)
$\chi''$ vs temperature with temporary stops at (a) $\Tsun$ = 30~K and $\Tsdeux$ = 24~K and (b) $\Tsun$ = 20~K and $\Tsdeux$ = 16~K; $t_{s_1}=t_{s_2}$ = 3000 s.
These values of $\Ts$ correspond to $T/\Tf = 0.9$ and 0.7 for both samples, where $\Tf$ is the freezing temperature.\cite{foot_Tf}
Filled symbols are used to represent the curves measured
on cooling, and open symbols the curves measured on heating. The
reference curves (without stops) are shown as lines, the arrows
indicate the cooling respective the heating curve. The insets show
the difference plots for double stop experiments (open symbols)
and associated single stop experiments (lines) measured on
heating. $\omega / 2 \pi$ = 510~mHz and $v_{\rm cool}^{\rm eff}
\approx v_{\rm heat}^{\rm eff} \sim 0.005$. }
\label{double_ac_mem}
\end{figure}

Results of ac-susceptibility memory experiments are shown in
Fig.~\ref{double_ac_mem}; the ac susceptibility is measured on
cooling and on the subsequent reheating. Measurements are made both
with and without temporary stop(s) at constant temperature on
cooling. In the figure,  $\chi''(T)$ is plotted vs. temperature for
the two samples. It is interesting to note that even without a
temperature stop, we observe differences between \agmn and \ising
as regard the relative levels of $\chi''(T)$ measured on cooling and
reheating (the "direction" of measurement is indicated with arrows
on the figure). In the case of \agmn, the heating curve lies
significantly above the cooling curve (except close to the lowest
temperature). Such a behavior can only be explained by strong
rejuvenation processes during the cooling and reheating procedure.
In the case of \ising, on the other hand, the susceptibility curve
recorded on reheating lies below the corresponding cooling curve,
indicating that some equilibration of the system is accumulative in
the cooling/reheating process. By making one or two temporary stops
during the cooling processes, memory dips are introduced in the
susceptibility curves on reheating
 as seen in
the main frames and illustrated as difference plots in the insets
of Fig.~\ref{double_ac_mem}. These figures illustrate one more
marked differences in the behavior of the two
samples: the memory dips are wider for the \ising than for the \agmn sample.

\subsection{ZFC memory}

\begin{figure}[hbt]
\includegraphics[width=0.45\textwidth]{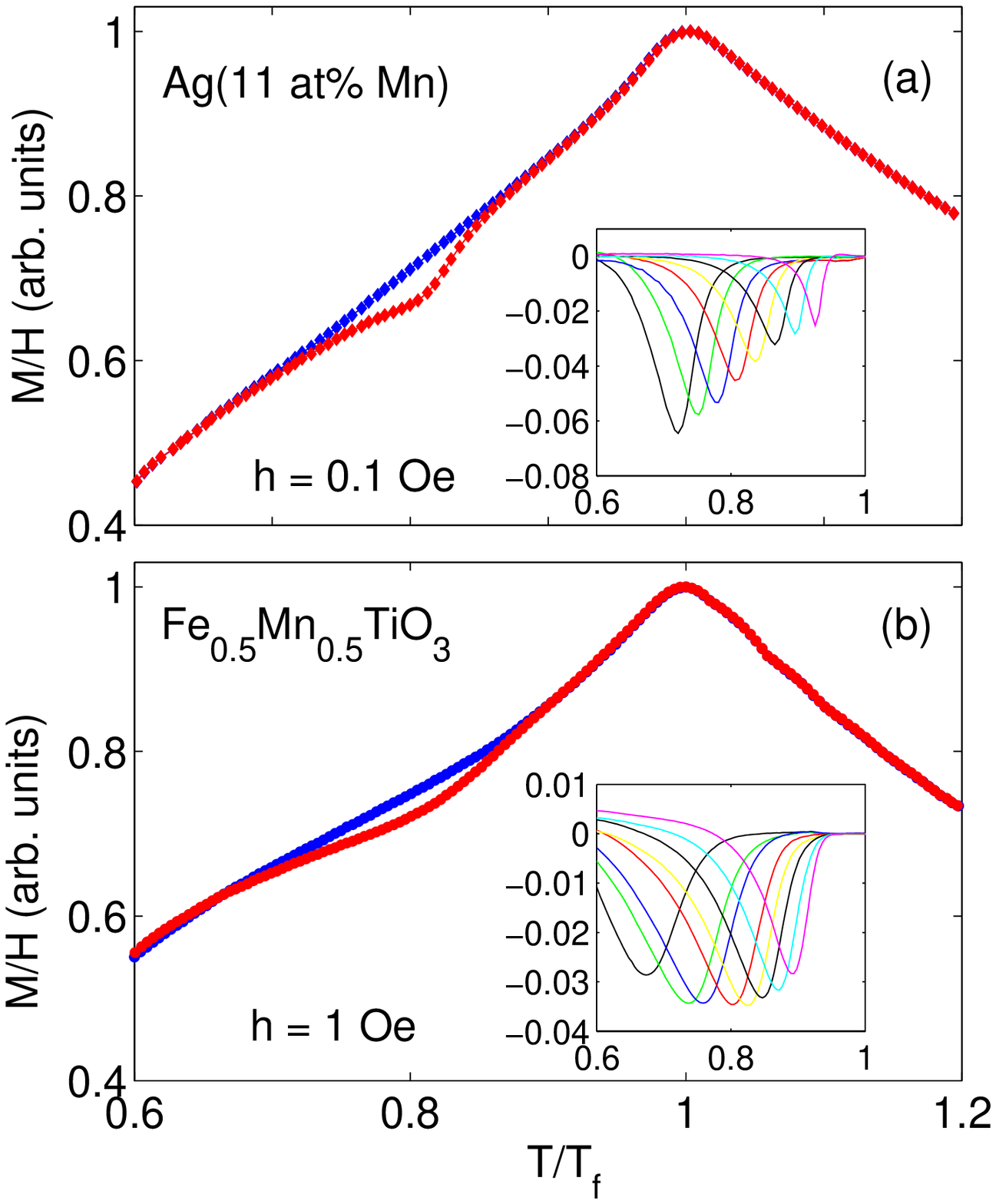}
\caption{(Color online) ZFC susceptibility vs $T/\Tf$ for (a) the \AgMn sample
and (b) the \ising sample. The ZFC susceptibility is measured after
continuous cooling (filled symbols) and cooling with a temporary
stop (open symbols) of 3000 s at $\Ts/\Tf$ $\sim$ 0.8 ($\Ts$ = 27~K for \AgMn and $\Ts$ = 19~K for \ising). The insets show the
corresponding difference plots ("stop curve" - reference curve),
and additional results for 3000 s stops made at (a) $\Ts$ = 24-31~K
and (b) $\Ts$ = 16-21~K. $v_{\rm cool}\sim 0.05$~K/s and
$v_{\rm heat}^{\rm eff} \sim 0.005$. \label{single_dc_mem}}
\end{figure}

ZFC memory experiments \cite{matetal2001,matetal2002} will here be used to further elucidate the differences between the two samples.
Results of single stop ZFC memory experiments for the \AgMn and
\ising spin glasses are shown in Fig.~\ref{single_dc_mem}. In both
cases, the sample is cooled  in zero magnetic field and the
cooling is temporary stopped at $\Ts/\Tf$ = 0.8 for $\ts$ = 3000 s
(main frame).
Here $\Tf$ is the freezing temperature of the ZFC magnetization.\cite{foot_Tf}
The cooling is subsequently resumed and the ZFC
magnetization recorded on reheating in a small magnetic field. A
reference curve measured using the same protocol but without the
stop is also recorded. The difference between corresponding
reference and single memory curves are plotted in the inset,
together with results obtained from similar measurements with
stops at some other temperatures. As seen in the main frames, the
curves corresponding to the single stops lie significantly below
their reference curves in a limited temperature range around
$\Ts/\Tf$. In the associated difference plots, this appears as
"dips" of finite width around the stop temperatures. It has been
argued that such dips directly reflect the memory
phenomenon.\cite{matetal2001}

\begin{figure}[htb]
\includegraphics[width=0.4\textwidth]{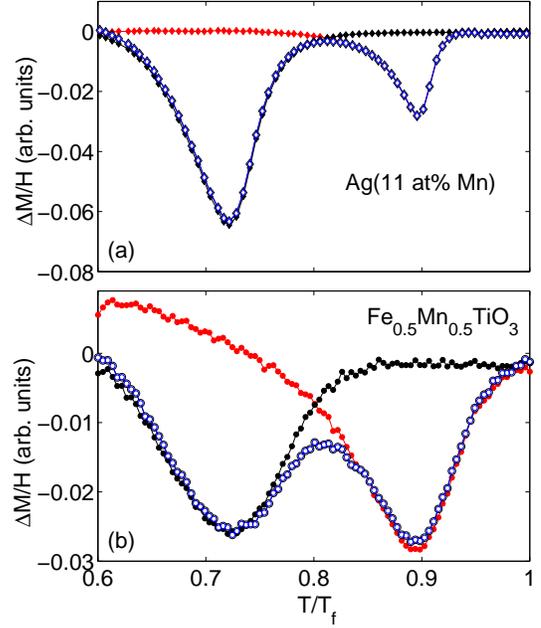}
\caption{(Color online) Difference plots of ZFC susceptibility vs $T/\Tf$
corresponding to single and double stop experiments, with one (or
two) temporary stops for 3000 s at (a) $T_{s_1}$ = 30~K or/and $T_{s_2}$ = 24~K
and (b) $T_{s_1}$ = 21~K or/and $T_{s_2}$ =
17~K. The single stop
experiments are represented using filled markers, while open
markers are used to represent the double stops experiments.
$v_{\rm cool}\sim 0.05$~K/s and  $v_{\rm heat}^{\rm eff} \sim
0.005$.} \label{double_dc_mem}
\end{figure}

The memory dips appear much broader for the \ising sample than for the
\AgMn sample. This can be seen even more clearly in the double stop
dc-memory experiments shown in Fig.~\ref{double_dc_mem}. Two 3000~s
stops are performed at $\Tsun/\Tf$ = 0.9 and $\Tsdeux/\Tf$ =
0.72 during the cooling to the lowest temperature. The results of
the corresponding single stop experiments at $\Tsun$ and $\Tsdeux$
are included. For both systems, the sum of the two single stop
curves (not shown here) is nearly equal to the double stop
curve.\cite{matetal2002} But, while for \agmn the two peaks are
well separated, for the same separation in $T/\Tf$, the single
stop curves of \ising overlap, due to the larger width of the
memory dips.

\subsection{Accumulative vs chaotic processes}
\label{subsec-acc-chaos}

In the ghost domain picture, the size of domains at a given
temperature $T$ can be increased by aging at nearby temperatures
(see Fig.~\ref{fig-leff}). However, noise is induced on the domains
at $T$ by the growth of domains at other temperatures outside the
overlap region (c.f. Sec.~\ref{subsubsec-cycle-strong}). In
particular the spin configuration subjected to continuous
temperature changes with a certain rate $v_{T}$ is expected to
attain a domain size $\Lmin(v_{T},T)$ at temperature $T$, due to the
competition between accumulative aging and  chaotic rejuvenation
processes (as discussed in Sec.~\ref{sec-heating-cooling}).
$\Lmin(v_{T},T)$ is reflected on the magnetic response of the
systems under continuous temperature changes; the larger
$\Lmin(v_{T},T)$ the smaller $\chi''(T)$. In the data of the ac
susceptibility measured under continuous cooling and reheating shown
in Fig.~\ref{double_ac_mem}, we found that for the \agmn sample the
$\chi''(T)$  curve  on reheating lies above the one measured on
cooling, while the opposite applies to the \ising sample. These
features suggest strong chaotic rejuvenation in \agmn and much
weaker in the \ising sample. This is qualitatively consistent with
the observations made in the temperature-shift experiments presented
in Sec.~\ref{sec-T-shift} that the overlap length for the \agmn
decays faster with increasing $\Delta T$ than that of the \ising
system.

\subsection{An apparent hierarchy of temperatures}

\begin{figure}[b]
\includegraphics[width=\columnwidth]{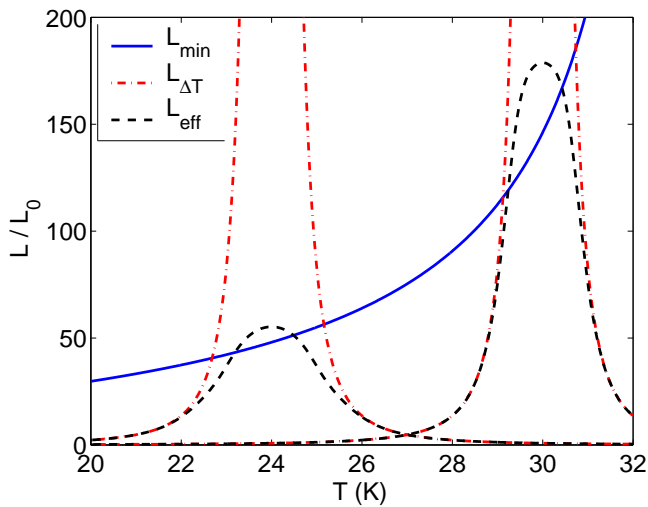}
\caption{
(Color online) A schematic illustration of the length scales
explored in the \agmn sample by continuous cooling (and heating)
with temporary halts at $T_{s1}=30$~K and  $T_{s2}=24$~K for $t_{s1}=t_{s2}=10^4$~s.
The effective domain size grown during the temporary halts at $T_s$ for a time $t_s$ is given by
$\Leff=L_{\Delta T=|T-T_{s}|}  F(L_{T_{s}}(t_s)/L_{\Delta T=|T-T_{s}|})$
[Eq.~(\ref{eq-scaling-leff-1}-\ref{eq-scaling-leff-4})].
It is bounded from above
by the overlap length  $L_{\Delta T=|T-T_{s}|}$.
The scaling function $F(x)=\tanh(x)(1-c x^{2\zeta}/\cosh(x))$
interpolates between the fit to the weakly
perturbed regime
(shown in Fig.~\ref{scale-leff-AgMn}) and the strongly
perturbed regime $\lim_{x \to \infty} F(x)=1$.
Here all parameters for the growth law $L_T(t)$
[see Fig.~\ref{fig-lt}] and the overlap length
$L_{\Delta T}$ are the same as in Sec.~\ref{subsubsec-Tchaos}.
The continuous cooling and heating yields the
minimum effective domain size $\Lmin$ at all temperatures.
Here it is simply assumed to be $L_{\rm min}=L_{T}(\tmin)$ with
$\tmin=100$~s.\cite{foot_lmin}
\label{fig-summary}}
\end{figure}

In a memory experiment, during a stop made at $T_{s}$ for a time $\ts$ the size of the domains  becomes $L_{\Ts}(\ts)$.
The ghost domains at nearby temperatures  will also grow but only up to
$\Leff$ [Eq.~(\ref{eq-scaling-leff-1})], as illustrated in
Fig.~\ref{fig-summary}.
The min($L_{T_s}(t_w),L_{\Delta T=|T_s-T|}$) is the upper bound for $\Leff$.
When the cooling is
resumed, the ac susceptibility curve merges with the reference
curve because  the overlap length becomes smaller than
the minimum domain size $\Lmin$, which depends on the cooling rate.\cite{foot_lmin}
On the subsequent reheating the domain sizes
at the temperatures around $\Ts$ are larger than $\Lmin$
so that the susceptibility curve (ac or dc) measured on heating
makes a ``dip'' with respect to the reference curve without any
stop.

Let us compare the schematic picture of length scales for the \AgMn
sample shown in Fig.~\ref{fig-summary} with the corresponding 2-stop
memory experiment shown  in Fig.~\ref{double_ac_mem} and
\ref{double_dc_mem}. We can see that the widths of the dips around
$T_{s}$ in the memory experiment correspond roughly to the widths of
the temperature-profile of the effective domain size $\Leff$ at
around $T_{s}$. The width of the memory dips becomes broader at
lower temperatures for all memory experiments both for the \AgMn and \ising samples. In the length scale picture it is also seen that the
temperature profile of $\Leff$ around $T_{s}$ becomes broader at
lower temperature, due to the temperature dependence of the growth
law  (see Fig.~\ref{fig-lt}). From this scenario it is clear that a
larger overlap length makes the weak chaos regime larger and hence
the memory dips broader. The observed differences between the width
of the dips of the \ising and \agmn samples may be attributed to
such a difference in overlap length in consistency with the results
shown in Fig.~\ref{fig_tapp}.

The distinct multiple memories shown in Figs.~\ref{double_ac_mem}
and \ref{double_dc_mem} could suggest some sort of ``hierarchy of
temperatures''. However, it is not necessary to invoke neither the
traditional hierarchical phase space picture \cite{vinetal96} nor
the hierarchical length scale picture
\cite{komyostak2000A,bouchaud2000,bouetal2001,berbou2002,berhol2002,takhuk2002,beretal} to understand
this multiple memory effect. Indeed, the domain sizes grown at the
two stop temperatures are almost of the same order of magnitude
since thermally activated dynamics does not allow significant
hierarchy of length scales explored at different temperatures (as
discussed in Appendix~\ref{app-growthlaw}). Thus the scenarios of
the memory effect which strongly relies on the assumption of
hierarchy of length scales \cite{
komyostak2000A,bouchaud2000,bouetal2001,berbou2002,berhol2002,takhuk2002,beretal} cannot explain the distinct
memory dips. Without temperature-chaos, the temperature profile of
$\Leff$ becomes so broad that the memory dips strongly overlap. Our
estimates suggest that the sizes of domains grown at the two stop
temperatures  shown in Fig. \ref{fig-summary} are much larger than
the overlap length between the equilibrium spin configurations at
the two temperatures within the experimental time scales. Thus  the
memories imprinted at the two stop temperatures are significantly
different from each other. Also, retrieval of such memories under
strong temperature-chaos effect is possible within the ghost domain
picture as discussed in Sec~\ref{subsec-theory-cycle}. A crucial
ingredient in experiments is the finiteness of the heating and
cooling rate which yields the characteristic length scale $\Lmin$,
schematically shown in Fig.~\ref{fig-summary}. As discussed in
section Sec. \ref{sec-heating-cooling}, $\Lmin$ plays the role of a
renormalized overlap length which leads to substantial reductions of
the recovery times of memories.

It should however be noted that much broader memory dips are
observed in experimental systems such as superspin glasses, for
which experiments probe short time (length) scales so that the
effect of temperature-chaos is negligible, and the memory dip is
determined by freezing of smaller and smaller domains on cooling in
a  fixed energy landscape\cite{jonetal2004} (see also
Ref.~\onlinecite{bouetal2001}).
Lastly, we note that the width of the memory dips give an indication
of the strength of the temperature-chaos effect, but a better
estimation is given by the twin T-shift experiments discussed in
Sec.~\ref{sec-shift}.

\section{Summary and conclusion}
\label{sec-summary}

Rejuvenation (chaos) and memory effects have been investigated after
temperature perturbations in
two model spin glass samples, the \ising Ising system and the
\AgMn Heisenberg system,
as well as after bond perturbations in the 4 dimensional EA Ising model.
These effects are discussed in terms of the ghost domain picture presented in Sec.~\ref{sec-theory}.

\begin{itemize}

\item
The ZFC relaxation is measured after a $T$-shift for the \ising and
\agmn samples. By analyzing the peak positions of the relaxation
rate $S(t)$ using the twin $T$-shift protocol introduced in Ref.
\onlinecite{jonyosnor2002}, evidences of both accumulative and
non-accumulative aging regimes in both samples were found (cf.
Fig.~\ref{fig_tapp}). The \agmn sample was found to exhibit stronger
deviations from fully accumulative aging with increasing $|\Delta
T|/\Tg$ than the \ising sample. A scaling analysis, performed on the
data of the \agmn sample (Fig.~\ref{scale-leff-AgMn}), reveals
increasing rejuvenation effects with increasing $\Delta T$. These
rejuvenation effects can consistently be understood in terms of
partial chaos at length scales smaller than the overlap length
$\LovlpT$, as first proposed in Ref.~\onlinecite{jonyosnor2003}.
However, data derived using larger perturbations $|\Delta T|> 0.6$~K
could not be used in the scaling analysis because the rejuvenation
effect saturates due to the slow cooling/heating rates. This
suggests the emergence of the strongly perturbed regime in the sense
that $\LovlpT < \Lmin$, where $\Lmin$ is the smallest observable
domain size due to the slow cooling/heating. The overlap length can
never be directly observed in experiments since it is hidden behind
$\Lmin$ as can be seen in the schematic Fig.~\ref{fig-summary}. On
the other hand some properties of the overlap length are obtained
indirectly by the scaling in the weakly perturbed regime
(Fig.~\ref{scale-leff-AgMn}).
\item

In the case of bond-shift simulations, the relaxation rate $S(t)$ is
found to show a similar behavior as in the $T$-shift experiments as
shown in Fig.~\ref{s_4dea.fig}; $S(t)$ is initially broadened with
increasing strength of the perturbation $p$, where also the scaling
analysis presented in Fig.~\ref{scale-lteff-4dea.fig} exhibits the
emergence of corrections to the fully accumulative aging. The
scaling ansatz Eq.~(\ref{eq-scaling-leff-1}-\ref{eq-scaling-leff-3})
was shown to hold with only the chaos exponent $\zeta$ being a
fitting parameter, a result that also gives further support to the
corresponding analysis from $T$-cycling experiments. However, again
the data with stronger perturbation (larger $p$) could not be used
in the scaling because the peak positions saturated to $O(1)$ (MCS),
which is the minimum time scale. In such a regime, $S(t)$ exhibits
considerable narrowing with the peak at $O(1)$ (MCS). This can again
be understood to be due to the emergence of the strongly perturbed
regime in the sense that $\LovlpT \sim  L_{0}$ where $L_{0}$ is the
unit of lattice spacing.

\item

For the \agmn sample temperature changes $|\Delta T| \gtrsim 1$~K
were shown to cause strong perturbations to the system. One and
two-step negative $T$-cycling protocols were used in order to
investigate the healing after such strong perturbations. The
recovery times after one-step cyclings shown in
Figs.~\ref{AgMn_cycl}  and \ref{AgMn-Tcycl-ac} were found to be
anomalously larger than the trivial recovery times $\taurecw$ given
in \eq{eq-tau-rec-weak} (which could be evaluated using the growth
law confirmed in Sec.~\ref{subsubsec-Tchaos}). Furthermore, the
two-step cyclings data displayed in Fig.~\ref{AgMn_2step-cycl} show
that the effects of perturbations made at two different temperatures
can give rise to much stronger rejuvenation than only an additive
increase of the effects of single perturbations. These striking
features are in agreement with ghost domain scenario
(Sec.~\ref{subsec-theory-cycle}).

\item

In the bond-cycling simulations, a large strength of the
perturbation, $p=0.2$, was used to focus on the strongly perturbed
regime. The healing process after one-step cycling simulations
displayed in Figs.~\ref{bond-cycle} and
and the two-step bond-cycling shown in Fig.~\ref{bond-two-stepcycle} were found to require anomalously large
recovery times, in consistency with the results from $T$-cycling
experiments and the predictions of the ghost domain scenario. It
is also demonstrated in Fig.~\ref{bond-two-stepcycle} that a
``memory'' of the history before the perturbation cannot be erased
completely even if the duration of the perturbation is effectively
longer than the duration of the preceding history.

\item It is proposed that the effects of finite heating/cooling
rates $v_{T}$ can be parameterized by a renormalized overlap length
$\Lmin(v_{T},T)$. In combination with a strong separation of time
scales, this can lead to an apparent  ``hierarchy of temperatures'',
as has been suggested earlier from experimental observations e.g. in
Refs.  \onlinecite{ledetal91,lefetal92}. In the ghost-domain
scenario however, the temperature chaos effect is the mechanism
which allows distinctness of memories at each level of the
``hierarchy'' as illustrated  in Fig. \ref{fig-summary}. This
explains the two well separated memory dips seen in e.g. the memory
experiment shown in  Fig.~\ref{double_ac_mem}.

\end{itemize}

To conclude, dynamical properties of the Ising \ising and the Heisenberg \agmn
spin glasses subjected to temperature changes and the 4 dimensional
Edwards Anderson Ising spin glass model subjected to bond changes
are examined and compared with each other in detail. The
temperature-shift/cycling experiments and bond-shift/cycling
simulations show remarkably similar features suggesting a common
mechanism of the rejuvenation and memory effects. The detailed
features of the rejuvenation effects after temperature/bond shifts
are found to agree with the anticipated crossover from weakly to
strongly perturbed regimes of the chaos effects. Quantitative
differences in different systems can be attributed to differences
of the magnitude of the overlap length. Anomalously large recovery
times of the memories are found to be required in the one and
two-step temperature/bond cycling  experiments when performed into
the strongly perturbed regime, in agreement with the ghost domain
scenario.

\acknowledgments We thank  Jean-Philippe Bouchaud, Ian Campbell,
Koji Hukushima, Hikaru Kawamura, Philipp Maas, Falk Scheffler,
Tetsuya Sato, Hajime Takayama  and Eric Vincent for stimulating
discussions. This work was financially supported by the Swedish
Research Council (VR). H.Y. is supported by the Ministry of
Education, Culture, Sports, Science and Technology of Japan,
Grant-in-Aid for Scientific Research 14740233.

\appendix

\section{Separation of length and time scale}
\label{app-growthlaw}

\begin{figure}[htb]
\includegraphics[width=0.9\columnwidth]{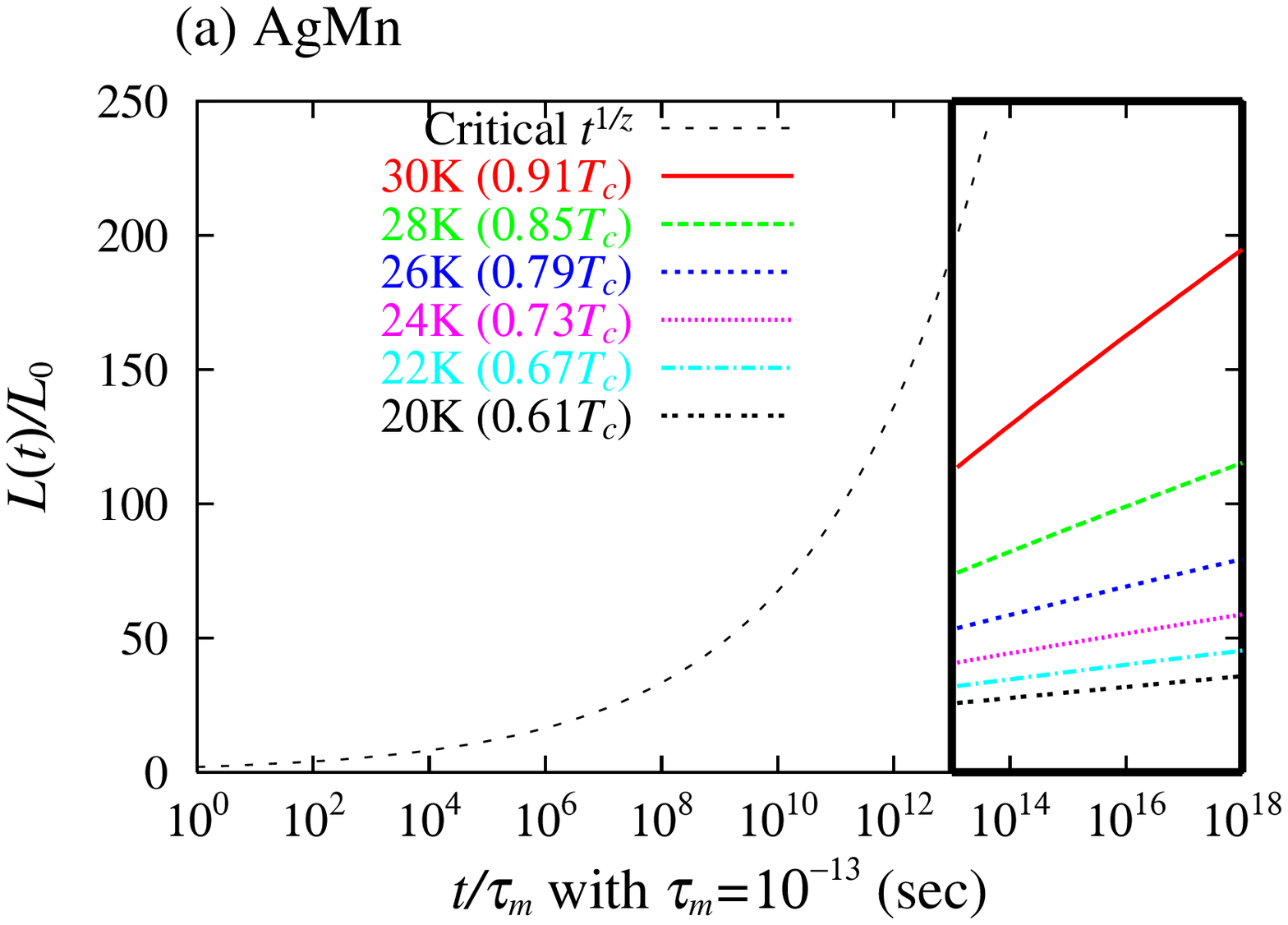}
\includegraphics[width=0.9\columnwidth]{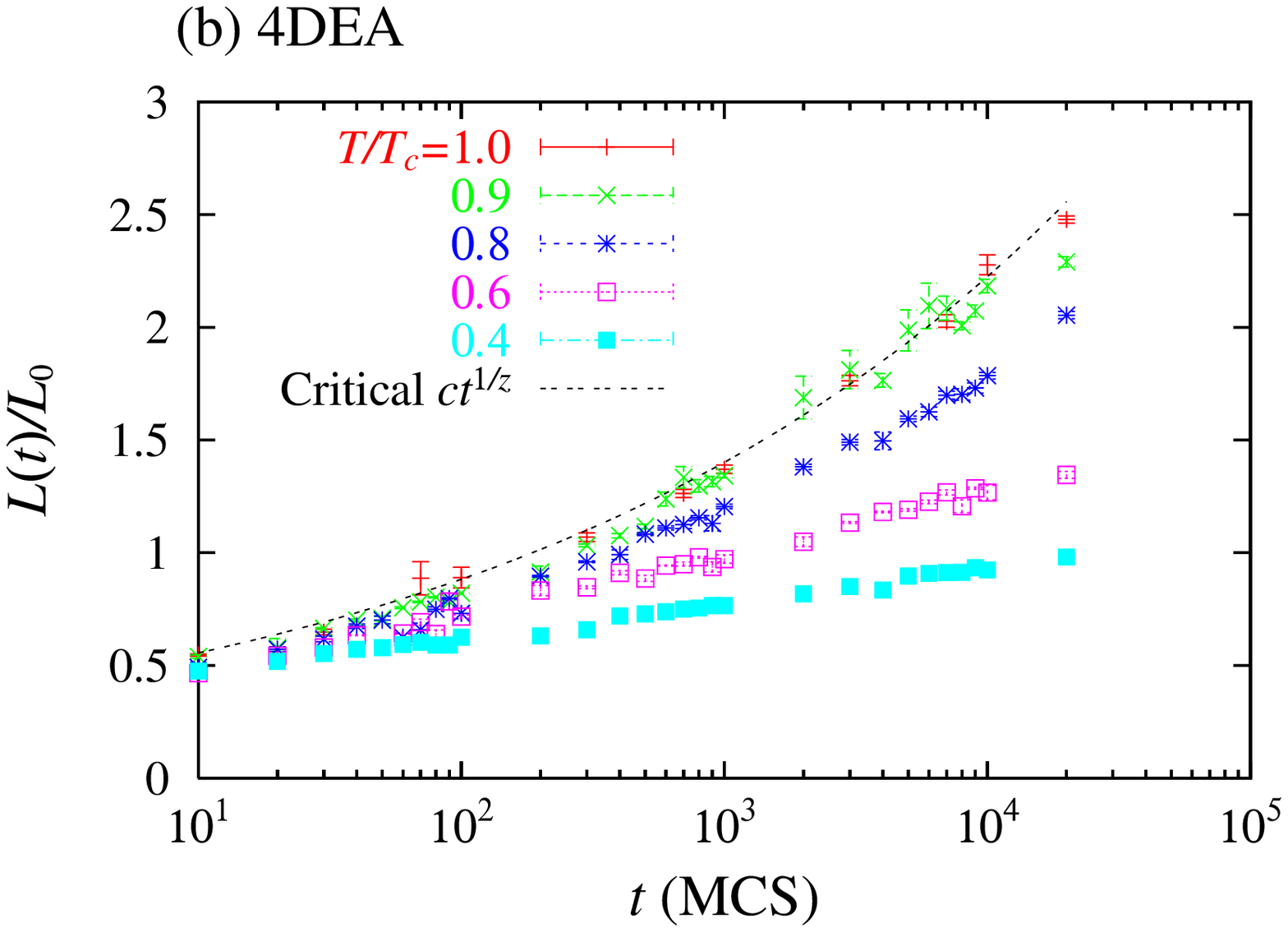}
\caption{(Color online) The dynamical length scales relevant in our experiments
and simulations. (a) A plot of the growth law \eq{eq-growth} with
parameters determined for a \AgMn sample in
Ref.~\onlinecite{jonetal2002PRL} from a  scaling analysis of  the
relaxation of the ac susceptibility in isothermal aging in the
range $20-30$ K. The growth law at the critical temperature
$L(t)/\Lo = (t/\tau_{m})^{1/z}$ ($z=6.5$) is also shown. (The
possible $O(1)$ prefactor in front of \eq{eq-growth} and the
critical growth law is not known.) The unit of time is chosen to
be  $\tau_{\rm m} \sim \hbar/J \sim 10^{-13}$~s. The experimental
time-length scale is the area surrounded by the box. (b) The size
of the domains of a 4 dimensional EA Ising spin glass model
determined directly in a previous numerical simulation
\cite{hukyostak2000}  monitoring the spatial correlation function
between two real replicas undergoing isothermal aging. The dotted
line is the power law for the critical regime $L(t)/\Lo =
(t/\tau_{m})^{1/z}$ with $z=4.98$.}
\label{fig-lt}
\end{figure}

In this appendix we summarize important properties of activated dynamics that governs
the growth of domains in spin glasses.
At mesoscopic levels, the dynamics of a spin glass at temperatures below $T_{g}$ is governed by
nucleation of droplet like excitations.\cite{fishus88noneq}
The energy barrier associated with a droplet of size $L$ is assumed to
scale as $E_{\rm b} \sim \Delta (T)(L/\Lo)^{\psi}$ with $\psi > 0$.
Then the time needed to nucleate a droplet by thermal activation becomes
\begin{equation}
\tau_{L}(T)=\tau_{0}(T)\exp\left[ \frac{\Delta (T)}{\kb T}
\left( \frac{L}{\Lo}\right)^{\psi} \right]
\label{eq-tau-activated}
\end{equation}
where the effects of critical fluctuations are taken into account
in a renormalized way in the characteristic energy scale
$\Delta(T)$ for the free-energy barrier and the characteristic
time scale $\tau_{0}(T)$ for the thermally activated processes.
They scale as $\Delta(T)/J\sim \epsilon^{\psi\nu}$ and
$\tau_{0}(T)/\tau_{\rm m} \sim  (\xi(T)/L_{0})^{z} \sim
|\epsilon|^{-z\nu}$ with  $\epsilon = T/T_g - 1$. The microscopic
time scale is $\tau_{\rm m} \sim \hbar/J \sim 10^{-13}$ s in atomic spin
glasses and  one  Monte Carlo Step (MCS) in Monte Carlo simulations.
$J \sim T_g$ sets the energy unit, $z$ is the dynamical critical
exponent and $\nu$ the exponent for the divergence of the
correlation length. The importance of accounting for critical
fluctuations has been realized in analyzes of recent simulations
\cite{hukyostak2000,berbou2002,yoshuktak2002} and
experiments.\cite{bouetal2001,jonetal2002PRL}

From \eq{eq-tau-activated} it follows that within a time scale $t$
a droplet excitation as large as
\begin{equation}
L_{T}(t)
\sim L_{0}\left[\frac{\kb T}{\Delta(T)}
\ln \left(\frac{t}{\tau_{0}(T)}\right) \right]^{1/\psi} \,
\label{eq-growth}
\end{equation}
can be thermally activated. At finite time scales there must be
corrections to the asymptotic form \eq{eq-growth} due to critical
fluctuations \cite{bouetal2001,yoshuktak2002} as well as
logarithmic correction terms to the energy barriers
\cite{mikdrokar95} but the actual forms of the corrections are not
yet known in spin-glasses.

An immediate consequence of activated dynamics is an extremely wide
separation of time scales with temperature. On the other hand the
variation of length scales with temperature is very mild (typically
less than one order in magnitude). Let us consider two arbitrarily
close temperatures $T$ and $T+\Delta T$. From \eq{eq-tau-activated}
it follows that the ratio of time scales at the two temperatures
associated with a given length $L$ for  $\Delta T/T \ll 1$ becomes,
\begin{equation}
\frac{\tau_{L}(T)}{\tau_{L}(T+\Delta T)} \sim
\frac{\tau_{0}(T)}{\tau_{0}(T+\Delta T)} \exp \left[\left(
\frac{L}{L_{*}(\Delta T)}\right)^{\psi} \right]
\label{eq-separation-time}
\end{equation}
where we introduced {\it time-separation length}
\begin{equation}
L_{*}(\Delta ) \sim \Lo  \left |
(\partial (\Delta (T)/\kb T)/\partial T) \Delta T \right  |
^{-1/\psi}. \label{eq-l-separation-time}
\end{equation}
The above expression means that two times become {\it
exponentially} separated if one observes dynamics on length scales
large enough compared with the time-separation length
$L_{*}(\Delta T)$. Note that $L_{*}(\Delta T)$ remains finite as
long as $\Delta T$ is non-zero. Looking at the ratio of length
scales for a given time scale $t$ at the two temperatures one
finds from \eq{eq-growth},
\begin{equation}
\frac{L_{T+\Delta T}(t)}{L_{T}(t)} \sim O(1)
\label{eq-separation-length}
\end{equation}
independently of the chosen time scale $t$. Thus, the magnitude of
the length scales explored at two close temperatures appears quite
similar as long as the same time scales are used. However, we
emphasize, that thermally activated dynamics brings about a
significant hierarchy of time scales, in spite of an apparent mild
hierarchy in length scales.

Assuming that \eq{eq-growth} is applicable to the Heisenberg like
\AgMn spin glass, Fig.~\ref{fig-lt}(a) shows the domain growth at
different temperatures in this system using earlier derived
experimental values for the parameters. \cite{jonetal2002PRL} In
Fig.~\ref{fig-lt}(b) the domain growth at different temperatures
is drawn for a 4d EA spin glass derived from previous numerical
simulations. \cite{hukyostak2000}

In spin glass materials the microscopic time is $\tau_{\rm m} \sim
10^{-13}$~s and relaxation experiments are performed typically in
the range of time scales $1-10^5$~s or $t/\tau_{\rm
m}=10^{13}-10^{18}$. On the other hand, in usual Monte Carlo
simulations of EA Ising spin glass models the microscopic time is
$\tau_{\rm m}=1$ Monte Carlo Step (MCS) and the range of time
scales  is $1-10^6$ (MCS) or $t/\tau_{\rm m}=1-10^{6}$. Because of
the slow dynamics, there is a common problem that the investigated
4-6 decades in time only cover a quite limited length scale as
shown in Fig.~\ref{fig-lt} (see also Fig 1 of
Ref.~\onlinecite{berbou2002}). Due to the much shorter time scales
probed in numerical simulations, the influence of critical
dynamics are much stronger than in experiments for the same
reduced temperature $T/\Tg$. It can also be seen in
Fig.~\ref{fig-lt} that the length scales are rather different:
simulations explore length scales of order $L_{T}(t)/L_{0} \sim
O(1)$ while experiments explore $L_{T}(t)/L_{0} \sim
O(10)-O(10^{2})$.

\section{Experimental details}
\label{app-exp}

\subsection{Non-commercial SQUID magnetometer}

The experiments were performed in non-commercial SQUID
magnetometers,\cite{magetal97} designed for low-field measurements
and optimum temperature control. The magnetic field is in these
set-ups generated by a small superconducting solenoid always working
in persistent mode. The time constant of the superconducting magnets
is of order 1 ms, but the whole procedure to change the field from
$h$ to $h+\Delta h$ and re-establish persistent mode in the magnet
takes $\sim$ 0.3 s which also determines the shortest observation
time in the ZFC relaxation experiments. In the case of the \ising
sample, cut into a rectangular shape ($2 \times 2 \times 4$ mm$^3$)
with the c-axis along the longest side, the probing field was
applied along the $c$-axis of the single crystal, and for the \agmn
spin-glass, turned into a cylindrical shape (4 mm long and 2 mm in
diameter), the field was applied along the symmetry axis of the
cylinder. The sample space is magnetically shielded with cans of
$\mu$-metal and niobium and the remaining background field is $<
1$~mOe, while the measurement field used are $h \approx 0.1 - 1$ Oe.
The sample is glued to a 50 mm long sapphire rod with a diameter of
2 mm, and the magnetization is recorded while keeping the sample
stationary in the center of the third coil of a third order
gradiometer. The sapphire rod is connected to a solid copper
cylinder on which the thermometer and a resistive heater are placed.
The materials in the experimental setup are chosen so that the heat
transfer between the sample and copper cylinder is maximized while
the heat exchange with the surroundings is kept quite low being
mainly carried by thermal conduction through the sample holder rod,
a thin walled stainless steel tube, via Cu(Be) holding springs into
the surrounding helium bath. The temperature can be kept fixed with
a stability better than 100 $\rm \mu K$ using an ac-bridge based
temperature control system and with a stability better than 5 mK
using a commercial temperature controller. The maximum cooling rate
in these magnetometers is $v_{\rm cool} \sim 0.05$~K/s in the
temperature ranges of these studies and, to maintain a good
temperature control, the maximum heating rate is limited to $v_{\rm
heat} \sim 0.5$~K/s. The maximum heating and cooling rates were
always employed when changing the temperature in the current
experiments. In ac measurements, a coil generating the ac field is
wound around the sample and a similar compensation coil is wound on
the sample rod above the sample and positioned so that it becomes
centered in the second pick-up coil of the gradiometer.

When measuring  both the ac and dc susceptibility as a function of
temperature, the magnetization is recorded on heating by elevating
the temperature in steps of 0.1 - 0.2 K. The measurement of the
magnetization at constant temperature (including temperature setting
and stabilization, and actual measurement) takes $\sim$ 30 - 60 s,
yielding an effective heating rate of $v_T^{\rm eff} \sim
0.005$~K/s. The magnetization is also recorded on cooling in ac
experiments yielding a cooling rate similar to the heating rate,
while the sample is cooled using the maximum cooling rate in dc
experiments.

Since the same thermal protocols are employed in both
dc-magnetization and ac-susceptibility measurements,  the same
nonequilibrium processes are probed in both methods, with the
difference that the ac susceptibility can be measured at all times
and temperatures during the thermal protocols, while the zero field
cooled (ZFC) magnetization can only be recorded as a function of
temperature (resp. time) after applying the probing field at the
lowest temperature (resp. the measurement temperature). The magnetic
fields employed in the experiments are small enough to ensure a
linear response of the system, and do hence not affect the
nonequilibrium dynamics.

Some complementary magnetization measurements were performed using a
Quantum Design MPMS5 SQUID magnetometer in order to check the
absolute susceptibility values.

\subsection{Effective age of a spin glass system measured by ZFC relaxation}
\label{app-teff}

\begin{figure}[htb]
\includegraphics[width=0.9\columnwidth]{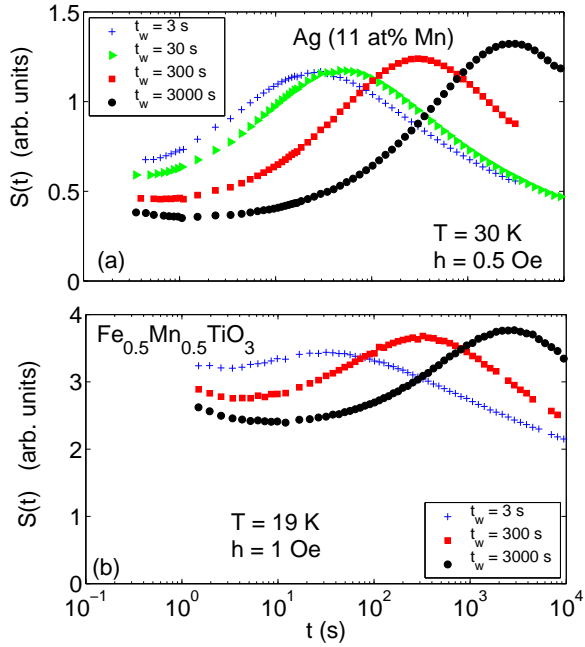}
\caption{(Color online) $S(t)$ of ZFC relaxation measurements
employing different wait times on (a) \agmn and (b) \ising. The
reduced temperature is 0.9 for both samples.} \label{iso}
\end{figure}

The relaxation rate $S(t)$ determined from isothermal ZFC relaxation
experiments is shown in Fig.~\ref{iso}. The sample is quenched
(cooled with the maximum cooling rate) from above $\Tg$ to the
measurement temperature $\Tm$. For both samples $S(t)$ exhibits a
peak at $\tpeak \approx \tw$ except for the shortest waiting time
$\tw=\tws=3$~s (which is the time needed in order to stabilize the
temperature\cite{footstab}). Due to the slow cooling, the system has
already been aged for a  certain effective time $\tmin$ when
reaching $\Tm$.\cite{rodkenorb2003} From the measurements with
$\tw=\tws$ we can determine $\tmin = \tpeak \approx 20-40$~s. The
effective age of the system after a $T$-shift is determined from the
time $\tpeak$ corresponding to the maximum of $S(t)$, as
$\teff(\tw+\tmin) = (\tpeak )^{1/\mu}$ with $\mu =1$ for the \AgMn
sample at $T\approx 30$~K and $\mu \approx 0.98-0.99$ at $T=17-19$~K
for the \ising sample.\cite{foot_mu} We notice that $\tmin$ can only
be determined from isothermal ZFC relaxation measurements, but not
from the corresponding ac relaxation measurement.

\begin{figure}[htb]
  \includegraphics[width=0.9\columnwidth]{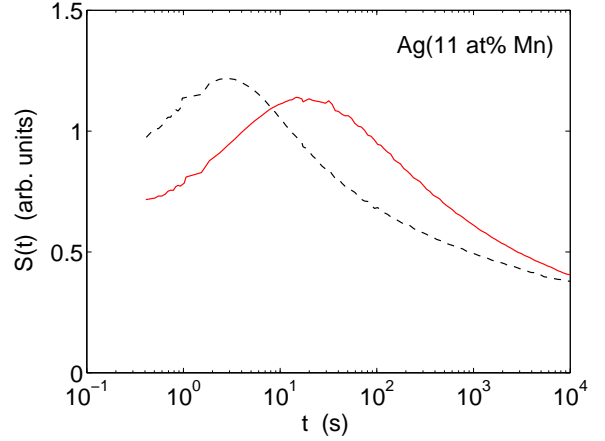}
\caption{(Color online) $S(t)$ curves obtained for \agmn after a rapid cooling
( $v_{\rm cool}\sim 0.05$~K/s)
to $\Tm= 30$~K (solid line) and after a rapid cooling to 29~K, where the
 sample was kept for 3000~s and subsequently heated rapidly ( $v_{\rm heat}\sim 0.5$~K/s) to $\Tm= 30$~K (dashed line).}
\label{AgMn_Smin}
\end{figure}

In Fig.~\ref{AgMn_Smin}, $S(t)$ is shown after a temperature quench
($v_{\rm cool} \sim 0.05$~K/s) to $\Tm$ and after a $T$-shift to a
temperature $\Ti=29$~K where it is aged for $3000$~s before the
temperature is rapidly raised ($v_{\rm heat} \sim 0.5$~K/s) to
$\Tm$. The maximum of the $S(t)$ curve of the $T$-shift experiment
appears at  $\tmin \sim 3$~s  which is considerably shorter time
than $\tmin \sim 30$~s in the case of quench. The fact that $\tmin$
is smaller after a positive $T$-shift than after a direct quench
implies the system is completely dominated by the strong chaos
effect. Namely, $|\Delta T|$ is so large that the intrinsic overlap
length $\LovlpT < \Lmin$, where $\Lmin$ is the domain size
corresponding to the effective age $\tmin$. We cannot evaluate the
value of $\LovlpT$ itself since it is masked by $\Lmin$ (on
heating). $\tmin$ ($\Lmin$) is smaller on heating than on cooling
due to the faster heating than cooling rate and the temperature
dependence of the domain growth (c.f. Appendix~\ref{app-growthlaw}).
The $\Lmin$ corresponding to $\tmin$ determined here gives an upper
limit for the renormalized overlap length $\Lmin(v_T,T)$ introduced
in Sec.~\ref{sec-heating-cooling}, since the time needed to
stabilize the temperature at $\Tm$ also contributes to $\tmin$
($\Lmin$). It is also possible to obtain a random initial
configuration with a small $\tmin$  by using a magnetic field making
a ``field quench''.\cite{noretal87} Throughout this article, we have
however only used the ordinary temperature quench to initialize the
system.

\section{Simulational Details}
\label{app-simulation}

The Monte Carlo simulations were performed on the 4 dimensional EA
Ising model whose critical temperature is $T_{g}/J=2.0$
\cite{bercam97}. The simulations were made at the temperatures
$T/J=1.2, 0.8$ ($T/T_{g}=0.6, 0.4$), which are well below the
critical temperature so that numerical time/length scales are {\it
not} strongly affected by critical fluctuations. Systems with
$24^{4}$ spins were used, which are large enough to get rid of
finite size effects within the numerical time
window.\cite{yoshuktak2002} The simulations were done using the
standard heat-bath Monte Carlo method up to a time scale of $10^5$
Monte Carlo Steps (MCS) starting with random spin configurations in
order to mimic aging after ideal temperature quenches. The average
over at least $32$ different realizations of random bonds were
taken. For the measurements of the ZFC susceptibilities $320$
samples were used.

The advantage of the 4 dimensional model compared with the 3
dimensional EA Ising model, which might appear slightly more
realistic, is that a rich amount of knowledge about the equilibrium
and dynamical properties of the model has been accumulated by
independent studies. This includes the critical properties
associated with the fixed point at $T_{g}$,\cite{bercam97} the $T=0$
glassy fixed point \cite{hukushima99,hartmann99} and the dynamical
length $L_{T}(t)$ obtained in
Ref.~\onlinecite{hukyostak2000,yoshuktak2002}. We note that a recent
Monte Carlo simulation \cite{yoshuktak2002} on the same model
confirmed, to a certain extent, certain predictions from the
fundamental dynamical scaling ansatz due to the droplet theory
\cite{bramoo87,fishus88eq,fishus88noneq} (including that for the
decay of TRM (thermo remanent magnetization) which are all expressed
in terms of the dynamical length scale $L_{T}(t)$. It is also worth
to mention that in equilibrium Monte Carlo simulations of the model
the anticipated signatures of the chaos effect have been observed
both in the case of temperature changes \cite{hukiba2003} (exchange
Monte Carlo method), and bond perturbation.\cite{ney98}

Unfortunately within the available computational power it is
difficult to observe even some weak traces of rejuvenation by
temperature changes by the standard heat-bath dynamical Monte Carlo
simulations.
\cite{rieger94,komyostak2000A,picricrit2001,berbou2002,takhuk2002,berbou2003}
Within feasible CPU times, one is forced to work at temperatures
rather close to $T_{g}$, such as $0.8-0.9 T_{g}$. In this
temperature range, the dynamics in the numerical time window is
strongly affected by critical fluctuations,
\cite{hukyostak2000,yoshuktak2002,berbou2002,jonyosnor2003} while
the experimental time window lies well outside the critical regime
\cite{jonyosnor2003} (see Fig. \ref{fig-lt}). Thus we limit
ourselves with bond-shift and cycling simulations.


\begin{thebibliography}{10}
\bibitem{foot_tpeak_tw}
The values of $\tpeak$ are obtained for $\tw=10,30,100,300,\ldots,10^{5}$ (MCS)
  by simulating isothermal aging. Some data are shown in
  Fig.~\ref{s_4dea.fig}(a)). Then $\tw$ is viewed as a function of $\tpeak$,
  and it was fitted to a polynomial of $\log(\tpeak)$ and the resulting smooth
  fitting function was considerd as the model function of $\tw=\teff(\tpeak)$.
\bibitem{foot_trec}
Factors that can affect the estimation of $t_{\rm rec}$ are the noise in the
  ac-signal and the time needed to stabilize the temperature (and the
  ac-signal) after the cycling.
\bibitem{foot_Tf}
In ac experiments, it is possible to define a frequency dependent freezing
  temperature $\Tf$ from a rather sharp maximum in the in-phase component of
  the susceptibility. Below this temperature, the longest relaxation time of
  the system exceeds the observation time ($t_{obs} = 1/\omega$) employed and a
  finite out phase component of the susceptibility appears. $\Tf$ approaches
  $\Tg$ for low frequencies, and increases with increasing frequency. In low
  field dc magnetization measurements, a corresponding maximum is observed in
  the temperature dependence of the ZFC magnetization. Such an experiment looks
  at the system on a certain observation time, that by comparison to
  ac-susceptibility results can be estimated to 10-100 s. By this analogy, we
  denote this temperature $\Tf$, which was found in our experiments to be equal
  to 23.6 K for \ising and 33.5 K for \agmn. In the dc memory experiments, the
  stop temperatures ($\Ts$) were chosen at similar $\Ts/\Tf$ values for both
  systems.
\bibitem{foot_lmin}
In Fig.~\ref{fig-summary}, $\Lmin$ for the continuous cooling/heating protocol
  is for simplicity given as $L_T(100\,{\rm s})$. This is based on the
  observation that in an isothermal aging experiments after temperature
  quenches down to $\Tm$ we find the effective times $\teff \approx 30$~s (see
  Fig.~\ref{iso} and \ref{fig-S2-AgMn}). In a memory experiments the
  cooling/heating rate is lower than the maximum one and we therefore expect
  $\Lmin$ to be a bit larger. The exact value of $\Lmin$ are different on
  cooling and heating as discussed in Sec.~\ref{subsec-acc-chaos}.
\bibitem{foot_mu}
The values of $\mu$ were determined from isothermal ZFC relaxation experiments,
  where $\tpeak=(\tw+\tmin)^\mu$.
\end{thebibliography}
\end{document}